\documentclass{LMCS}

\usepackage{amssymb}  \usepackage{amstext} \usepackage{amsmath}
\usepackage{latexsym} \usepackage{xspace}  \usepackage{hyperref}

\title{Adventures in Time and Space } 
\author[N.~Danner]{Norman Danner\rsuper a}
\address{{\lsuper{a}}Department of Mathematics and Computer Science, 
           Wesleyan University, Middletown, CT 06459 USA}
\email{ndanner@wesleyan.edu}
\author[J.~S.~Royer]{James S. Royer\rsuper b}
\address{{\lsuper b}Department of Electrical Engineering and Computer Science, 
         Syracuse University, 
         Syracuse, NY 13210 USA}	
\email{royer@ecs.syr.edu} 

\theoremstyle{plain}

\newtheorem{theorem}{Theorem}
\newtheorem{corollary}[theorem]{Corollary}
\newtheorem{lemma}[theorem]{Lemma}
\newtheorem{proposition}[theorem]{Proposition}

\theoremstyle{definition}

\newtheorem{definition}[theorem]{Definition}
\newtheorem{remark}[theorem]{Remark}

\newtheorem{scholium}[theorem]{Scholium}

\newtheorem{example}[theorem]{Example}

\newcommand{\sqr}[2]{{\vcenter{\hrule height.#2pt 
	\hbox{\vrule width.#2pt height#1pt \kern#1pt
	   \vrule width.#2pt}
	\hrule height.#2pt}}}

\newcommand{\ProoF}[1]{\smallskip\noindent{\it Proof #1.}}

\newcommand{\topic}[1]{\medskip\noindent{\textsc{#1.{}}}\enspace}

\newcommand{\newfontobj}[2]{
  \newcommand{#1}[1]{
    \expandafter\def\csname##1\endcsname{{#2 ##1}}}}

\newcommand{\newthingie}[2]{
  \newcommand{#1}[1]{
    \expandafter\def\csname##1\endcsname{{#2 ##1}}}}

\newfontobj{\class}{\rm}
\newfontobj{\category}{\bf}
\class{P}
\class{PF}

\class{NP}
\class{coNP}
\class{Mon}
\class{TCont}
\class{TMon}
\class{BCont}
\class{Len}
\class{TLen}
\newcommand{\BFF}{\mathrm{BFF}}

\newcommand{\Strut}[1]{\rule{0pt}{#1}}

\renewcommand{\gets}{\ensuremath{\mathrel{\colon=}}\xspace}

\newcommand{\preimg}{{\rm preimage}}
\newcommand{\set}[1]{\{\,#1\,\}}
\newcommand{\Set}[1]{\left\{\,#1\,\right\}}
\newcommand{\pair}[1]{\mathopen{\langle}#1\mathclose{\rangle}}

\newcommand{\clen}[1]{\langle\kern-2.5pt| #1 |\kern-2pt\rangle_{r}}

\newcommand{\forallae}{\mathord{\stackrel{\kern 0.1em\infty}{\forall}}}

\newcommand{\entails}{\vdash}

\renewcommand{\phi}{\varphi}

\newcommand{\suchthat}{\mathrel{\,\stackrel{\rule{0.03em}{0.5ex}}%
{\rule[-.1ex]{0.03em}{0.5ex}}\,}}

\newcommand{\bo}{\mathbf{0}}
\newcommand{\bl}{\mathbf{1}}

\newcommand{\dash}{\hbox{-}}

\newcommand{\defeq}{\mathrel{\stackrel{\text{def}}{=}}}

\newcommand{\sqdot}{\rule{0.5mm}{0.5mm}}
\newcommand{\lam}[1]{\lambda #1\,\sqdot\,}

\newcommand{\POLY}{{\cal P}\kern-2pt{}oly}

\newcommand{\SF}{\mathcal{F}}

\newcommand{\SP}{\mathcal{P}}

\newcommand{\SR}{\mathcal{R}}

\newcommand{\SV}{\mathcal{V}}
\newcommand{\SU}{\mathcal{U}}

\newcommand{\bll}{\mathopen{\ll\kern -11.5pt\ll}}
\newcommand{\bgg}{\mathclose{\gg\kern -11.5pt\gg}}

\newcommand{\basic}[1]{\mathopen{\langle\kern -3pt
	\langle}#1\mathclose{\rangle\kern -3pt \rangle}}

\newcommand{\of}{\colon}

\newcommand{\Quad}[1]{\hspace*{#1em}}

\newcommand{\refeq}[1]{(\ref{#1})}

\newcommand{\Nat}{{\rm N}}

\renewcommand{\today}{\number\day \space \ifcase\month\or
	January\or February\or March\or April\or May\or June\or
	July\or August\or September\or October\or November\or
	December\fi \space \number\year}

\newcommand{\bb}{\mathbf{b}}

\newcommand{\bp}{\mathbf{p}}
\newcommand{\bq}{\mathbf{q}}
\newcommand{\br}{\mathbf{r}}

\newcommand{\calS}{\mathcal{S}}

\newcommand{\calF}{\mathcal{F}}

\newfontobj{\fnct}{\it}
\newfontobj{\syst}{\sf}
\newfontobj{\vari}{\it}
\newfontobj{\vars}{\sf}

\fnct{iter}
\fnct{destr}
\fnct{next}
\fnct{init}

\syst{HTP}
\syst{DILL}
\newcommand{\PCF}{\ensuremath{\mathsf{PCF}}\xspace}

\newcommand{\textsfb}[1]{{\rm\bf\textsf{#1}}}

\newcommand{\Iif}{\textsfb{If}\xspace}
\newcommand{\Ithen}{\textsfb{then}\xspace}
\newcommand{\Ielse}{\textsfb{else}\xspace}

\newcommand{\is}{\mathrel{\colon\colon=}}

\newcommand{\irule}[2]{\ensuremath{{\frac{\strut\textstyle #1}%
  {\strut\textstyle #2}}}} 

\newcommand{\depth}{\textit{depth}}
\newcommand{\semantics}[1]{\ensuremath{[\![ #1 ]\!]}}

\newcommand{\tally}[1]{\ensuremath{\underline{#1}}}

\newcommand{\introrule}{\textit{-I}\xspace}

\newcommand{\figrule}{\par\vspace{1ex}\hrule\par}

\newcommand{\ol}[1]{\overline{#1}}
\newcommand{\nat}{\mathbb{N}}

\renewcommand{\PF}{\mathcal{P\kern-2pt F}}
\newcommand{\concat}{\mathop{\oplus}}
\newcommand{\synsep}{\;\;|\;\;}
\newcommand{\level}{\textit{level}}
\newcommand{\Tally}{\mathsf{T}}
\renewcommand{\Nat}{\mathsf{N}}
\newcommand{\Verb}[1]{\;\;#1\;\;}
\newcommand{\unbounded}{\dagger}

\newcommand{\conso}{{\sf c}_{\bo}}
\newcommand{\consl}{{\sf c}_{\bl}}
\newcommand{\consa}{{\sf c}_{\mathbf{a}}}
\newcommand{\tsto}{{\sf t}_{\bo}}
\newcommand{\tstl}{{\sf t}_{\bl}}

\newcommand{\tsta}{{\sf t}_{\ba}}
\newcommand{\cdr}{{\sf d}}
\newcommand{\op}{{\rm \textsfb{op}}}
\newcommand{\suc}{\mathop{{\sf s}}}
\newcommand{\Rep}{{\mathsf{R}}}
\newcommand{\bop}{\mathsf{op}}
\newcommand{\dn}{\mathsf{dn}}
\newcommand{\down}{\mathsf{down}}
\newcommand{\fix}{\mathsf{fix}}
\newcommand{\prn}{\ensuremath{\mathsf{prn}}\xspace}
\newcommand{\crec}{\mathsf{crec}}

\newcommand{\Iin}{\mathop{\textsf{in}}\xspace}

\renewcommand{\Iif}{\mathop{\mathsf{if}}\xspace}
\renewcommand{\Ithen}{\mathop{\mathsf{then}}\xspace}
\renewcommand{\Ielse}{\mathop{\mathsf{else}}\xspace}
\newcommand{\Ilet}{\mathop{\mathsf{let}}\xspace}
\newcommand{\Iletrec}{\mathop{\mathsf{letrec}}\xspace}
\newcommand{\emptycont}{\underline{\ }}
\newcommand{\emptyenv}{\ensuremath{\{\}}}
\newcommand{\rulelabel}[1]{\hbox{\it #1:}\Quad1}
\newcommand{\sidecond}[1]{\Quad{0.5} \left(#1\right)}

\newcommand{\state}[1]{\left(\strut#1\right)}

\newcommand{\tcapprox}[1]{\mathrel{\sqsubseteq^{\mathrm{tc}}_{#1}}}
\newcommand{\potapprox}[1]{\mathrel{\sqsubseteq^{\mathrm{pot}}_{#1}}}

\newcommand{\cekeval}{\mathsf{eval}_{\text{CEK}}}
\newcommand{\cekcost}{\mathrm{cost_{\text{CEK}}}}
\newcommand{\cektime}{\mathrm{CEK\dash{}time}}

\newcommand{\val}{\mathsf{val}}
\newcommand{\Cost}{\textit{cost}}
\newcommand{\Pot}{\textit{pot}}
\newcommand{\basepot}{\mathsf{Pot}}
\newcommand{\basecost}{\mathsf{Cost}}
\newcommand{\costprj}{\textit{cost}}
\newcommand{\potprj}{\textit{pot}}
\newcommand{\evalsto}{\downarrow}

\newcommand{\fun}{\mathsf{fun}\xspace}
\renewcommand{\arg}{\mathsf{arg}\xspace}
\newcommand{\halt}{\mathsf{halt}\xspace}
\newcommand{\test}{\mathsf{test}\xspace}
\newcommand{\currytime}{\Lambda_{\star}}
\newcommand{\aptime}{\mathbin{\star}}

\newcommand{\TC}{\mathbf{TC}}
\newcommand{\MC}{\mathbf{MC}}

\newcommand{\subty}{\mathrel{\leq\colon}}
\newcommand{\supty}{\mathrel{\colon\!\geq}}
\newcommand{\suptyneq}{\mathrel{\colon\!\gneq}}
\newcommand{\subtyneq}{\mathrel{\lneq\colon}}

\newcommand{\shape}{\textit{shape}}
\renewcommand{\depth}{\textit{depth}}
\newcommand{\side}{\textit{side}}
\newcommand{\undo}{\textit{undo}}
\newcommand{\shiftsto}{\mathrel{\propto}}
\newcommand{\dally}{\textit{dally}}
\renewcommand{\colon}{\mathpunct{:}}
\newcommand{\longvec}[1]{\overrightarrow{\Strut{1.2ex}#1}}

\newcommand{\lollipop}{\multimap}

\newcommand{\Lennwf}{\mathcal{L}_{\rm nwf}}
\newcommand{\Lenwt}{\mathcal{L}_{\rm wt}}
\newcommand{\semlen}[1]{\mathcal{L}\semantics{#1}}
\newcommand{\eqnwf}{\mathrel{=_{{\rm nwf}}}}
\newcommand{\eqwt}{\mathrel{=_{{\rm wt}}}}
\newcommand{\leqnwf}{\mathrel{\leq_{{\rm nwf}}}}

\newcommand{\leqwt}{\mathrel{\leq_{{\rm wt}}}}
\newcommand{\lwtleq}{\leq_{{\rm wt}}}
\newcommand{\semlenwt}[1]{\mathcal{L}_{\rm wt}\semantics{#1}\xspace}

\newcommand{\Valwt}{\mathcal{V}_{\rm wt}}
\newcommand{\semval}[1]{\mathcal{V}\semantics{#1}}

\newcommand{\semvalnwf}[1]{\mathcal{V}_{\rm nwf}\semantics{#1}}
\newcommand{\semlennwf}[1]{\mathcal{L}_{\rm nwf}\semantics{#1}}
\newcommand{\semvalwt}[1]{\mathcal{V}_{\rm wt}\semantics{#1}}
\let\lwtequiv\eqwt

\newcommand{\Time}{\mathcal{T}}
\newcommand{\semtime}[1]{\mathcal{T}\semantics{#1}}
\newcommand{\sempot}[1]{\mathcal{P}\semantics{#1}}

\newcommand{\semtimeval}[1]{\mathcal{T}_{\text{val}}\semantics{#1}}

\newcommand{\semgen}[1]{\calS\semantics{#1}}
\newcommand{\lamr}[1]{\lambda_r #1\,\sqdot\,}
\newcommand{\ba}{\textbf{a}}
\newcommand{\uses}{\textit{uses}}
\newcommand{\tail}{\textit{tail}}
\newcommand{\bmax}{\mathbin{\vee}}
\newcommand{\bigmax}{\mathop{\bigvee}}

\newcommand{\hatp}{\widehat{p}}

\newcommand{\hatGamma}{\widehat{\Gamma}}

\newcommand{\barrho}{\ol{\rho}}
\newcommand{\bart}{\ol{t}}

\newcommand{\baru}{\ol{u}}
\newcommand{\barw}{\ol{w}}
\newcommand{\barr}{\ol{r}}

\newcommand{\hatq}{\widehat{q}}
\newcommand{\hatr}{\widehat{r}}

\newcommand{\hatt}{\widehat{t}}

\newcommand{\barGamma}{\ol{\Gamma}}
\newcommand{\gr}{\ensuremath{\mathsf{GR}}}
\newcommand{\atr}{\ensuremath{\mathsf{ATR}}\xspace}

\newcommand{\len}[1]{\left|#1\right|}

\newcommand{\tcx}[1]{\left\Vert#1\right\Vert}
\newcommand{\pot}[1]{\langle\!\langle#1\rangle\!\rangle}

\newcommand{\SL}{\mathcal{L}}

\newcommand{\SLwt}{\mathcal{L}_{{\rm wt}}}
\newcommand{\ST}{\mathcal{T}}

\newcommand{\bcl}{\ensuremath{\mathsf{BCL}\xspace}}
\newcommand{\NatNorm}{\Nat_{\text{\rm norm}}}
\newcommand{\NatSafe}{\Nat_{\text{\rm safe}}}

\newcommand{\Prg}{{\mathord{\diamond}}}
\newcommand{\Orl}{{\mathord{\Box}}}

\newcommand{\Natl}[1]{\ensuremath{\Nat_{#1}}}
\newcommand{\Tallyl}[1]{\ensuremath{\Tally_{#1}}}

\newcommand{\TailPos}{\textit{TailPos}}

\newcounter{foo}
\renewcommand{\thefoo}{\alph{foo}}
\newcommand{\foocnt}{\refstepcounter{foo}(\thefoo)}
\newcommand{\successor}{\mathop{\textit{succ}}}
\newcommand{\lendy}{\textit{len}}
\newcommand{\BM}{\mathbf{M}}

\def\doi{3 (1:9) 2007}
\lmcsheading%
{\doi}
{1--53}
{}
{}
{Aug.~\phantom{0}7, 2006}
{Mar.~13, 2007}
{}   

\begin{document}
\nocite{Reynolds72} 
\bibliographystyle{amsalpha}


\begin{abstract}
  This paper investigates what is essentially a call-by-value
  version of \PCF under a complexity-theoretically motivated type
  system.  The programming formalism, \atr, has its first-order
  programs characterize the polynomial-time computable functions,
  and its second-order programs characterize the type-2 basic
  feasible functionals of Mehlhorn and of Cook and Urquhart.  (The
  \atr-types are confined to levels 0, 1, and 2.)  The type system
  comes in two parts, one that primarily restricts the sizes of
  values of expressions and a second that primarily restricts the
  time required to evaluate expressions.  The size-restricted part
  is motivated by Bellantoni and Cook's and Leivant's implicit
  characterizations of polynomial-time.  The time-restricting part
  is an affine version of Barber and Plotkin's DILL.  Two semantics
  are constructed for \atr.  The first is a pruning of the
  na\"{\i}ve denotational semantics for \atr.  This pruning removes
  certain functions that cause otherwise feasible forms of recursion
  to go wrong.  The second semantics is a model for \atr's time
  complexity relative to a certain abstract machine.  This model
  provides a setting for complexity recurrences arising from \atr
  recursions, the solutions of which yield second-order polynomial
  time bounds.  The time-complexity semantics is also shown to be
  sound relative to the costs of interpretation on the abstract
  machine.
\end{abstract}
\maketitle

\section{Introduction}\label{S:intro}

\begin{quote}
  \it A Lisp programmer knows the value of everything, but the cost
  of nothing.\\ \hbox{} \hfill --- \rm Alan Perlis
\end{quote}
\bigskip
Perlis' quip is an overstatement---but not by much.  Programmers in
functional (and object-oriented) languages have few tools for
reasoning about the efficiency of their programs.  Almost all tools
from traditional analysis of algorithms are targeted toward roughly
the first-order fragment of C.  What tools there are from formal
methods are interesting, but piecemeal and preliminary.

This paper is an effort to fill in part of the puzzle of how to
reason about the efficiency of programs that involve higher types.
Our approach is, roughly, to take $\PCF$ and its conventional
denotational semantics \cite{Plotkin:PCF,winskel:book} and, using
types, restrict the language and its semantics to obtain a
higher-type ``feasible fragment'' of both $\PCF$ and the $\PCF$
computable functions.  Our notion of higher-type feasibility is
based on the \emph{basic feasible functionals} (BFFs)
\cite{CookUrquhart:feasConstrArith,Mehlhorn76}, a higher-type
analogue of polynomial-time computability, and Kapron and Cook's
\cite{KapronCook:mach} machine-based characterization of the
type-level 2 BFFs.\footnote{Mehlhorn \cite{Mehlhorn76} originally
  discovered the class of type-2 BFFs in the mid-1970s.  Later Cook
  and Urquhart \cite{CookUrquhart:feasConstrArith} independently
  discovered this class and extended it to all finite types over the
  full set-theoretic hierarchy.  \textbf{N.B.} If one restricts
  attention to continuous models, then starting at type-level 3
  there are alternative notions of ``higher-type polynomial-time''
  \cite{IKR:II}.  Dealing with type-level 3 and above involves some
  knotty semantic and complexity-theoretic issues beyond the scope
  of this paper, hence our restriction of $\atr$ types to orders 2
  and below.}  Using a higher-type notion of computational
complexity as the basis of our work provides a connection to the
basic notions and tools of traditional analysis of algorithms (and
their lifts to higher types).  Using types to enforce feasibility
constraints on $\PCF$ provides a connection to much of the central
work in formal methods.

Our approach is in contrast to the work of 
\cite{BNS,Hofmann03,LeivantMarion93} which also involves 
higher-type languages and types that
guarantee feasibility.  Those programming formalisms  are 
feasible in the sense that they
have polynomial-time normalization properties and that
the type-level 1 functions expressible by these systems 
are guaranteed to be (ordinary) polynomial-time computable.
The higher-type constructions of these formalisms are 
essentially aides for type-level 1 polynomial-time programming.
As of this writing, there is scant analysis of what higher-type
functions these systems compute.\footnote{The work
  of \cite{BNS,Hofmann03}  and of this paper sit on different
  sides of an important divide in higher-type computability
  between notions of \emph{computation over computable data}
  (e.g., \cite{BNS,Hofmann03,LeivantMarion93})
  and notions of \emph{computation over continuous data}
  (e.g., this paper)
  \cite{longley:ubiq,Longley:notions:1}.}
      
For a simple example of a feasible higher-type function, 
consider $C \of (\Nat\to\Nat) \to (\Nat\to\Nat) \to 
(\Nat\to\Nat)$ with $C\, f\, g = f \circ g$.
(\emph{Convention:} $\Nat$ is always interpreted 
as $\set{\bo,\bl}^*$, i.e., $\bo$-$\bl$-strings.)
In our setting, a reasonable implementation of $C$
has a run-time bound that is a second-order polynomial (see
\S\ref{S:defs:2orderpoly}) in the complexities of arbitrary $f$ and $g$; in
particular, if $f$ and $g$ are polynomial-time computable, 
so is $C\,f\,g$.  Such a combinator $C$ can be considered
as part of the ``feasible glue'' of a programming 
environment---when used  with other components, 
its complexity contribution is (higher-type) 
polynomially-bounded in
terms of the complexity of the other components \emph{and} the 
combined complexity can be expressed in a natural, 
compositional way.   More elaborate examples of feasible
functionals include many of the deterministic black-box 
constructions from cryptography.  Chapter 3 in Goldreich 
\cite{Goldreich:I} has detailed examples, but a typical
such construct  takes one pseudo-random generator, $g$, and builds 
another, $\tilde{g}$, with better cryptographic properties but 
with not much worse complexity properties than the original $g$.  
Note that these $g$'s and $\tilde{g}$'s may be feasible only in
a probabilistic- or circuit-complexity sense.\footnote{See
\cite{KapronCook:mach,IKR:I} for a more extensive justification
that the BFFs provide a sensible type-2 analogue of the 
polynomial-time computable functions.}

While our notion of feasibility is based on the 
BFFs, our semantic models allow our formalism to 
compute more than just the standard BFFs.  For example,
consider $\textsf{prn}\of (\Nat\to\Nat\to\Nat)\to\Nat\to\Nat$
with:
\begin{gather} \label{e:prn}
\left.\begin{array}{rcl}
        \prn \; f\; \epsilon &\;\longrightarrow\; &
             f \;\epsilon\;\epsilon. \\
        \prn \; f\; (\ba\concat y) &\,\longrightarrow\,&
           f \;(\ba\concat y) \; (\prn \; f\; y).
\end{array}\right\}\end{gather}
(\emph{Conventions:} $\oplus$ denotes string concatenation
and  $\ba\in\set{\bo,\bl}$.)  So, $\prn$ is a version of 
Cobham's \cite{Cobham65} \emph{primitive recursion on notation}
(or alternatively, a string-variant of 
\textsf{foldr}). It is well-known that $\prn$ is \textbf{not} 
a BFF: starting with polynomial-time primitives, $\prn$ can be used to 
define any primitive recursive function. However as Cobham 
noted, if one modifies \refeq{e:prn} by adding  the side-condition:
\begin{gather*}
	(\exists p_f,\,\hbox{a polynomial})(\forall x)
	[\,|\prn\; f\; x| \leq p_f(|x|)\,],
\end{gather*}
this modified $\prn$ produces definitions of just polynomial-time
computable functions from polynomial-time computable
primitives. Bellantoni and Cook \cite{BellantoniCook} showed how get
rid of \emph{explicit} use of such a side condition through what
amounts to a typing discipline.  However, their approach (which has
been in large part adopted by the implicit computational complexity 
community, see Hofmann's survey \cite{Hofmann:survey}), requires
that $\prn$ be a ``special form'' and that the $f$ in \refeq{e:prn}
must be ultimately given by a purely syntactic definition.  We, on
the other hand, want to be able to define $\prn$ within $\atr$ (see
Figure~\ref{fig:atr:prn}) and have the definition's meaning given by
a conventional, higher-type denotational semantics.  We thus use
Bellantoni and Cook's \cite{BellantoniCook} (and Leivant's
\cite{Leivant:FM2}) ideas in both syntactic \emph{and semantic}
contexts.  That is, we extract the growth-rate bounds implicit in
the aforementioned systems, extend these bounds to higher types, and
create a type system, programming language, and semantic models that
work to enforce these bounds. As a consequence, we can define $\prn$
(with a particular typing) and be assured that, whether the $f$
corresponds to a purely syntactic term or to the interpretation a
free variable, $\prn$ will not go wrong by producing something of
huge complexity.  The language and its model thus implicitly
incorporate side-conditions on growth via
types.\footnote{Incorporating    \label{fn:fp} 
  side-conditions in models is nothing new.  A fixed-point combinator 
  has the implicit side-condition that its argument is continuous (or 
  at least monotone) so that, by Tarski's fixed-point theorem
  \cite{winskel:book}, we know the result is meaningful.  Models of
  languages with fixed-point combinators typically have continuity
  built-in so the side-condition is always implicit.}  Handling
constructs like $\prn$ as first class functions is important because
programmers care more about such combinators than about most any
standard BFF.

\subsection*{Outline} 
Our $\atr$ formalism is based on Bellantoni and Cook
\cite{BellantoniCook} and Leivant's \cite{Leivant:FM2} ideas on
using ``data ramification'' to rein in computational complexity.  \S
\ref{S:bcl} puts these ideas in a concrete form of $\bcl$, a simple
type-level~1 programming formalism, and sketches the proofs of three
basic results on $\bcl$: (i) that each $\bcl$ expression is
\emph{polynomial size-bounded}, (ii) that computing the value of a
$\bcl$ expression is \emph{polynomial time-bounded}, and (iii) each
polynomial-time computable function is denoted by some
$\bcl$-expression.  Most of this paper is devoted to showing the
analogous results for $\atr$.  \S \ref{S:better} discusses how one
might change $\bcl$ into a type-2 programming formalism, some of the
problems one encounters, and our strategies for dealing with these
problems. $\atr$, our type-2 system, is introduced in \S\ref{S:atr}
along with its type system, typing rules, and basic syntactic
properties.  The goal of \S\S\ref{S:polys}--\ref{S:pbnd} is to show
(type-2) polynomial size-boundedness for $\atr$.  This is
complicated by the fact (described in \S\ref{S:polys}) that the
na\"{i}ve semantics for $\atr$ permits exponential blow-ups.
\S\S\ref{S:impred}--\ref{S:flat} show how to prune back the
na\"{\i}ve semantics to obtain a setting in which we can prove
polynomial size-boundedness, which is shown in \S\ref{S:pbnd}.  The
goal of \S\S\ref{S:cek}--\ref{S:tcplus} is to show (type-2)
polynomial time-boundedness for $\atr$.  Our notion of the cost of
evaluating $\atr$ expressions is based on a particular abstract
machine (described in \S\ref{S:cek:mach}) that implements an
$\atr$-interpreter and the costs we assign to this machine's steps
(described in \S\ref{S:cek:cost}).  \S\ref{S:time} and
\S\ref{S:tcminus} set up a time-complexity semantics for $\atr^{-}$
expressions (where $\atr^{-}$ consists of $\atr$ without its
recursion construct) and establish that this time-complexity
semantics is: (i) sound for the abstract machine's cost model (i.e.,
the semantics provides upper bounds on these costs), and (ii)
\emph{polynomial time-bounded}, that is that the time-complexity
each $\atr$ expression $e$ has a second-order polynomial bound over
the time-complexities of $e$'s free variables.  \S\ref{S:comp}
shows that $\atr$ can compute each type-2 basic feasible functional.
\S\ref{S:finis} considers possible extensions of our work.  We begin
in \S \ref{S:defs} which sets out some basic background definitions 
with \S\S \ref{S:tc}--\ref{S:defs:bff} covering the more exotic 
topics.

\subsection*{Acknowledgments}
Thanks to Susan Older and Bruce Kapron for repeatedly listening to
the second author describe this work along its evolution.  Thanks to
Neil Jones and Luke Ong for inviting the second author to Oxford for
a visit and for some extremely helpful comments on an early draft of
this paper. Thanks to Syracuse University for hosting the first
author during September 2005.  Thanks also to the anonymous referees
of both the POPL version of this paper \cite{DR:ATS:Popl} and the
present paper for many extremely helpful comments.  Finally many
thanks to Peter O'Hearn, Josh Berdine, and the Queen Mary theory
group for hosting the second author's visit in the Autumn of 2005
and for repeatedly raking his poor type-systems over the coals until
something reasonably simple and civilized survived the ordeals.
This work was partially supported by EPSRC grant GR/T25156/01 and
NSF grant CCR-0098198.

\section{Notation and conventions}\label{S:defs}

\subsection{Numbers and strings} \label{S:nums}
We use two representations of the natural numbers: dyadic and
unary.  Each element of $\nat$ is identified with its dyadic
representation over $\set{\bo,\bl}$, i.e., $0\equiv \epsilon$, \
$1\equiv \bo$, \ $2\equiv \bl$, \ $3\equiv\bo\bo$, etc.  We freely
pun between $x\in\nat$ as a number and a $\bo$-$\bl$-string.  Each
element of $\omega$ is identified with its unary representation over
$\set{\bo}$, i.e., $0\equiv \epsilon$, \ $1\equiv \bo$, \ $2\equiv
\bo\bo$, \ $3\equiv\bo\bo\bo$, etc.  The elements of $\nat$ are used
as numeric/string values to be computed over.  The elements of
$\omega$ are used as tallies to represent lengths, run times, and
generally anything that corresponds to a size measurement.
\emph{Notation:} For each natural number $k$, \ $\tally{k}$ =
$\bo^k$.  Also $x\concat y$ = the concatenation of strings $x$ and
$y$.

\subsection{Simple types}
Below, $\bb$ (with and without decorations) ranges over 
base types and $B$ ranges over nonempty sets of base types. 
The \emph{simple types} over  $B$ 
are given by: 
$
  T \;\is\; B   \synsep (T\to T).
$
As usual, $\to$ is right associative and unnecessary parentheses
are typically dropped in type expressions, e.g., 
$(\sigma_1\to(\sigma_2 \to\sigma_3)) = 
\sigma_1\to\sigma_2\to\sigma_3$.  A type  $\sigma_1 \to 
\cdots\to\sigma_k\to \bb$  is often written as $(\sigma_1, \ldots, 
\sigma_k) \to \bb$ or, when $\sigma = \sigma_1 = \dots = \sigma_k$, 
as $(\sigma^k)\to\bb$.  
The \emph{simple product types} over  $B$
are given by:
$
  T \;\is\; B  \synsep T\to T
       \synsep () \synsep T \times T,
$
where $()$ is the type of the empty product. 
As usual,  $\sigma^1 = \sigma$, 
$\sigma^{k+1} = \sigma^k\times \sigma$, 
$\times$ is left associative, and
unnecessary parentheses are typically dropped in type expressions.
The \emph{level} of a simple (product) type  is given by:
$\level(\bb)$ = $ \level(\,()\,)$ = 0; 
$\level(\sigma\times\tau)$ =
$\max(\level(\sigma),\level(\tau))$; and 
$\level(\sigma\to \tau)$ = $\max(1+\level(\sigma),\level(\tau))$.
In this paper types are always interpreted over cartesian closed 
categories; hence, the two types 
$(\sigma_{1},\ldots,\sigma_{k})\to\tau$ and 
$\sigma_{1}\times\dots\times\sigma_{k}\to\tau$ may be identified.
By convention, we identify $()\to\tau$ with $\tau$
and $\lam{()}e$ with $e$.

\subsection{Subtyping} Suppose $\subty$ is a reflexive partial 
order on  $B$.  Then $\subty$ can be extended to 
a reflexive partial order over the simple types over $B$ by closing 
under:
\begin{gather}  \label{e:subty}
  \sigma_1\subty \sigma_0 \;\;\&\;\; \tau_0\subty \tau_1
  \;\;\iff\;\;
  \sigma_0\to \tau_0 \subty \sigma_1\to\tau_1.
\end{gather}
We read ``$\sigma\subty\tau$'' as 
$\sigma$ is a \emph{subtype} of $\tau$;
and write 
$\tau\supty\sigma$ for $\sigma\subty\tau$
and 
$\sigma\subtyneq\tau$ for $[\,\sigma\subty\tau$ and 
$\sigma\not=\tau]$ .

\subsection{Type contexts and judgments}

A \emph{type context} $\Gamma$ is a finite (possibly empty) mapping
of variables to types; these are usually written as a list:
$v_1\of\sigma_1,\, \ldots, \, v_k \of \sigma_k$.
$\Gamma, \Gamma'$ denotes the union of two type contexts with
disjoint preimages. $\Gamma\cup\Gamma'$ denotes the union of two
\emph{consistent} type contexts, that is, $\Gamma(x)$ and
$\Gamma'(x)$ are equal whenever both are defined.  
The \emph{type judgment} $\Gamma\entails_{\SF} e\of\sigma$ asserts
that the assignment of type $\sigma$ to expression $e$ follows
from the type assignments of $\Gamma$ under the typing rules for 
formalism $\SF$.  We typically omit the subscript in $\entails_{\SF}$
when $\SF$ is clear from context. 

\subsection{Semantic conventions} For a particular semantics $\calS$
for a formalism $\calF$, \ $\semgen { \,\cdot\,}$ is the 
\emph{semantic map} 
that takes an $\calF$-syntactic object  to its $\calS$-meaning. 
$\semgen{\tau}$ is the collection of 
things named by a type $\tau$ under $\calS$.  
For a type context $\Gamma= x_1\of \tau_1,\ldots,x_n\of\tau_n$, \
$\semgen{\Gamma}$ is the set of all finite maps 
$\set{ x_1 \mapsto a_1, \ldots, x_n \mapsto a_n}$, where $a_1 \in 
\semgen{\tau_1}, \ldots, a_n\in\semgen{\tau_n}$; i.e., 
\emph{environments}.  
\emph{Convention:} $\rho$ (with and without decorations) ranges
over environments and $\emptyenv$ = 
the empty environment.
$\semgen{\Gamma\entails e\of\tau}$ is the map from $\semgen{\Gamma}$ 
to $\semgen{\tau}$ such that $\semgen{\Gamma\entails e\of\tau}\,\rho$ 
denotes the element of $\semgen{\tau}$ that is the $\calS$-meaning of 
expression $e$  when $e$'s free-variables  have the meanings
given by $\rho$.  
\emph{Conventions:}  $\semgen{e}$ is typically written in  place of $\semgen{\Gamma\entails e\of\tau}$ since the 
type judgment is  usually understood from context.  
When $e$ is closed, $\semgen{e}$ is sometimes written in place of
$\semgen{e}\,\emptyenv$. 
Also, $e_{0}=_{\mathcal{S}}
e_{1}$ means $\semgen{e_{0}}=\semgen{e_{1}}$.

\subsection{Syntactic conventions} 
Substitutions  (e.g., $e[x\gets e']$) are always assumed to be 
capture avoiding.  Terms of the form $(x\,e_1\ldots e_k)$ 
are sometimes written as $x(e_1,\ldots, e_k)$.

\begin{figure}
\begin{minipage}{\textwidth}\small
\begin{align*}
  E &\;\;\is \;\;  K 
        \synsep (\consa\; E) 
        \synsep (\cdr\; E)
        \synsep (\tsta\; E)
        \synsep (\down\; E\; E) 
        \\ 
        & \Quad{4.25}
        \synsep X
        \synsep (E \; E)
        \synsep (\lam{X}E)
        \synsep (\Iif \; E\;\Ithen\; E \; \Ielse\;E )
        \synsep (\fix \; E) \\[1ex]
  K &\;\;\is \;\;  \set{\bo,\bl}^{*}\Quad3
  T  \;\;\is \;\; \hbox{the simple types over $\Nat$}
\end{align*}
\caption{$\PCF$ syntax}\label{fig:pcf:syn}
\end{minipage}
\figrule 
\begin{minipage}{\textwidth}\small
\begin{gather*}
    \rulelabel{Const-I}
    \irule{ 
                }{\Gamma\entails k\of\Nat} 
    \Quad{3}
    \rulelabel{{\rm$\op$}-I}
    \irule{ 
                \Gamma\entails e\of\Nat}{
                \Gamma\entails (\op\;e)\of\Nat}                
    	\Quad{3}
    \rulelabel{$\down$-I}
    \irule{ 
                \Gamma_0 \entails e_0\of\Nat \Quad{1.5}
		\Gamma_1 \entails e_1\of\Nat}{
        \Gamma_0\cup\Gamma_1 \entails(\down\;e_0\;e_1)\of\Nat}         
    \\[1ex]  
    \rulelabel{Id-I}
    \irule{ 
                }{\Gamma,\,x\of\sigma\entails x\of\sigma} 
    \Quad{3}    
    \rulelabel{$\to$-I}
    \irule{ 
                \Gamma,\,x\of\sigma\entails e\of\tau}{
        \Gamma\entails (\lam{x} e) \of\sigma\to\tau} 
        \Quad3
    \rulelabel{$\to$-E}\;
    \irule{ 
                \Gamma_0\entails e_0\of\sigma\to\tau \Quad{1.5}
        \Gamma_1\entails e_1\of\sigma}{
                \Gamma_0\cup\Gamma_1\entails (e_0\;e_0)\of \tau}
        \\[1ex] 
     \rulelabel{If-I}
     \irule{ 
                \Gamma_0\entails e_0\of\Nat \Quad{1.5}
                \Gamma_1\entails e_1\of\Nat \Quad{1.5}
                \Gamma_2\entails e_2\of\Nat                }{
                \Gamma_0\cup\Gamma_1\cup\Gamma_2
                \entails 
                    (\Iif \,e_0 \, \Ithen \,e_1 \, \Ielse\, 
                    e_2)\of\Nat} 
        \Quad{3} 
    \rulelabel{$\fix$-I}
 \irule{ 
   \Gamma\entails 
     \left(\lam{x}e\right)
                \of\sigma\to\sigma}{
   \Gamma\entails 
      \left( \fix\; (\lam{x} e)\right) 
                 \of \sigma}
\end{gather*}
\caption{The $\PCF$ typing rules}\label{fig:pcf:typing}
\end{minipage}
\figrule 
\begin{minipage}{\textwidth}\small
\begin{gather*}
  (\consa\;v) \;\longrightarrow\; \ba\concat v. 
  \Quad3
  (\cdr\;(\ba\concat v)) \;\longrightarrow\; v. 
  \Quad3
  (\cdr\;\epsilon) \;\longrightarrow\; \epsilon. 
        \\[1ex] 
  (\tsta\;v) \;\longrightarrow\; 
  \begin{cases}
    \bo, & \hbox{if $v$ begins with $\ba$;}\\
    \epsilon, & \hbox{otherwise.}
  \end{cases}
        \Quad3
  (\down\;v_0\;v_1) \;\longrightarrow\; \begin{cases} 
                        v_0, & \hbox{if $|v_0|\leq|v_1|$;}\\
                        \epsilon, & \hbox{otherwise.}
                        \end{cases}\\[1ex]
  (\Iif\,v_0\,\Ithen\,v_1\,\Ielse\,v_2) \;\longrightarrow\;
  \begin{cases}
    v_1, & \hbox{if $v_0\not=\epsilon$;}\\ 
    v_2, & \hbox{if $v_0=\epsilon$.}
  \end{cases}                        \Quad{3}
   \fix\;(\lam{x}e) \;\longrightarrow\;e[x\gets(\fix\; (\lam{x} e)].
\end{gather*}
\caption{The $\PCF$ reduction rules for $\consa$, $\cdr$,
$\tsta$, $\down$, 
$\Iif$-$\Ithen$-$\Ielse$, and $\fix$}
\label{fig:pcf:red}
\end{minipage}
\figrule 
\end{figure}

\subsection{Call-by-value $\PCF$}\label{S:defs:pcf}
The syntax of our version of
$\PCF$ is given in Figure~\ref{fig:pcf:syn}, where the syntactic
categories are: constants ($K$),  raw-expressions ($E$),
variables ($X$), and type-expressions ($T$)  and where
$\ba\in\set{\bo,\bl}$.  Figure~\ref{fig:pcf:typing} states $\PCF$'s 
typing rules, where $\op$ stands for any of $\conso$,
$\consl$, $\cdr$, $\tsto$, and $\tstl$
and where 
$E\is e,e_0,e_1,e_2$, 
$K\is k$, 
$T\is \sigma,\tau$, and
$X\is x$.
For emphasis 
we may write $\lam{x\of \sigma}e$ instead of $\lam{x}e$, but the 
type of $x$ can always be inferred from any type judgement
in which $\lam{x}e$  occurs.
The intended interpretation
of $\Nat$ is $\nat$ ($\cong \set{\bo,\bl}^*)$.
The reduction rules are essentially the 
standard ones for call-by-value $\PCF$ (see 
\cite{Plotkin75,Pierce:types}).  In particular, the reduction rules 
for $\conso$, $\consl$, $\cdr$, $\tsto$, $\tstl$,
$\down$, $\Iif$-$\Ithen$-$\Ielse$, and $\fix$
are given in Figure~\ref{fig:pcf:red}.  
Note that in $\Iif$-$\Ithen$-$\Ielse$ tests, 
$\epsilon$ corresponds to \emph{false} and elements of 
$(\nat-\set{\epsilon})$ correspond to \emph{true}.  
In tests,  we use $x\not=\epsilon$ for syntactic sugar for $x$ 
and use $|e_0| \leq |e_1|$ as syntactic sugar for 
$(\down\; \conso(e_0)\;  \conso(e_1))$.\footnote{We will see in
\S \ref{S:better} and \S \ref{S:atr}
why $\down$ (as opposed to $|\cdot|\leq|\cdot|$) 
is a primitive.}   An operational semantics for  $\PCF$ is provided 
by the CEK-machine given in \S \ref{S:cek:mach}.  
We take  $\SV$ (for \emph{value}) to be a conventional denotational 
semantics for  $\PCF$ \cite{winskel:book}.  Standard arguments 
show that our operational semantics corresponds to $\SV$.

\subsection{Total continuous functionals}  \label{S:tc}
Let $\sigma$ and $\tau$  be simple product types over base type 
$\Nat$.
Inductively define:
$\TC_{()}$ = $ \star$;
$\TC_{\Nat}$ = $\nat$; 
$\TC_{\sigma\times\tau}$ = $\TC_{\sigma}\times\TC_{\tau}$;
$\TC_{\sigma\to\tau}$ = the \emph{Kleene/Kreisel total continuous 
functions} 
from $\TC_{\sigma}$ to $\TC_{\tau}$; 
the $\TC_{\sigma}$'s together form a cartesian closed category 
$\TC$.\footnote{For background on 
the Kleene/Kreisel total continuous functions and 
$\TC$, see the historical survey of  Longley 
\cite{Longley:notions:1} and  the technical surveys of Normann  
\cite{Normann99}  and Schwichtenberg \cite{Schwichtenberg:cont}.} 
This paper is concerned with only the type-level 0, 1, and 2 portions 
of $\TC$ from which we construct models of our programming
formalisms.

\subsection{Total monotone continuous functionals.}  
\label{S:defs:MC}
Let $\sigma$ and $\tau$ be  simple product types over base type $\Tally$ 
(for  \emph{tally}). Inductively define the $\MC_{\sigma}$ sets and 
partial orders $\leq_\sigma$ by:
$\MC_{\Tally} = \omega$ and $\leq_{\Tally}$ 
= the usual ordering on $\omega$; 
$\MC_{()} = \star$  and $\star\leq_{()}\star$; 
$\MC_{\sigma\times\tau}=\MC_\sigma\times\MC_\tau$ and 
$(a,b)\leq_{\sigma\times\tau}(a',b')\iff
a\leq_\sigma a'$ and $b\leq_\tau b'$; and 
$\MC_{\sigma\to\tau}$ = the  
Kleene/Kreisel total continuous functions from $\MC_\sigma$ 
to $\MC_\tau$ that are monotone (w.r.t.~$\leq_\sigma$ 
and $\leq_\tau$), and $\leq_{\sigma\to\tau}$ is the point-wise 
ordering on $\MC_{\sigma\to\tau}$.  (E.g., $\MC_{\Tally\to\Tally} 
= \{\,f\of\omega\to\omega \suchthat 
f(0)\leq f(1) \leq f(2) \leq \cdots \,\}$.)
The $\MC_{\sigma}$'s turn out to form a 
cartesian closed category $\MC$.  
As with $\TC$, our concern is with only the type-level 
0, 1, and 2 portions of $\MC$ from which we construct our
models of size and time bounds. 
\emph{Convention:} We typically omit the subscript in $\leq_\sigma$
when the $\sigma$ is clear from context.

\subsection{Lengths} \label{S:defs:length}
For $v\in\nat$, let $|v| = \tally{k}$, where $k$ is the length of 
the dyadic representation of $v$ (e.g., $|\bl\bl\bo| = \tally{3} 
=\bo\bo\bo$).
For $f\in\TC_{(\Nat^k)\to\Nat}$, define $|f| \in 
\MC_{(\Tally^k)\to\Tally}$ by:
\begin{gather} \label{e:flen}
  |f|(\vec{\ell\,}) 
  \;\;=\;\;
  \max\Set{|f(\vec{v})| \suchthat \;\strut
        \hbox{$|v_1|\leq \ell_1,\ldots, |v_{k}|\leq \ell_{k}$}}. 
\end{gather}
(This is Kapron and Cook's \cite{KapronCook:mach} definition.)
For each $\sigma$, a simple type 
over $\Nat$, let $\len{\sigma}=\sigma[\Nat\gets\Tally]$
(e.g., $|\Nat\to\Nat|=\Tally\to\Tally$).  So
by the above, $|v|\in \MC_{\len{\sigma}}$ when
$\level(\sigma)\leq 1$ and $v\in\TC_{\sigma}$.
Here is a type-level 2 notion of length that suffices for this paper.  
For $\gamma = (\sigma_{1},\ldots,\sigma_{k})\to\Nat$ of  level-2,
$F\in\TC_{\gamma}$, and $\ell_{1}\in\MC_{\len{\sigma_{1}}}, 
\ldots,\ell_{k}\in \MC_{\len{\sigma_{k}}}$, define 
\begin{gather} \label{e:flen2}
  |F|(\vec{\ell}\,)  \;\; = \;\;
  \max\Set{|F(\vec{v})| \suchthat |v_1| \leq_{|\sigma_1|} \ell_1, 
     \,\dots\,,|v_k| \leq_{|\sigma_k|}\ell_k}.
\end{gather}
$|F|$ as defined above turns out to be an element of 
$\MC_{\len{\gamma}}$.\footnote{When 
  $\gamma$ is type-level 2 and $\ell\in\MC_{|\gamma|}$, 
  generally $\set{ F\in\TC_\gamma \suchthat |F|\leq_{|\gamma|} 
  \ell}$ fails to be compact in the appropriate topology.  
  Consequently, the type-3 analogue of \refeq{e:flen2} fails to 
  yield lengths that are total.  There are alternative notions of 
  type-2 length that avoid this problem;  \cite{IKR:II} investigates
  two of these.
  }

\subsection{Maximums and  polynomials} 
Let $v_{1} \bmax v_{2}$ $= \max(\set{v_{1},v_{2}})$ and let 
$\bigmax_{i=1}^k v_i = \max(\set{v_1, \dots,v_k})$ for  
$v_1,\dots,v_k\in\omega$.
By convention, $\max(\emptyset)=0$.
We allow $\bmax$ as another arithmetic operation in 
polynomials; $\bmax$ binds 
closer than either multiplication or addition.
Coefficients in polynomials will 
always be nonnegative; hence polynomials denote
monotone nondecreasing functions, i.e., 
type-level 1 elements of $\MC$.

\begin{figure}
\begin{minipage}{\textwidth}\small
\begin{align*}
  P &\;\;\is \;\;      K 
                \synsep (\bmax\;P\;P)
                \synsep (+\;P\;P)
                \synsep (\ast\; P \;P)                 \synsep V
                \synsep (P\; P)
                \synsep (\lam{V}P )\\[1ex]
  K &\;\;\is \;\; 
      \tally0 \synsep \tally1 \synsep \tally2 \synsep \dots 
    \Quad3
  T \;\;\is\;\; 
     \hbox{the level 0, 1, and 2 simple types over $\Tally$}   
\end{align*}
\caption{The syntax for second-order polynomials and their
standard types}
\label{fig:2opolys:syn}
\end{minipage}
\figrule 
\begin{minipage}{\textwidth}\small
\begin{gather*}
    \rulelabel{Const-I}
    \irule{ 
                }{\Sigma\entails k\of\Tally} 
                \Quad{3}
    \rulelabel{{\rm$\odot$}-I}
    \irule{ 
                \Sigma\entails p_0\of\Tally\Quad1  
                \Sigma\entails p_1\of\Tally}{
                \Sigma\entails (\odot\;p_0\;p_1)\of\Tally}    
                \sidecond{\odot = \ast,\, +,\, \bmax}           
\end{gather*}
\caption{The additional typing rules for the second-order polynomials}
\label{fig:2opolys:ty}
\end{minipage}
\figrule 
\end{figure}

\subsection{Second-order polynomials} \label{S:defs:2orderpoly}
We define the \emph{second-order polynomials} \cite{KapronCook:mach} 
as a type-level 2 fragment of the simply typed $\lambda$-calculus 
over base type $\Tally$ with arithmetic operations $\bmax$, 
$+$, and $\ast$.  Figure~\ref{fig:2opolys:syn} gives the syntax,
where the syntactic categories are: constants ($K$), raw 
expressions ($P$), and type expressions ($T$).
We often write $\bmax$-, $+$-, and $\ast$-expressions in infix form.
The typing rules are \emph{Id-I}, 
$\to$-\emph{I}, and $\to$-\emph{E} from Figure~\ref{fig:pcf:typing} 
together with the rules in Figure~\ref{fig:2opolys:ty}.  Moreover, 
the only variables allowed are those of of type levels~0 and~1.
Our semantics $\mathcal{L}$ (for \emph{length}) for second-order 
polynomials is: $\semlen{\sigma}=\MC_{\sigma}$
for each $\sigma$, a simple type over $\Tally$, and 
$\semlen{\Sigma\entails p \of\sigma}$ = the standard definition.
The \emph{depth} a second-order polynomial $q$ is the maximal depth 
of nesting of applications in $q$'s $\beta$-normal form, e.g., 
$g_0((g_0(2 \ast y \ast g_1(y^2))\bmax6)^3)$ has depth 3.  
There is a special case for variables of higher type: type-level 
$\ell$ variables are assigned depth $\ell$.\footnote{Since, 
  for example, for $f\of\Tally\to\Tally$, \
  $f\equiv_\eta \lam{x}f(x)$ and $\depth(\lam{x}f(x))=1$.}
For second-order polynomials,  depth plays something 
like the role degree does for ordinary polynomials.

\subsection{Time complexity}\label{S:defs:complex}
The CEK machine (\S \ref{S:cek:mach}) provides an operational 
semantics for $\PCF$ as well as for the formalisms $\bcl$ 
(\S\ref{S:bcl}) and $\atr$ (\S\ref{S:atr}).
Since this paper concerns the evaluation of 
expressions and the associated costs, we use
the CEK machine as our standard model of computation and use the 
CEK cost model (\S \ref{S:cek:cost}) as  our standard notion of 
time complexity.   
As discussed in \S \ref{S:cek:cost}, Sch\"{o}nhage's \emph{storage 
modification machine} \cite{schonhage80} is roughly the 
standard complexity-theoretic model of computation and cost 
underlying our CEK model.  Storage modification machines 
and Turing machines are polynomially-related models of computation 
\cite{schonhage80}.  Our CEK machine handles oracles  
(type-1 functions over $\nat$) as the values of particular
variables in the initial environment for an evaluation.   
As with Kapron and Cook's answer-length cost model for oracle
Turing  machines \cite{KapronCook:mach}, part of the CEK-cost of 
querying an oracle includes the length of the answer.

\subsection{Basic feasibility}\label{S:defs:bff}
Suppose 
$\tau = (\sigma_1,\dots,\sigma_k)\to\Nat$ is a
simple type over $\Nat$ of level 1 or 2 and 
that $f\in\semval{\tau}$.
($\semval{\,\cdot\,}$ was introduced in \S \ref{S:defs:pcf}.)
We say that $f$ is a \emph{basic feasible functional} (or BFF)
when
there is a closed, type-$\tau$ $\PCF$-expression $e_f$ and a 
second-order polynomial function $q_f$ such that 
(i) $\semval{e_f}=f$ and 
(ii) for all $v_i\in\semval{\sigma_1},\dots,v_k\in\semval{\sigma_k}$, 
  $\cektime(e_f,v_1,\dots,v_k) \;\leq\; q_f(|v_1|,\dots,|v_k|)$,
where $\cektime$ is introduced in Definition~\ref{d:cekcosts} of \S 
\ref{S:cek:cost}. 
For level-1 $\tau$, this gives us the usual notion of 
type-1 polynomial-time computability.  The original definitions
and characterizations of the type-2 BFFs
\cite{Mehlhorn74,CookUrquhart:feasConstrArith,CookKapron:FM}
were all in terms of programming formalisms.  The definition here
is based on Kapron and Cook's  machine-based characterization of the
type-2 BFFs \cite{KapronCook:mach}.

\begin{figure} \small
\begin{minipage}{\textwidth}
\begin{gather*}
  E \;\;\is \;\;  \dots 
         \synsep (\prn\;  E) \Quad3
  T \;\;\is \;\; 
  \hbox{the level 0 and 1 types over $\NatNorm$ and $\NatSafe$}  
\end{gather*}
\caption{$\bcl$ syntax}\label{fig:bcl:syn}
\end{minipage}
\figrule 
\begin{minipage}{\textwidth}
\begin{gather*}
    \rulelabel{Zero-I}
    \irule{ 
                }{\Gamma\entails \epsilon\of\NatNorm} 
                \Quad{3}
    \rulelabel{Subsumption}
    \irule{ 
                \Gamma\entails e\of\sigma}{
                \Gamma \entails e\of\tau} 
                \sidecond{\sigma \subty \tau}
        \\[1ex]  
    \rulelabel{{\rm $\cdr$}-$I'$}
    \irule{ 
       \Gamma \entails e\of\NatNorm}{
       \Gamma \entails (\cdr \; e) \of\NatNorm}
    \Quad3
    \rulelabel{{$\prn$}-I}
    \irule{ 
      \Gamma \entails e\of\NatNorm\to\NatSafe\to\NatSafe
      }{
                                \Strut{2.8ex}
          \Gamma\entails 
                \left(
                \prn\;e
                \right)
                \of \NatNorm\to\NatSafe}                  
\end{gather*}
\caption{Additional $\bcl$ typing rules} \label{fig:bcl:types}
\figrule 
\newcommand{\Com}[1]{\hfill\emph{// \makebox[7cm][l]{#1}}}
\begin{verse} {\ }\\
\emph{cat$\of\NatNorm\to\NatSafe\to\NatSafe$} =
 		\Com{cat w x = $w\concat x$.  
      	So, $|\textit{cat}\; w\; x| = |w|+|x|$.}\\
	\Quad{1}  
	$\lam{w,x}$ $\Ilet$ $f\of\NatNorm\to\NatSafe\to\NatSafe=$ 
      \\
	\Quad{5.5} $\lam{y,z}$  $\Iif$ $\tsto(y)$ $\Ithen$ $\conso(z)$ 
   		$\Ielse$ $\Iif$ $\tstl(y)$ $\Ithen$ $\consl(z)$
	   	$\Ielse$ $x$\\
	\Quad{4.25}    
		$\Iin$ $\prn$ $f$ $w$\\
\emph{dup}$\of\NatNorm\to\NatNorm\to\NatSafe =$ 
		\Com{dup w x = 
		$\overbrace{x\concat \cdots \concat x}^{|w| \text{ many}}$.
		So, $|\textit{dup}\;w \; x| = |w|\cdot |x|$.}\\
	\Quad{1} $\lam{w,x}\Ilet$ $g\of\NatNorm\to\NatSafe\to\NatSafe=$ 
		 $\lam{y,z}$ $\Iif$
          $y\not=\epsilon$ $\Ithen$ (\emph{cat x z}) $\Ielse$ 
          $\epsilon$ \\
	\Quad{4} $\Iin$ $\prn$ $g$ $w$\\[-3ex]
\end{verse}
\caption{Two sample $\bcl$ programs} \label{fig:bcl:ex}
\figrule
\end{minipage}
\end{figure}

\section{The $\bcl$ formalism}\label{S:bcl}

The programming formalisms of this paper are built on work of
Bellantoni and Cook \cite{BellantoniCook} and Leivant \cite{Leivant:FM2}. 
Bellantoni and Cook's paper 
takes a programming formalism for the primitive recursive functions,
imposes certain intensionally-motivated constraints, and obtains 
a formalism for the polynomial-time computable functions.  To explain 
these constraints and how they rein in 
computational strength, we sketch both $\bcl$, a simple type-1 
programming formalism based on Bellantoni and Cook's and Leivant's 
ideas, and $\bcl$'s properties.\footnote{$\bcl$ is much closer to 
Leivant's formalism \cite{Leivant:FM2}, which uses a ramified type 
system, than Bellantoni and Cook's, which does not use a 
conventional type system.}    This sketch provides an initial
framework for this paper's formalisms.

$\bcl$ has the same  syntax as
$\PCF$ (\S \ref{S:defs:pcf}) with three changes:
(i) $\fix$ is replaced with $\prn$ (for \emph{primitive recursion
on notation} \cite{Cobham65}) that has the reduction rule given by
\refeq{e:prn}, 
(ii) the only variables allowed are those of base type, and 
(iii) the type system is altered as described below. 
If we were to stay with the simple types
over $\Nat$ and the $\PCF$-typing rules (Figure~\ref{fig:pcf:typing}
and with $\prn\of(\Nat \to \Nat \to\Nat) \to\Nat\to\Nat$), the 
resulting formalism would compute exactly the primitive recursive 
functions.  Instead we modify the types and typing as follows.  
$\Nat$ is replaced with two base types, $\NatNorm$ (\emph{normal 
values}) and $\NatSafe$ (\emph{safe values}),
subtype ordered $\NatNorm \subty \NatSafe$.
The $\bcl$ types are just the type-level 
0 and 1 simple types over $\NatNorm$ and $\NatSafe$.  
Both base types have intended interpretation $\nat$.
The point of the two base types is to separate the roles of 
$\nat$-values: a $\NatNorm$-value can be used to drive a recursion, 
but cannot be the result of a recursion, whereas a $\NatSafe$-value 
can be the result of a recursion, but cannot be used to drive a 
recursion.  These intentions are enforced by the $\bcl$ typing rules,
consisting of: \emph{ID-I}, \emph{$\to$-I}, and \emph{$\to$-E} 
from Figure~\ref{fig:pcf:typing}; \emph{Const-I}, $\conso$-\emph{I},
$\consl$-\emph{I}, $\cdr$-\emph{I}, $\tsto$-\emph{I},
$\tstl$-\emph{I}, $\down$-\emph{I}, and \emph{If-I} also from
Figure~\ref{fig:pcf:typing} where each $\Nat$ is changed to
$\NatSafe$; and the rules in
Figure~\ref{fig:bcl:types}. (\emph{Zero-I} and $\cdr$-$I'$
are needed to make the $\prn$ reduction rules type-correct.)
Figure~\ref{fig:bcl:ex} contains two sample $\bcl$ programs.  
For the sake of readability, we use the $\Ilet$ construct as 
syntactic sugar.\footnote{Where 
  \label{fn:let}
  $(\Ilet$ $x$ = $e'$ $\Iin$ $e)$ \; 
  $\stackrel{\text{def}}{\equiv}$ \; 
  $e[x\gets e']$. This permits naming defined functions.
} 

Propositions~\ref{p:bc:size:polybnd}, \ref{p:bc:time:polybnd},
and~\ref{p:bc:pt:complete} state the key computational limitations
and capabilities of $\bcl$.  In the following $\vec{x}\of\NatNorm$
abbreviates $x_1\of\NatNorm,\allowbreak \dots,x_m\of\NatNorm$ 
and $\vec{y}\of\NatSafe$ abbreviates
$y_1\of\NatSafe,\dots,y_n\of\NatSafe$.  Recall from
\S\ref{S:defs:complex} that our standard notion of time complexity
is the time cost model of the CEK-machine 
(Definition~\ref{d:cekcosts}(a)).

\begin{proposition}[$\bcl$ polynomial size-boundedness] 
  \label{p:bc:size:polybnd} Suppose 
  $\vec{x}\of\NatNorm, \vec{y}\of\NatSafe\entails e\of \bb$.
  
  (a) If\/ $\bb=\NatNorm$, then   
  for all values of $\vec{x},\,\vec{y}$,  \ $|e| \leq \bigmax_{i=1}^m|x_i|$.
  
  (b) If $\bb=\NatSafe$, then 
  there is a polynomial $p$ over over $|x_1|,\dots,\allowbreak |x_m|$ 
  such that, for all values of $\vec{x},\,\vec{y}$, \ $|e| \leq p + \bigmax_{j=1}^n|y_j|$.
\end{proposition}

Proposition~\ref{p:bc:size:polybnd}'s proof is an induction on $e$'s
syntactic structure, where the $\prn$-case is the crux of the
argument.  Here is a sketch of a mild simplification of that case.
(This sketch is the model for several key  subsequent arguments.)
Suppose $e= \prn \; e' \; x$, where $x_0\of\NatNorm, \vec{x}
\of\NatNorm, \allowbreak y_0\of\NatSafe, \vec{y} \of\NatSafe
\entails (e'\; x_0 \; y_0) \of\NatSafe$ and $x \in \set{x_1, \dots,
x_m}$. Also suppose that, for all values of $x_0,\dots,\allowbreak
x_m,y_0,\dots,y_n$,
$
  |e'\;x_0\;y_0| \leq 
      p'(|x_0|)+ {\textstyle \bigmax_{j=0}^n|y_j|}
$
where $p'$ is a polynomial over $|x_0|$ (explicitly) and
$|x_1|,\dots,|x_m|$ (implicitly).  Fix the values of
$x_1,\dots,y_n$, where in particular 
$x$ has the value $\ba_1\dots \ba_k$ for 
$\ba_1,\dots,\ba_k\in\set{\bo,\bl}$. We determine bounds for
$|\prn\; e'\; \epsilon|$, \ 
$|\prn\; e'\; \ba_k|$,
$|\prn\; e'\; \ba_{k-1}\ba_k|, \dots,
|\prn\; e'\; \ba_1\dots\ba_k|$ in turn.
First, 
$
|\prn\; e'\; \epsilon| 
   \;=\;    |e'\;\epsilon\;\epsilon|
   \;\leq\; p'(\tally0) + \textstyle \bigmax_{j=1}^n|y_j|.  
$
Next,
\begin{gather*} 
|\prn\;e'\; \ba_k| 
  \;\;=\;\; |e' \; \ba_k\; (\prn\; e'\;\epsilon)| 
  \;\;\leq\;\; p'(\tally1) + |\prn\; e'\;\epsilon|\bmax    
      \textstyle\bigmax_{j=1}^n|y_j| \;\;\leq \\
  p'(\tally1) + (p'(\tally0)+ 
  \textstyle\bigmax_{j=1}^n|y_j|)
      \bmax \textstyle\bigmax_{j=1}^n|y_j|     
  \;\;\leq\;\; p'(\tally0)+p'(\tally1)+
  \textstyle\bigmax_{j=1}^n|y_j|.
\end{gather*}
Continuing, we end up with $|\prn \; e'\; x| \leq
p'(\tally0)+p'(\tally1)+\dots+p'(\tally{k})+\bigmax_{j=1}^n|y_j| \leq 
(|x|+\tally1)\ast p'(|x|) + \bigmax_{j=1}^n|y_j|$.
So, $p=(|x|+1)\ast p'(|x|)$ suffices for this case.

\begin{proposition}[$\bcl$ polynomial time-boundedness]
 \label{p:bc:time:polybnd}
  Given
  $\vec{x}\of\NatNorm, \vec{y}\of\NatSafe\entails e
  \of (\bb_1,\allowbreak \dots,\allowbreak\bb_\ell)\to\bb$,
  there is a polynomial $q$ over over 
  $|w_1|,\dots,\allowbreak |w_\ell|$,
  $|x_1|,\dots,\allowbreak |x_m|$, 
  $|y_1|,\dots,\allowbreak |y_n|$
  such that, for all values of $w_1,\dots, \allowbreak
  y_n$, \ $q$ bounds   
  the CEK-cost 
  of evaluating $(e\; w_1\;\dots\;w_\ell)$.
\end{proposition}

Proposition~\ref{p:bc:time:polybnd}'s proof rests on three 
observations: 
(i) evaluating
$(\prn \; e \; e')$ 
takes $|e'|$-many (top-level) recursions,
(ii) by the first observation and the details of CEK costs, the 
time-cost of a CEK evaluation of a $\bcl$ expression can be bounded 
by a polynomial over the lengths of base type values involved, and 
(iii) Proposition~\ref{p:bc:size:polybnd} provides polynomial bounds 
on all these lengths.  Proposition~\ref{p:bc:time:polybnd} thus 
follows through a straightforward induction on the syntactic 
structure of $e$.  Proposition~\ref{p:bc:pt:complete}'s proof 
is mostly an exercise in programming.

\begin{proposition}[$\bcl$ polynomial-time completeness]
 \label{p:bc:pt:complete}
 For each poly\-nomial-time computable $f\in((\nat^\ell)\to\nat)$, 
 there is an 
 $\entails e_f\of(\NatNorm^\ell)\to\NatSafe$ such that 
 $\semval{e_f}=f$.
\end{proposition}

$\bcl$ is $\subty$-predicative in the sense that no
information about a $\NatSafe$-value can ever make its way into 
a $\NatNorm$-value.  For example:

\begin{proposition}\label{p:bcl:inhab}
 Suppose  $\entails e \of (\NatNorm,\NatSafe)\to\NatNorm$.
 Then $e\equiv_{\alpha\beta} \lam{w,x} e'$ with
 $e'=\epsilon$ 
 or else
 $e'=(\cdr^{(k)}\,w)$ for some $k\geq 0$,
 where $(\cdr^{(0)}\,w) = w$ and $(\cdr^{(k+1)}\,w) = 
 (\cdr\; (\cdr^{(k)}\,w))$.
\end{proposition}

$\bcl$'s $\subty$-predicativity plays a key role in proving
the polynomial size-bounds of Proposition~\ref{p:bc:size:polybnd}, but
plays no direct (helpful) role in the other proofs.

\section{Building a better $\bcl$}\label{S:better}

Our definition of $\atr$ in the next section can be thought of
as building an extension of $\bcl$ that: 
(i) computes the type-2 $\BFF$s,
(ii) replaces $\prn$ with something closer to $\fix$, and
(iii) admits  reasonably direct complexity theoretic analyses.
This section motivates some of the differences between
$\bcl$ and $\atr$.

\subsection*{Types and depth}  
We want to extend $\bcl$'s type system to allow 
definitions of functions as such
$
  F_0 = \lam{f\in\nat\to\nat,\, x\in\nat}f(f(x)),
$
a basic feasible functional.  A key question then is 
how to assign types to functional parameters such as $f$ above. 
Under  $f\of\NatNorm\to\NatSafe$, \ $F_0$
fails to have a well-typed definition.  Under any of 
$f\of\NatNorm\to\NatNorm$, $f\of\NatSafe\to\NatSafe$, and 
$f\of\NatSafe\to\NatNorm$, \ $F_0$ has a well-typed definition, 
but then so does
$
  F_1 = \lam{f\in\nat\to\nat,\,x\in\nat}f^{(|x|)}(x) 
$
which is \emph{not} basic feasible.  Thus some nontrivial 
modification of the $\bcl$ types seems necessary for any 
extension to type-level 2.

We sketch a na\"{\i}ve extension that uses of an informal notion of 
the depth of an expression (based on second-order polynomial 
depth, see \S \ref{S:defs:2orderpoly}).  Let the 
\emph{na\"{\i}ve depth} of an expression 
(in normal form) 
be the depth of nesting of applications of type-level 1 variables.  
For example, given $f\of\Nat\to\Nat\to\Nat$, then 
$
 f(\conso(f(\conso(x),\, y)),\, \consl(\cdr(y)))
$
has na\"{\i}ve depth 2.  We can regard  the values of $x$ and $y$ as 
(depth-0) inputs and the values of $\conso(x)$ and $\cdr(y)$ as 
the results of polynomial-time computations over those inputs.  Taking 
the type-level 1 variables as representing oracles, the value of 
$f(\conso(x),\, y)$ can then be regarded as a depth-1 input (that is 
an input that is in response to a depth-0 query); hence,
$\conso(f(\conso(x),\, y))$ is the result of a polynomial-time computation 
over a depth-1 input.  Similarly, the value of 
$f(\conso(f(\conso(x),\, y))$, $\consl(\cdr(y)))$  can be regarded 
as a depth-2 input.  Thus, our na\"{\i}ve extension 
amounts to having, for each $d\in\omega$, depth-$d$ versions of 
both $\NatNorm$ and $\NatSafe$ and treating all arrow types as 
``depth polymorphic'' so, for instance, the type of $f$ as above 
indicates that $f$ takes depth-$d$ safe values to depth-$(d+1)$ 
normal values, for each $d\in\omega$.  This permits a well-typed
definition for $F_0$, but not for $F_1$. 

The na\"{\i}vete of  the above is shown by another example.
Let 
\begin{align}\label{e:f2}
  F_2 &=
    \lam{f\in\nat\to\nat,\,y\in\nat}[\,
     g^{(|y|)}(y), \;
     \hbox{where }g =  \lam{w\in\nat} \left(f(w)\bmod (y+1)\right)\,].
\end{align}
$F_2$ is basic feasible, $|F_2(f,y)|  \leq |y|$, but 
it is reasonable to think of 
$F_2(f,y)$ having unbounded na\"{\i}ve depth.  

\begin{figure}[t] \small
\begin{minipage}{\textwidth}
\begin{gather*}
    \rulelabel{$\down$-$I'$}
    \irule{ 
                \Gamma_0 \entails e_0\of\NatSafe \Quad{1.5}
                \Gamma_1 \entails e_1\of\NatNorm}{
        \Gamma_0\cup\Gamma_1\entails(\down\;e_0\;e_1)\of\NatNorm}  
        \\[1ex] 
     \rulelabel{If-$I'$}
     \irule{ 
                \Gamma_0\entails e_0\of\NatSafe \Quad{1.5}
                \Gamma_1\entails e_1\of\NatNorm \Quad{1.5}
                \Gamma_2\entails e_2\of\NatNorm            }{
                \Gamma_0\cup\Gamma_1\cup\Gamma_2
                 \entails 
                    (\Iif \;e_0 \; \Ithen \;e_1 \; 
                        \Ielse\; e_2)\of\NatNorm}                
\end{gather*}
\caption{Additional rules for $\bcl'$} \label{fig:bcl':down}
\figrule
\end{minipage}
\end{figure}
Our solution to this problem is to use a more relaxed version of 
$\subty$-predictivity than that of $\bcl$.  
To explain this let us consider $\bcl'$,  which is the result of 
adding rules of Figure~\ref{fig:bcl':down} to $\bcl$.  (The rewrite 
rule for $\down$ is given in Figure~\ref{fig:pcf:red}.)  These 
typing rules allow information about $\NatSafe$ values to flow into
$\NatNorm$ values, but only in very controlled ways.  
In \emph{$\down$-$I'$}, the controlling condition is 
that the length of this $\NatSafe$ information is bounded by the 
length of some prior $\NatNorm$ value.  In 
\emph{If-$I'$}, essentially 
only one bit of information about a $\NatSafe$ value is allowed to 
influence the $\NatNorm$ value of the expression.  
Because of these controlling conditions, the
proofs of Propositions~\ref{p:bc:size:polybnd},
\ref{p:bc:time:polybnd}, and~\ref{p:bc:pt:complete} go through for
$\bcl'$ with only minor changes, but in place of 
Proposition~\ref{p:bcl:inhab} we have:

\begin{proposition} \label{p:bcl':inhab}
  $\set{\semval{e} \suchthat\; \entails_{\bcl'}
  e\of\NatNorm\to\NatSafe\to\NatNorm} = $ the set of 
  polynomial-time computable 
  $f\in\nat\to\nat\to\nat$ such that
  $|f(x,y)| \leq |x|$ for all $x$ and $y$.
\end{proposition}

Each $\bcl'$ type $\gamma$ has a quantitative meaning in 
the sense that every element of  $\set{\semval{e} \suchthat\; 
\Gamma \entails_{\bcl'} e\of\gamma}$ has a polynomial size-bound 
of a particular form.  $\atr$ has rules  analogous to 
\emph{If-$I'$} and $\down$-$I'$ and, consequently, functions 
such as $F_2$ have well-typed definitions.  Moreover, each $\atr$ 
type $\gamma$ has a quantitative meaning in the sense that 
$\set{\semval{e} \suchthat\; \entails_{\atr} e\of\gamma} = $ 
the set of all $\atr$-computable functions having second-order 
polynomial size-bounds of a form dictated by $\gamma$.    
In particular, for each $\gamma$,  a $d_\gamma\in\omega$ can be 
read off such that all the bounding polynomials for 
type-$\gamma$ objects can be of depth $\leq d_\gamma$. 
This is the (non-na\"{\i}ve) connection of $\atr$'s type-system
to the notion of depth.
The above  glosses over the issue of the ``depth polymorphic'' 
higher types which are discussed in \S\ref{S:atr}.

\subsection*{Truncated fixed points} For $\PCF$, 
$\fix$ is thought as
expressing general recursion.  It would be ever so
convenient if one could replace $\fix$ with some higher-type
polynomial-time construct and obtain ``the'' feasible version of
$\PCF$ in which  all (and only) the 
polynomial-time recursion schemes are expressible. 
However, because of some basic
limitations of subrecursive programming formalisms 
\cite{March72,Royer87a}, it is unlikely that there is \emph{any} 
finite collection of constructs through which one can 
express all and only such recursion schemes.

Our goals are thus more modest. We make use of the programming
construct $\crec$, for \emph{clocked recursion}.  The $\crec$ construct is a 
descendant of Cobham's \cite{Cobham65} bounded recursion on 
notation and \emph{not} a true fixed-point constructor.  The 
reduction rule  for $\crec$ is:
\begin{gather}
      \crec \; a \;  (\lamr{f}e) 
\label{e:lbfix:red}
     \;           \longrightarrow  
     \; \lam{\vec{x}} 
      \left(\Iif \; |a|\leq |x_{1}|\;
      \Ithen 
         \;\left(e'\;\vec{x}\,\right) \;
      \Ielse \; \epsilon\right) \\
      \nonumber
       \hbox{with } e' = e[f\gets 
      \left(\strut \crec \;\; (\bo\concat a)\;
                (\lamr{f}e)\right) ],
\end{gather}
where $a$ is a constant and $\vec{x}=x_1, \ldots, x_k$ is a sequence 
of variables.  Roughly, $|a|$ acts as the tally of the number of 
recursions thus far and $\bo\concat a$ is the result of a tick of 
the clock.  The value of $x_1$ is the program's estimate of the 
total number of recursions it needs to do its job.  Typing 
constraints will make sure that each $\crec$-recursion terminates 
after polynomially-many steps.  Without these constraints, $\crec$
is essentially equivalent to $\fix$.
Clocking the fixed point process is a strong restriction.  
However, results on clocked programming systems
(\cite[Chapter~4]{RC94}) suggest that clocking, whether explicit or
implicit, is needed  to produce programs for which one can
determine explicit run-time bounds.  
\smallskip

Along with clocking, we impose two other restrictions on recursions.

\subsection*{One use} In any expression of the form $(\crec \; a\;
(\lamr{f}e))$,  we require that
$f$ has at most one \emph{use} in $e$.  Operationally
this means that, in any
possible evaluation of $e$, at most one application of $f$
takes place.  One  consequence of this restriction is
that no free occurrence of $f$ is allowed within any inner $\crec$
expression. (Even if $f$ \emph{occurs} but once in an inner 
$\crec$, the presumption is that $f$ may be \emph{used} many times.)  
Affine typing constraints enforce this one-use restriction.
Note that $\prn$ is a one-use form of recursion.

The motivation for the one-use restriction stems from the recurrence
equations that come out of time-complexity analyses of 
recursions.  Under the one-use restriction, bounds on the cost of
$m$ steps of a $\crec$ recursion are provided by recurrences of the
form $T(m,\vec{n}) \leq T(m-1,\vec{n})+q(\vec{n})$, where $\vec{n}$
represents the other parameters and $q$ is a (second-order)
polynomial.  Such $T$'s grow polynomially in $m$.  Thus, a
polynomial bound on the depth of a $\crec$ recursion implies a
polynomial bound on the recursion's total cost.  If, say, two uses
were allowed, the recurrences would be of the form $T(m,\vec{n})
\leq 2\cdot T(m-1,\vec{n})+q(\vec{n})$ and such $T$'s can grow
exponentially in $m$.

\subsection*{Tail recursions} We  restrict $\crec$ terms to expressing 
just tail recursions.  \emph{Terminology:} The \emph{tail terms} 
of an expression $e$ consist of: 
(i) $e$ itself,
(ii) $e'$, when $(\lam{x}e')$ is a tail term, and  
(iii) $e_1$ and $e_2$, when $(\Iif\;e_0\;\Ithen\;e_1\Ielse\;e_2)$ 
is a tail term.  A \emph{tail call} in $e$ is a tail
term of the form $(f\; e_1\; \dots \; e_k)$.  Informally, a tail
recursive definition is a function definition in which
every recursive call is a tail call.  Formally, we say that
$(\crec\; a \; (\lamr{f}e))$ \emph{expresses a tail
recursion} when each occurrence of $f$ in $e$ is as the head
of a tail call in $e$.\footnote{Because of the one-use restriction,
this simple definition of tail recursion suffices for this
paper.  For details on the more general notion see 
\cite{Reynolds:def2:98,EOPL:2}.} 

Simplicity is the foremost motivation for the restriction 
to tail recursions as they are easy to work with from both 
programming and  complexity-theoretic standpoints.  Additionally,  
tail recursion is a well-studied  and widely-used 
universal form of recursion: there are 
\emph{continuation passing style} translations of many program 
constructs into pure tail-recursive programs. (Reynolds 
\cite{Reynolds:cont:hist} provides a nice historical introduction.)
Understanding the complexity theoretic properties of tail-recursive 
programs should lead to an understanding of a much more general set 
of programs.

\section{Affine tiered recursion} \label{S:atr}

\subsection*{Syntax} 

$\atr$ (for \emph{affine tiered recursion}) has the same syntax
as $\PCF$ with three changes:
(i) $\fix$ is replaced with $\crec$ as discussed in the previous
section,
(ii) the only variables allowed are those of type-levels 0 and 1, 
and
(iii) the type system is altered as described below.

\begin{figure}[t] \small
\begin{minipage}{\textwidth}
\begin{gather*}
  E \;\; \is \;\;  \dots
        \synsep (\crec\;  K\; (\lamr{X}E) )\Quad3
  L \;\; \is \;\; (\Orl\Prg)^*  \synsep \Prg (\Orl\Prg)^*
  \\
  T_0 \;\;\is\;\; 
  \Natl{L}   \Quad3
  T \;\;\is\;\; 
    \hbox{the level 0, 1, and 2 simple types over $T_0$}
\end{gather*}
\caption{$\atr$ syntax} \label{fig:atr1:syn}
\end{minipage}
\figrule 
\end{figure}

\subsection*{Types} 
The $\atr$ types consist of \emph{labeled base types} ($T_0$ from
Figure~\ref{fig:atr1:syn}) and the level 1 and 2 simple types over
these base types.  We first consider labels ($L$ from 
Figure~\ref{fig:atr1:syn}).

\subsubsection*{Labels} Labels are strings of alternating $\Prg$'s 
and $\Orl$'s in which the rightmost symbol of a nonempty label is
always $\Prg$.  A label $\ba_k\dots\ba_0$ can be thought of as
describing program-oracle conversations: each symbol $\ba_i$
represents an action ($\Orl=$ an oracle action, $\Prg=$ a program
action) with the ordering in time being $\ba_0$ through $\ba_k$.
\emph{Terminology:} $\varepsilon$ = the empty label, $\ell\leq \ell'$
means label $\ell$ is a suffix of label $\ell'$, and $\ell\bmax \ell'$ is the
$\leq$-maximum of $\ell$ and $\ell'$. Also let $\successor(\ell) = $ the
successor of $\ell$ in the $\leq$-ordering, $\depth(\ell)=$ the number of
$\Orl$'s in $\ell$, and, for each $d\in\omega$, \ $\Orl_d =
(\Orl\Prg)^d$ and $\Prg_d = \Prg(\Orl\Prg)^d$.  Note:
$\depth(\Orl_d)=\depth(\Prg_d)=d$.

\subsubsection*{Labeled base types} The $\atr$ base types are all of the form $\Natl{\ell}$, where $\ell$ is a label.  
These base types are subtype-ordered by: 
$\Natl{\ell} \subty \Natl{\ell'}$ $\iff$ 
$\ell\leq \ell'$.  We thus have the linear ordering:
$
  \Natl{\varepsilon}                  \subty
  \Natl{\Prg}                         \subty
  \Natl{\Orl\Prg}                     \subty
  \Natl{\Prg\Orl\Prg}                 \subty
  \cdots\,,
$
or equivalently, 
$
  \Natl{\Orl_0}      \subty
  \Natl{\Prg_0}         \subty
  \Natl{\Orl_1}      \subty
  \Natl{\Prg_1}         \subty
  \cdots\;$.
Define $\depth(\Natl{\ell})=\depth(\ell)$.  $\Natl{\Orl_d}$ and
$\Natl{\Prg_d}$ are the depth-$d$ analogues of the $\bcl'$-types
$\NatNorm$ and $\NatSafe$, respectively.  These types can be
interpreted as follows.
\begin{itemize}
  \item A $\Natl{\varepsilon}$-value is an ordinary base-type input
        or else is bounded by some prior 
        (i.e., previously computed) $\Natl{\varepsilon}$-value.
  \item A $\Natl{\Prg_d}$-value is the result of a
        (type-2) 
        polynomial-time computation over 
        $\Natl{\Orl_{d}}$-values or else is bounded by some
        prior $\Natl{\Prg_d}$-value.
  \item A $\Natl{\Orl_{d+1}}$-value is the answer to a query
        made to a type-1 input on $\Natl{\Prg_d}$-values
        or else is bounded by some
        prior $\Natl{\Orl_{d+1}}$-value.
\end{itemize} 
The $\Natl{\Orl_d}$ types are called \emph{oracular}
and the $\Natl{\Prg_d}$'s are called 
\emph{computational}.

\subsubsection*{The $\atr$ arrow types} These are just the level 1
and 2 simple types over the $\Nat_\ell$'s.  The subtype relation
$\subty$ is extended to these arrow types as in \refeq{e:subty}.
\emph{Terminology:} Let $\shape(\sigma)=$ the simple type over
$\Nat$ resulting from erasing all the labels.  The \emph{tail} of a
type is given by:
\begin{gather*}
        \tail(\Natl{\ell}) \;=\; \Natl{\ell}.   \Quad2
        \tail(\sigma\to\tau) \;=\; \tail(\tau).
\end{gather*}
Let $\depth(\sigma) = \depth(\tail(\sigma))$.  When $\tail(\sigma)$
is oracular, we also call $\sigma$ oracular and let $\side(\sigma) =
\Orl$.  When $\tail(\sigma)$ is computational, we call $\sigma$
computational and let $\side(\sigma)=\Prg$.

\begin{definition}[Predicative, impredicative, 
flat, and strict types]  \label{d:impred-flat} 
  An $\atr$ type $\gamma$ is \emph{predicative} when $\gamma$ is a 
  base type or when $\gamma=(\sigma_1,\dots,\sigma_k)\to
  \Natl{\ell}$ and $\tail(\sigma_i) \subty\Natl{\ell}$ for each $i$.  
  A type is \emph{impredicative} 
  when it fails to be predicative.
  An $\atr$ type $(\sigma_{1},\dots,\sigma_{k})\to\Natl{\ell}$ is
  \emph{flat} when $\tail(\sigma_{i})=\Natl{\ell}$ for some $i$.  A
  type is \emph{strict} when it fails to be flat. 
\end{definition}

\emph{Examples:} $\Natl{\varepsilon} \to \Natl{\Prg}$ is predicative
whereas $\Natl{\Prg}\to \Natl{\varepsilon}$ is impredicative, and
both are strict.  Both $\Natl{\Prg}\to \Natl{\Prg}$ and
$\Natl{\Prg}\to\Natl{\Orl\Prg}\to \Natl{\Prg}$ are flat, but the
first is predicative and the second impredicative.  Recursive
definitions tend to involve flat types.

Example~\ref{e:disasters} below illustrates that values of both
impredicative and flat types require special restrictions in any
sensible semantics of $\atr$.  Our semantic restrictions for these
types are made precise in \S\ref{S:impred} and \S\ref{S:flat}
below. Here we give a quick sketch of these restrictions as they
figure in definition of $\shiftsto$, the \emph{shifts-to} relation,
used in the typing rules.  \emph{For each impredicative type
$(\vec{\sigma})\to\Natl{\ell}$}: if $\entails 
f\of(\vec{\sigma})\to\Natl{\ell}$,
then the value of $|f(\vec{x})|$ is essentially independent of the
values of the $|x_i|$'s with $\tail(\sigma_i) \suptyneq \Natl{\ell}$.
\emph{For each flat type $(\vec{\sigma})\to\Natl{\ell}$} (that for
simplicity here we further restrict to be a level-1 computational
type): if $\entails f\of(\vec{\sigma})\to\Natl{\ell}$, then
$|f(\vec{x})|\leq p + \bigmax\set{ |x_i|\suchthat \tail(\sigma_i) =
\Natl{\ell}}$, where $p$ is a second-order polynomial over elements of
$\set{ |x_i|\suchthat \tail(\sigma_i)\subtyneq \Natl{\ell}}$.  (Compare
this to the bound of Proposition~\ref{p:bc:size:polybnd}(b).)

\begin{figure}[t] \small
\begin{minipage}{\textwidth}
\vspace{-2ex}
\begin{gather*}
    \rulelabel{Zero-I}
    \irule{ 
        }{\Gamma;\Delta\entails \epsilon\of\Natl{\varepsilon}} 
        \Quad{3}
    \rulelabel{Const-I}
    \irule{ 
        }{\Gamma;\Delta\entails k\of\Natl{\Prg}} 
        \\[-0.5ex]
    \rulelabel{Int-Id-I}
    \irule{ 
        }{\Gamma,\,x\of\sigma;\Delta\entails x\of\sigma} 
        \Quad3
    \rulelabel{Aff-Id-I}
    \irule{ 
        }{\Gamma;x\of{\gamma}\entails x\of\gamma} 
        \\[1ex] 
    \rulelabel{Shift}
    \irule{ 
        \Gamma;\Delta\entails 
          e\of \sigma}{
        \Gamma;\Delta\entails e\of\tau}
        \sidecond{\sigma \shiftsto \tau}
        \Quad{3}
    \rulelabel{Subsumption}
    \irule{ 
        \Gamma;\Delta\entails e\of\sigma 
          }{
        \Gamma ; \Delta \entails  e\of\tau} 
        \sidecond{\sigma \subty \tau}
        \\[1ex] 
    \rulelabel{{\rm$\op$}-I}
    \irule{ 
        \Gamma;\Delta\entails e\of\Natl{\Prg_d}}{
        \Gamma;\Delta\entails (\op\;e)\of\Natl{\Prg_d}}  
    \Quad3        
    \rulelabel{$\down$-I}
    \irule{ 
                \Gamma;\Delta_0 \entails e_0\of\Natl{\ell_0} 
                \Quad{1.5}
                \Gamma;\Delta_1 \entails e_1\of\Natl{\ell_1} }{
        \Gamma;\Delta_0,\Delta_1 \entails
        (\down\;e_0\;e_1)\of\Natl{\ell_{1}}}         
        \\[1ex]  
    \rulelabel{$\to$-I}
    \irule{ 
        \Gamma,\,x\of\sigma;\Delta\entails e\of\tau}{
        \Gamma;\Delta\entails (\lam{x} e) 
           \of\sigma\to\tau} 
        \Quad3
    \rulelabel{$\to$-e}
    \irule{ 
        \Gamma;\Delta\entails e_0\of\sigma\to\tau 
                \Quad{1.5}
        \Gamma;\emptycont\entails e_1\of\sigma}{
                \Gamma;\Delta\entails (e_0\;e_1)\of \tau}
        \\[1ex] 
     \rulelabel{$\Iif$-I}
     \irule{ 
                \Gamma;\emptycont\entails e_0\of\Natl{\ell}  
                  \Quad{1.5} 
                \Gamma;\Delta_{1}\entails e_1\of\Natl{\ell'} 
                  \Quad{1.5}
                \Gamma;\Delta_{2}\entails e_2\of\Natl{\ell'}
                         }{
                \Gamma;\Delta_{1}\cup\Delta_{2} \entails 
                (\Iif \;e_0 \; \Ithen \;e_1 \; \Ielse\; e_2)\of
                \Natl{\ell'}}
        \\[1ex] 
    \rulelabel{$\crec$--I}
    \irule{ 
      \entails K\of\Natl{\Prg}   
        \Quad{1.5}
        \Gamma;f\of\gamma \entails e\of\gamma  
      }{
          \Gamma;\emptycont\entails 
                \left(
                \crec\;K \;(\lamr{f} e) 
                \right)
                \of \gamma}
      \sidecond{\gamma\in\mathcal{R} 
                     \strut
                     \hbox{ and }
                     \TailPos(f,e)}
  \\  
  \hbox{where:\hspace*{.9\textwidth}} 
  \\[-0.5ex]
  \mathcal{R} \;\;\defeq\;\;
    \Set{ (\bb_1,\bb_2,\dots,\bb_k)
        \to \bb \;\suchthat\;
        \hbox{$\bb_1$ and each $\bb_i\subty \bb_1$
        is oracular}}.\\[0.5ex]
  \TailPos(f,e) \;\;\defeq\;\;
  \left[\;\hbox{Each occurrence of $f$ in $e$ is as the
        head of a tail call}\;\right].                
\end{gather*}
\end{minipage}
\caption{$\atr$ typing rules}\label{fig:ltr:types}
\figrule 
\end{figure}
\subsection*{Typing rules}

The $\atr$-typing rules are given in Figure \ref{fig:ltr:types}.
The rules \emph{Zero-I, Const-I, Int-Id-I}, \emph{Subsumption},
$\op$-\emph{I}, $\to$-\emph{I}, and $\to$-\emph{E} are essentially
lifts from $\bcl$ (with one subtlety regarding $\to$-\emph{E} 
discussed below).  The $\Iif$-\emph{I} and $\down$-\emph{I} 
rules were motivated in \S\ref{S:better}.  The remaining three 
rules \emph{Aff-Id-I} and
$\crec$-\emph{I} (that relate to recursions and the split type
contexts) and \emph{Shift} (that coerces types) 
require some discussion.

\subsubsection*{Affinely restricted variables and {\rm $\crec$}} Each
$\atr$ type judgment is of the form $\Gamma;\Delta\entails
e\of\gamma$ where each type context is separated into two parts: a
\emph{intuitionistic zone} ($\Gamma$) and an \emph{affine}
\emph{zone} ($\Delta$).  $\Gamma$ and $\Delta$ are simply finite
maps (with disjoint preimages) from variables to $\atr$-types.  By
convention, ``$\emptycont$'' denotes an empty zone.  Also by
convention we shall restrict our attention to $\atr$ type
judgments in which each affine zone consists of at most one type
assignment.  (See Scholium~\ref{sch:atr:types}(a).)  In reading the
rules of Figure~\ref{fig:ltr:types}, think of a variable in an
affine zone as destined to be the recursor variable in some $\crec$
expression.  An intuitionistic zone can be thought of as assigning
types to each of the mundane variables.

\emph{Terminology:} A variable $f$ is said to be \emph{affinely
restricted} in $\Gamma;\Delta\entails e\of\sigma$ if and only if $f$
is assigned a type by $\Delta$ or is $\lambda_{r}$-abstracted over
in $e$.

The use of split type contexts is adapted from Barber and Plotkin's
DILL \cite{barber:dill:96,bp:dill:97},\footnote{The discussion of
DILL in \cite{ohearn:bunched:03} is quite helpful.}
a linear typing scheme that
permits a direct description of $\to$, the intuitionistic arrow of
the conventional simple types.  The key rule borrowed from DILL is
$\to$-\emph{E} which forbids free occurrences of affinely restricted
variables in the operand position of any intuitionistic application.
This precludes the typing of $\crec$-expressions containing subterms
such as $\lamr{f}( \lam{g} (g\,(g\,\epsilon))\, f)$ $\equiv_{\beta}
\lamr{f} (f\, (f \,\epsilon))$ where $f$ is used multiple times.

The $\crec$-\emph{I} rule forbids any free occurrence of an affinely
restricted variable; if such a free occurrence was allowed, it could
be used any number of times through the $\crec$-recursion.
The $\crec$-\emph{I} rule requires that the recursor variable have a
type $\gamma\in\SR$ which in turn becomes the type of the
$\crec$-expression.  The restrictions in $\SR$'s definition (in
Figure~\ref{fig:ltr:types}) are a more elaborate version of the
typing restrictions for $\prn$-expressions in $\bcl$.  When $\gamma
= (\Natl{\Orl_d}, \bb_2,\dots, \bb_k) \to \bb \in \SR$, it turns out
that $\SR$'s restrictions limit a type-$\gamma$ $\crec$-expression
to at most $p$-many recursions, where $p$ is some fixed, depth-$d$
second-order polynomial (Theorem~\ref{t:polybnd}).  
Excluding $\Natl{\Prg},\dots,
\Natl{\Prg_{d-1}}$ in $\gamma$ forbids depth $0,\dots,d-1$ analogues
of $\NatSafe$-parameters  from figuring in the recursion, and
consequently, the recursion cannot accumulate information that could
change the value of $p$ unboundedly.

\begin{scholium}\label{sch:atr:types} {\ }

(a)  Judgments with with multiple type assignments 
in their affine zone are derivable.  However, such a judgment is a
dead end in the sense that \emph{$\crec$-I}, the only means to
eliminate an affine-zone variable, requires a singleton affine zone.

(b) $\atr$ has no explicit $\lollipop$-types. \emph{Implicitly}, a
$(\lamr{f}e)$ subexpression is of type $\gamma\lollipop\gamma$ and
$\crec$-\emph{I} plays roles of both $\lollipop$-\emph{I} and
$\lollipop$-\emph{E}.  $\atr$'s very restricted use of affinity
permits this $\lollipop$-bypass.

(c)
As mentioned in \S \ref{S:better}, the restriction to tail recursions in 
\emph{$\crec$-I} is in the interest of simplicity. In a follow-up  
to the present paper,  we show how to relax this 
restriction to allow a broader range of affine recursions 
in $\atr$ programs \cite{Danner:Royer:twoalg}.
Dealing with this broader
range of recursions turns out to require nontrivial extensions of the 
techniques of \S\S \ref{S:time}--\ref{S:tcplus} below.

\end{scholium}

\subsubsection*{Shift}  The \emph{Shift} rule 
covariantly coerces the type of a term to be deeper.  Before stating
the definition of the shifts-to relation ($\shiftsto$), we first
consider the simple case of shifting types of shape $\Nat\to\Nat$.
The core idea is simply: $(\Natl{\ell_1}\to\Natl{\ell_0}) \shiftsto
(\Natl{\ell_1'}\to\Natl{\ell_0'})$ when $\depth(\Natl{\ell_0'}) -
\depth(\Natl{\ell_0}) = \depth(\Natl{\ell_1'}) - \depth(\Natl{\ell_1})\geq
0$.  The motivation for this is that if $p$ and $q$ are second-order
polynomials of depths $d_p$ and $d_q$, respectively, and $x$ is a
base-type variable appearing in $p$ that is treated as representing
a depth-$d_x$ value (with $d_x\leq d_q$), then $p[x\gets q]$ is, in
the worst case, of depth $d_p+(d_q - d_x)$.  The full story for
shifting level-1 types has to account of arbitrary arities, the
sides of the component types, and impredicative and flat types, but
even so it is still not too involved.  Shifting level-2 types
involves a new set of issues that we discuss after dealing with the
level-1 case.  Recall that $\max(\emptyset)=0$.

\begin{definition}[$\shiftsto$, the shifts-to relation] 
\label{d:shift} {\ }

(a)
We inductively define $\shiftsto$ by:
$\Natl{\Orl_d} \shiftsto \Natl{\Orl_{d'}}$ and
$\Natl{\Prg_d} \shiftsto \Natl{\Prg_{d'}}$
when $d\leq d'$; and 
$(\sigma_1\to\dots\to\sigma_k\to\Natl{\ell_0}) \shiftsto 
(\sigma_1'\to\dots\to\sigma_k'\to\Natl{\ell_0'})$ when

\quad(i) $\Natl{\ell_0} \shiftsto \Natl{\ell_0'}$,
$\sigma_1\shiftsto \sigma_1',\,\dots,\,\sigma_k\shiftsto \sigma_k'$,

\quad(ii) $\tail(\sigma_i)=\Natl{\ell_0}$
implies $\tail(\sigma_i')=\Natl{\ell_0'}$ for $i=1,\dots,k$, and 

\quad(iii) $\depth(\Natl{\ell_0'}) - \depth(\Natl{\ell_0}) \geq
D((\vec{\sigma})\to\Natl{\ell_0},\vec{\sigma}')$.

\smallskip
(b) $
D((\vec{\sigma})\to\Natl{\ell_0},\vec{\sigma}') 
  \defeq
        \max\set{\depth(\sigma_i')-\depth(\sigma_i) \suchthat
        \sigma_i\subty \Natl{\ell_0}},
$
for $\sigma_1\shiftsto \sigma_1',\,\dots,\,\sigma_k\shiftsto
\sigma_k'$ where each $\sigma_i$ and $\sigma_i'$ is a base type.
(See Definition~\ref{d:D2} for the general definition of $D$.) 
\end{definition}

For base types:
$\Natl{\ell}\shiftsto \Natl{\ell'}$ if and only if 
$\depth(\Natl{\ell})\leq\depth(\Natl{\ell'})$
and $\side(\Natl{\ell})=\side(\Natl{\ell'})$.
It follows from this and condition (i) that no type
(or component of a type) can change sides as a result
of a shift. 

For level-1 types: Condition (i) says that the component types on
the right are either the same as or else deeper versions of the
corresponding types on the left.  Condition (ii) preserves flatness
(which is critical in level-2 shifting).  Condition (iii) is just
the core idea stated above.  Note that the max in
Definition~\ref{d:shift}(b) includes only types $\subty \Natl{\ell_0}$.
This is because as remarked above, if $\sigma_i\suptyneq
\Natl{\ell_0}$, then the $i$-th argument has essentially no effect on
the size of the $\Natl{\ell_0'}$-result.

\emph{Example:} Consider the problem: $\Gamma; \emptycont \entails
f(f(x))\of{?}$, where $\Gamma = f \of \Natl{\Prg} \to
\Natl{\Orl\Prg},\, x \of \Natl{\Prg}$.  Using $\to$-\emph{E} and
\emph{Subsumption}, we derive $\Gamma;\emptycont\entails
f(x)\of\Natl{\Prg\Orl\Prg}$.  Using \emph{Shift} we derive
$\Gamma;\emptycont \entails f\of\Natl{\Prg\Orl\Prg} \to
\Natl{\Orl\Prg\Orl\Prg}$.  Using $\to$-\emph{E} again we obtain
$\Gamma; \emptycont \entails f(f(x))\of\Natl{\Orl\Prg\Orl\Prg}$ as
desired.

Now let us consider shifting level-2 types.  Suppose we want to
shift $(\Natl{\Orl_0}\to\Natl{\Orl_1})   \to \Natl{\Orl_3}$ to
some type of the form $(\Natl{\Orl_0}\to\Natl{\Orl_2})   \to 
\Natl{\Orl_d}$.  What  should the value of $d$ be?
Suppose $f \of\Natl{\Orl_0}\to\Natl{\Orl_1}$.  Without
using subsumption, building a term of type $\Natl{\Orl_3}$
from $f$ requires nesting applications of $f$ (using
type-1 shifts).  The longest chain of such depth-increasing
applications is 3.\footnote{Note
  that the outer two of these three applications must 
  involve shifting the type of the argument.  Also, informally, 
  in $f(f(f(\down(f(f(f(f(\epsilon)))),\epsilon))))$ only the 
  outer three applications of $f$ count as \emph{a chain of 
  depth-increasing applications} because of the drop in depth 
  caused by the $\down$. Formally, no shadowed 
  (Definition~\ref{d:shadowing}) application can be in a 
  depth-increasing chain.}  
When the argument type $\Natl{\Orl_0}\to\Natl{\Orl_1}$ is shifted to
$\Natl{\Orl_0}\to\Natl{\Orl_2}$, each application of this argument
now ups the depth by an additional $+1$.  So, the largest depth that
can result from the change is $d=3 + 3 \cdot 1 = 6$.  When shifting
$(\vec{\sigma})\to\Natl{\ell}$ to some $(\vec{\sigma}')\to\Natl{\ell'}$
with each $\sigma_i$ and $\sigma_i'$ a level-1 type, to determine
$\ell'$ we must:
(a) determine all the ways a $\Natl{\ell}$ value could be built by 
a chain of depth-increasing applications of arguments of the types 
$\vec{\sigma}$, (b) for each of these ways, figure the increase
in the depth of the $\Natl{\ell}$-value when each $\sigma_i$-argument 
is replaced by its $\sigma_i'$ version, and (c) compute the maximum
of these increases.   To help in this, we introduce
$\undo$ in Figure~\ref{fig:undo}.
\begin{figure}[t] \small
\begin{minipage}{\textwidth}
\begin{align*}
  &\undo((\Natl{\ell_1},\dots,\Natl{\ell_k})\to\Natl{\ell_0},
  \Natl{\ell})\;\defeq \\
  &\quad
  \begin{cases}
     \hbox{undefined}, 
       & \hbox{if $(\Natl{\ell_1},\dots,\Natl{\ell_k})\to\Natl{\ell_0}$ 
       is flat or $\ell_0 > \ell$;}\\
    \Natl{\ell' \concat \ell''}, &
    \hbox{otherwise, where $\ell' = \max\set{\ell_i \suchthat \ell_i<\ell_0}$
       and $\ell''$ is the}\\
       & \hbox{\Quad1 suffix of 
       $\ell$ following the leftmost occurrence of $\ell_0$ in $\ell$.}
   \end{cases}  
\end{align*} 
\caption{The definition of $\undo$} \label{fig:undo}
\end{minipage}
\figrule 
\end{figure} 
\emph{Example:} For $d>0$, \ 
$\undo(\Natl{\Orl_0}\to\Natl{\Orl_1},\Natl{\Orl_d}) = 
 \undo(\Natl{\Orl_0}\to\Natl{\Orl_1},\Natl{\Prg_d}) = 
 \Natl{\Orl_{d-1}}$.
To compute $\undo(\tau,\Natl{\ell})$, one determines if a type-$\tau$ 
argument could be used in a chain of depth-increasing  applications 
that build a $\Natl{\ell}$ value, and if so, one figures (in terms of 
$\ell$) where a leftmost application of such an argument could occur, 
and  returns the $\subty$-largest  type of the arguments of this 
application.  (It is straightforward to prove that $\undo$ behaves 
as claimed.)  
\textbf{N.B.} If $\undo(\gamma,\Nat_\ell)$ is defined, then
$\undo(\gamma,\Nat_\ell) \subtyneq \Nat_\ell$. 
We now define:

\begin{definition}[$D$ for level-2 types] \label{d:D2}
Suppose $\sigma_1\shiftsto \sigma_1',\dots,\sigma_k\shiftsto
\sigma_k'$.

(a) 
$D((\vec{\sigma})\to \Natl{\ell},\vec{\sigma}') \defeq
\max(\{\,\depth(\sigma_i')-\depth(\sigma_i) 
 + 
D\big((\vec{\sigma})\to\undo(\sigma_i,\Natl{\ell}),\vec{\sigma}'\big) 
\suchthat 
\undo(\sigma_i,\Natl{\ell})$ is defined$\,\})$,
when each $\sigma_i$ and $\sigma_i'$ is level-1.

(b) 
$D((\vec{\sigma})\to\Natl{\ell},\vec{\sigma}') \defeq$
$D((\vec{\sigma})_0\to\Natl{\ell},(\vec{\sigma}')_0) + 
D((\vec{\sigma})_1\to\Natl{\ell},(\vec{\sigma}')_1)$, 
when the $\sigma_i$'s contain both level-0 and level-1 types and 
where $(\vec{\gamma})_i$ denotes the subsequence of level-$i$ types
of $\vec{\gamma}$. 
\end{definition}

The recursion of Definition~\ref{d:D2}(a) determines maximum
increase in depth as outlined above.  Since applications 
amount to simultaneous substitutions, the contributions of 
the level-0 and level-1 argument shifts are independent.
Thus Definition~\ref{d:D2}(b)'s formula suffices for the 
general case.
\emph{Example:} See the discussion below of \emph{fcat}
from Figure~\ref{fig:atr:prn}.

Now let us consider the reason behind condition (ii) in
Definition~\ref{d:shift}(a).  A term of a flat type can be used an
\emph{arbitrary} number of times in constructing a value.
Consequently, if Definition~\ref{d:shift}(a) had allowed flat
level-1 types (which increase the depth by $0$) to be shifted to
strict level-1 types (which increase the depth by a positive
amount), then it would have been impossible to bound the depth
increase of shifts involving arguments of flat types.

\subsection*{Some examples} 

Figure~\ref{fig:atr:prn} contains five sample programs.  These
examples use the syntactic sugar of the $\Ilet$ and $\Iletrec$
constructs.\footnote{Where \label{fn:letrec}
  $(\Iletrec\;f = e' \;\Iin\; e)
  \; \stackrel{\text{def}}{\equiv} \; e[f \gets (\crec
  \;\tally0\; (\lamr{f}e')) ]$ and $\Ilet$ is as in
  foot\-note~\ref{fn:let}.}  
The first three programs and their typing are all straightforward.
For the typing of \emph{fcat}, \emph{cat}'s type is shifted to
$\Natl{\Orl_1} \to\Natl{\Prg_1} \to\Natl{\Prg_1}$ and \textit{prn}'s
type is shifted to $(\Natl{\Prg_0}\to \Natl{\Prg_1} \to
\Natl{\Prg_1})\to \Natl{\Orl_0}\to \Natl{\Prg_1}$.  The final
program computes
\begin{gather} \label{e:findk}
\lam{f\in(\nat\to\nat),x\in\nat}
\begin{cases}
  (\mu k<x)\left[k=\max_{i\leq k}\lendy(f(i))\strut\right],
     & \hbox{if such a $k$ exists;}\\
  x, & \hbox{otherwise;}  
\end{cases}
\end{gather}
where $\lendy(z)$ = the dyadic representation of the length of $z$.
This is a surprising and subtle example of a BFF due to
Kapron \cite{Kapron:thesis} and was a key example that lead to 
the Kapron-Cook Theorem \cite{KapronCook:mach}.  
In \emph{findk}, we assume we have: a
type-$(\Nat_{\Orl_1}\to\Nat_{\Orl_1}\to\Nat_{\Orl_1})$
definition of $(x,y)\mapsto[\tally1,$ if $x=y$; $\epsilon$,
otherwise], a type-$(\Nat_{\Orl_1}\to\Nat_{\Orl_1})$ definition
of $\lendy$, a type-$(\Nat_{\Orl_1} \to \Nat_{\Orl_1} \to
\Nat_{\Orl_1})$ definition of \textit{max}, and a
type-$(\Nat_{\Prg_0}\to \Nat_{\Prg_0})$ definition of $x\mapsto x+1$.
Filling in these missing definitions is a straightforward exercise.
A more challenging exercise is to define \refeq{e:findk} via
$\emph{prn}$'s.
\begin{figure}[t]\small
\newcommand{\Com}[1]{\hfill\emph{// \makebox[6cm][l]{#1}}}
\begin{verse}
$\textit{reverse}\of 
    \Natl{\varepsilon} \to\Natl{\Prg}$ \ = 
     \Com{reverse $\ba_1\dots\ba_k\;=\;\ba_k\dots\ba_1$.}\\
\Quad1 $\lam{w} \Iletrec$ 
          $f\of\Natl{\varepsilon}\to\Natl{\Prg}\to\Natl{\Prg}\to\Natl{\Prg}$ \ = \\
\Quad4     $\lam{b,\,x,\,r}\Iif$ $(\tsto\;x)$ \Quad{1.55}
              $\Ithen$ $f\; b\; (\cdr\; x)\; (\conso\; r)$\\
\Quad8       $\Ielse$ $\Iif$ $(\tstl\;x)$ 
              $\Ithen$ $f\; b\; (\cdr\; x)\; (\consl\; r)$ \\
\Quad8  $\Ielse$ $r$\\
\Quad{3}     $\Iin$ $f$ $w$ $w$  $\epsilon$ \\[2ex]
$\textit{prn}\of(
               \Natl{\Prg}\to
               \Natl{\Prg}
               \to\Natl{\Prg})\to
                  \Natl{\varepsilon}
                  \to\Natl{\Prg}$ \ = 
                  \Com{See~\refeq{e:prn}.}
                  \\
\Quad1 
  $\lam{e,\, y} \Iletrec$    
     $f\of\Natl{\varepsilon}\to
           \Natl{\Prg}\to
           \Natl{\Prg}\to
           \Natl{\Prg}
         \to\Natl{\Prg}$ \ = \\
\Quad5
    $\lam{b,\,
          x,\,
          z,\,
          r} \Iif$ $(\tsto\;x)$ \Quad{1.4}
          $\Ithen$ $f\;b \; (\cdr\; x)\; (\conso\; z)
  \; (e\; (\conso\;z) \;r)$\\
\Quad{10}
  $\Ielse$ $\Iif$ $(\tstl\;x)$ $\Ithen$ 
    $f\;b \; (\cdr\; x)\; (\consl\; z) \; 
         (e\; (\consl\;z) \;r)$\\
\Quad{10}
  $\Ielse$ $r$ \\
\Quad{3.9} $\Iin$ $f\;y\;(\textit{reverse}\; y)\;\epsilon
   \;(e\;\epsilon\;\epsilon)$  \\[2ex]
\emph{cat}$\of
     \Natl{\varepsilon} \to
     \Natl{\Prg} \to
     \Natl{\Prg} $ \ =
     \Com{cat w x = $w\concat x$ as before.}\\
\Quad1 $\lam{w,\,x}\Ilet$ $f\of\Natl{\Prg}\to
                       \Natl{\Prg}
                       \to\Natl{\Prg}$ \ = \\
\Quad5 $\lam{y,\,z}\Iif$ $(\tsto\;y)$ $\Ithen$ $(\conso\;z)$ 
   $\Ielse$ $\Iif$ $(\tstl\;y)$ $\Ithen$ $(\consl\;z)$
   $\Ielse$ $x$\\
\Quad{4.2}    $\Iin$ \textit{prn} $f$ $w$\\[2ex]
\emph{fcat}$\of
        (\Natl{\Prg} \to \Natl{\Orl\Prg}) \to
        \Natl{\varepsilon} \to
        \Natl{\Prg\Orl\Prg}$ \ = 
       \Com{fcat f $\,\ba_1\dots\ba_k = 
         (f\,\ba_1\dots\ba_k)\concat $ }\\ 
        \Quad1 $\lam{f,\,x}$ $\Ilet$ $e\of\Natl{\Prg}\to
                       \Natl{\Prg\Orl\Prg}
                       \to\Natl{\Prg\Orl\Prg}$ \ = 
     \Com{\Quad3 $(f\,\ba_2\dots\ba_k)\concat \cdots
     \concat (f\,\ba_k)\concat (f\,\epsilon)$}                       
                       \\
\Quad5                        
                        $\lam{y,\,r} (cat\; (f\; y) \; r)$ \\
\Quad{4.2}                        
                         $\Iin$ $prn$ $e$ $x$  \\[2ex]
\emph{findk}$\of(\Nat_\Prg\to\Nat_{\Orl\Prg})\to
\Nat_{\varepsilon}\to\Nat_{\varepsilon}$
    \ =  \Com{See \refeq{e:findk}} \\
\Quad1 $\lam{f,x}$ $\Iletrec$
   $h \of \Nat_{\Orl\Prg} \to \Nat_{\varepsilon} \to\Nat_{\varepsilon}$ =
   \Com{Invariant: $k\leq \lendy(m) $
   and $|m| \leq |f|(|x|)$}    \\
\Quad5 $\lam{m,k}$  
            $\Iif$ $k==x$ $\Ithen$ $k$ \\
\Quad{8.25} $\Ielse$ $\Iif$ $k==(\lendy\; m)$ 
                      $\Ithen$ $k$ \\
\Quad{8.25} 
                      $\Ielse$ $h$ $(\textit{max}$ 
                             $(f$ $(k+1))$ $m)$ 
                             $(\down \; (k+1)\; x)$ \\
\Quad4 $\Iin$ \ $h$ $(f$ $\epsilon)$  $\epsilon$
\end{verse}
\caption{$\atr$ versions of \emph{reverse}, \emph{prn},
\emph{cat}, \emph{fcat}, and \emph{findk}} \label{fig:atr:prn}
\figrule
\end{figure}

\subsection*{Semantics}

The CEK machine of (\S\ref{S:cek:mach}) provides an operational
semantics of $\atr$.  For a denotational semantics we
\emph{provisionally} take the obvious modification of $\PCF$'s
$\mathcal{V}$-semantics.  
($\mathcal{V}$ was introduced in \S \ref{S:defs:pcf}.)
Example~\ref{e:disasters} illustrates some
inherent difficulties with $\mathcal{V}$ as a semantics for
$\atr$.  We shall circumvent these difficulties by some selective
pruning of $\mathcal{V}$ in \S\ref{S:impred} and \S\ref{S:flat}.

\subsection*{Some syntactic properties}

\begin{definition}[Use] \label{d:use}
 If variable $x$ fails to occur free in expression $e$, 
 then $\uses(x,e)=0$; otherwise $\uses(x,e)$ is given by:
 \begin{gather*}
    \uses(x,x)=1. 
    \Quad2
    \uses(x,(\op\;e_0)) = 
    \uses(x,(\lam{y}e_0)) = \uses(x,e_0).
    \\
    \uses(x,(\down\;e_0\;e_1)) = \uses(x,(e_0\; e_1)) = 
    \uses(x,e_0)+\uses(x,e_1).
    \\
    \uses(x,(\Iif \; e_0\; 
          \Ithen\;e_1\;\Ielse\; e_2 )) 
          = \uses(x,e_0) + \uses(x,e_1)\bmax\uses(x,e_2).
    \\
    \uses(x,(\crec\;K\;(\lamr{f}e_0))) = \unbounded\;
    (\equiv \emph{ unbounded}).
 \end{gather*}
 By convention,   
 $a<\unbounded$ and 
 $a+\unbounded=\unbounded+a =a\bmax \unbounded
  = \unbounded\bmax a = \unbounded$ for each $a\in\nat$.
\end{definition}

\begin{lemma}[One-use]\label{l:1use}
  If
  $\Gamma;f\of\gamma\entails e\of\gamma$
  or $\Gamma;\emptycont\entails (\crec\; k \; (\lamr{f} e))
  \of \gamma$, then $\uses(f,e)\leq 1$.
\end{lemma}  

\begin{lemma}[Subject reduction]\label{l:ltr:subred}
  If $\Gamma;\Delta\entails e\of\gamma$ and 
  $e$ $\beta\eta$-reduces to $e'$,
  then $\Gamma;\Delta\entails e'\of\gamma$.
\end{lemma}

\begin{lemma}[Unique typing of subterms] \label{l:unique}
  If $\Gamma;\Delta\entails e\of\sigma$,  then each
  occurrence of a subterm in $e$ has a uniquely assignable
  type that is 
  consistent with $\Gamma;\Delta\entails e\of\sigma$.
\end{lemma}

\begin{lemma} \label{l:shifty}
  $\Gamma;\Delta\entails 
  \lam{\vec{x}}e\of (\vec{\sigma})\to\Nat_\ell$
  if and only if 
  $\Gamma,\vec{x}\of\vec{\sigma};\Delta\entails 
  e\of\Nat_\ell$.
\end{lemma}

Lemma~\ref{l:1use} follows from a straightforward structural
induction on judgment derivations.  The proof of
Lemma~\ref{l:ltr:subred} is an adaptation the argument for
\cite[Theorem~15.3.4]{Pierce:types}.  The proof of
Lemma~\ref{l:unique} is also an adaptation of standard arguments.
We make frequent, implicit use of Lemma~\ref{l:unique} below.
Lemma~\ref{l:shifty} is a  reality check on the definition of
$\shiftsto$.  The proof of this is a completely standard induction on
derivations \emph{except} in the case where the last rule used in
deriving $\Gamma;\Delta\entails \lam{\vec{x}}e\of
(\vec{\sigma})\to\Nat_\ell$ is \emph{Shift}.  The argument for this
case is an induction on the structure of $e$, where application is
the key subcase.  There one simply checks that our definition of
$\shiftsto$ correctly calculates upper bounds on the increase in
depth.

\subsection*{$\atr$'s computational limitations and capabilities}

The major goals of the rest of the paper are to establish type-level
2 analogues of Propositions~\ref{p:bc:size:polybnd},
\ref{p:bc:time:polybnd}, and~\ref{p:bc:pt:complete} for $\atr$.
We shall first prove Theorem~\ref{t:polybnd}, a polynomial 
size-boundedness result for $\atr$.  The groundwork for this result
will be the investigation of second-order size-bounds in the next
few sections.

\begin{remark}[Related work]\label{rem:ramtypes}
  As noted in \S \ref{S:intro}, ramified types based on Bellantoni 
  and Cook's ideas, higher types, and linear types are common 
  features of  work on implicit complexity (see Hofmann's survey 
  \cite{Hofmann:survey}),  but most of that work has focused on 
  guaranteeing complexity of type-level 1 programs.  The $\atr$ 
  type system is roughly a refinement of the type systems of 
  \cite{IKR:I,IKR:II} which were constructed to help study 
  higher-type complexity classes.  Also, the type systems of this 
  paper and \cite{IKR:I,IKR:II} were greatly influenced by 
  Leivant's elegant ramified type systems 
  \cite{Leivant:FM2,Leivant:poly}.
  We note that in \cite{Leivant03} Leivant proposes a formalism that
  uses intersection types to address the same problems dealt with
  by our \emph{Shift} rule (e.g., how to type $f(f(x))$).
\end{remark}

\begin{remark}[Pragmatic predicativity]\label{rem:defense}
Many of the formalisms based on Bellantoni and Cook's ideas are  
predicative in the sense of Proposition~\ref{p:bcl:inhab}---no 
information about ``safe values'' can influence ``normal values.''
Two principles followed in this paper are: (i)
The ramification of data (e.g., the normal/safe distinction) 
and the complexity it adds to the type system is 
something we will put up with to control the size of values;
(ii) however, if there is a good reason to cut through the 
ramification while still controlling sizes, then we will happily 
do so.
As a consequence of (i), our type system for second-order 
polynomial size-bounds is strictly predicative.
As a consequence of (ii), $\atr$'s type system includes the 
$\Iif$-\emph{I} and $\down$-\emph{I} rules and impredicative types
to handle examples like $F_2$ of \refeq{e:f2}.

There is a price for the $\down$ construct---its use tends 
to complicate correctness arguments
for algorithms.  For example, consider the subexpression 
$(\down \; (k+1)\; x)$ in the $\atr$-program for 
\emph{findk} in Figure~\ref{fig:atr:prn}.
The purpose of the $\down$ is to guarantee to the type system 
that the subexpression's value is small (e.g., $\leq|x|$).  
The correctness of the algorithm depends critically 
on the easy observation that, in any run of the program,
the value of the subexpression will always be $k+1$.
This is common in expressing algorithms in $\atr$---one knows 
that a value is small, but an application of $\down$
is needed to convince the type-system of this. As a result
the correctness proof needs a lemma showing that
original value is indeed small and the $\down$ expression
does not change the value.
Thus our use of $\down$ and (mild) impredicativity is
a compromise between the simplicity, but restrictiveness,
of predicative systems and the richer, but more complex,
type systems that permit finer reasoning about 
size.\footnote{Hofmann's work on non-size-increasing functions 
\cite{Hofmann03,Hofmann:strength} provides a nice example
of a type system for fine control of sizes, but 
that system is not helpful in dealing 
with the $F_2$ or \emph{findk} examples.}

\end{remark}

\section{Size bounds}\label{S:polys}

\subsection{The second-order polynomials under the size types}

To work with size bounds, we introduce the \emph{size types}
and a typing of second-order polynomials under these types.
The size types parallel the intuitionistic part of $\atr$'s 
type system.

\begin{figure}[t]\small
\begin{gather*}
    \rulelabel{Zero-I}  
    \irule{}{
        \Sigma\entails\tally{0}\of\Tallyl{\varepsilon} }
        \Quad3
    \rulelabel{Const-I}  
    \irule{}{
        \Sigma\entails\tally{k}\of\Tallyl{\Prg} }
        \Quad3
    \rulelabel{Subsumption} 
    \irule{
        \Sigma\entails p\of\sigma}{
        \Sigma\entails p\of\tau} \sidecond{\sigma \subty \tau}
        \\[0.25ex]
    \rulelabel{Shift} 
    \irule{
        \Sigma\entails p\of\sigma}{
        \Sigma\entails p\of\tau} \sidecond{\sigma \shiftsto \tau}
        \Quad{3}
    \rulelabel{$\bmax$-I}  
    \irule{
        \Sigma_0\entails p_0\of\Tallyl{\ell} \Quad{1.5}
        \Sigma_1\entails p_1\of\Tallyl{\ell}}{
        \Sigma_0\cup\Sigma_1\entails 
           (\bmax\; p_0\; p_1)\of\Tallyl{\ell}}
        \\[0.25ex]
    \rulelabel{$+$-I} 
    \irule{\Sigma_{1}\entails p_{1}\of\Tallyl{\Prg_d} \Quad{1.25} 
           \Sigma_{2}\entails p_{2}\of\Tallyl{\Prg_d}
    }{\Sigma_{1}\cup\Sigma_{2}\entails 
        (+\;p_{1}\; p_{2})\of\Tallyl{\Prg_d}}  
    \Quad{3}
    \rulelabel{$\ast$-I} 
    \irule{\Sigma_{1}\entails p_{1}\of\Tallyl{\Prg_d} \Quad{1.25} 
           \Sigma_{2}\entails p_{2}\of\Tallyl{\Prg_d}
    }{\Sigma_{1}\cup\Sigma_{2}\entails 
        (\ast\;p_{1}\; p_{2})\of\Tallyl{\Prg_d}}  	
\end{gather*}
\caption{Additional typing rules for the second-order polynomials under the size types}\label{fig:sizepolytyping}
\figrule 
\end{figure}

\begin{definition}\label{d:sizetypespolys}{\ }

  (a)
  For each $\atr$ type $\sigma$, let $|\sigma| =
  \sigma[\Nat\gets\Tally]$.  (E.g., $|\Nat_\epsilon \to\Nat_{\Prg}| 
  = \Tally_\epsilon \to\Tally_{\Prg}$.) 
  These $|\sigma|$'s are the \emph{size types}. 
  All the $\atr$-types terminology and operations  (e.g., 
  $\shape$,  $\tail$, $\subty$, $\shiftsto$, etc.) are defined 
  analogously for  size types.
  
  (b) 
  The typing rules for the second-order polynomials under the
  size types consist of \emph{Id-I}, \emph{$\to$-I}, and   
  \emph{$\to$-E} from Figure~\ref{fig:pcf:typing} 
  and the rules of Figure~\ref{fig:sizepolytyping}.  
\end{definition}

Recall the $\mathcal{L}$-semantics for second-order polynomials
introduced in \S \ref{S:defs:2orderpoly}.
We \emph{provisionally} take $\semlen{\sigma}=
\semlen{\shape(\sigma)}$ and define 
$\semlen{\Sigma\entails p\of\sigma}$ 
as before.  Later, a  pruned version 
of the $\mathcal{L}$-semantics will end up as our intended 
semantics for the second-order polynomials
to parallel our pruning of the $\SV$-semantics for $\atr$. 

The following definition formalizes what it means for an $\atr$ 
expression to be polynomially size-bounded.  \textbf{N.B.} This 
definition heavily overloads the ``length of'' notation, $|\cdot|$. 
In particular, if $x$ is an $\atr$ variable, we  treat $|x|$ 
as a size-expression variable.  Definition~\ref{d:sizebnd}(c) is 
based on a similar notion from \cite{IKR:II}.  

\begin{definition}\label{d:sizebnd}
Suppose $\Gamma;\Delta\entails e\of \sigma$ is an $\atr$-type 
judgment.

(a) $|\Gamma;\Delta| \defeq
\Set{ |x| \mapsto |\sigma| \suchthat 
(\Gamma;\Delta)(x)=\sigma}$.

(b)
For each $\rho\in\semval{\Gamma;\Delta}$,  define 
$|\rho| \in\semlen{|\Gamma;\Delta|}$ by $|\rho|(|x|) =
|\rho(x)|$.\footnote{\textbf{N.B.} 
  The $|\cdot|$ in ``$|x|$'' is syntactic, whereas the $|\cdot|$ 
  in ``$|\rho|$'' and ``$|\rho(x)|$'' are semantic.}

(c)
We say that  the second-order polynomial $p$ 
\emph{bounds the size of $e$} (or, $p$ is a 
\emph{size-bound for $e$}) with respect to $\Gamma;\Delta$
when
$|\Gamma;\Delta|\entails p\of|\sigma|$ and
$|\semval{e}\,\rho| \;\leq_{|\sigma|}\; \semlen{p}\,|\rho|$
for all $\rho\in \semval{\Gamma;\Delta}$.
(The ``with respect to'' clause is dropped when it is clear from 
context.)
\end{definition}

Lemmas~\ref{l:tt:subred}, \ref{l:labelsoundness}, 
and~\ref{l:0gamma} below note a few basic properties of the 
second-order polynomials under the size types.  
Lemma~\ref{l:labelsoundness} connects the depth of a 
second-order polynomial $p$  and the depths of 
the types assignable to $p$.  
Lemmas~\ref{l:tt:subred} and~\ref{l:shifty'} follow by 
proofs similar to those for Lemmas~\ref{l:ltr:subred}
and~\ref{l:shifty}.
Lemma~\ref{l:labelsoundness}'s proof is a 
straightforward induction on judgment derivations, and 
Lemma~\ref{l:0gamma} is just an observation.
\emph{Terminology:}  
Inductively define $\tally0_\gamma$ by:
$\tally{0}_{\Tallyl{\ell}} = \tally0$
and  $\tally{0}_{\,\sigma\to\tau} =
\lam{x}\tally{0}_{\tau}$.
By abuse of notation, we often write $\tally{0}_{\gamma}$ for 
$\semlen{\entails \tally0_{\gamma}\of\gamma} \emptyenv$.

\begin{lemma}[Subject Reduction]\label{l:tt:subred}
  Suppose $\Sigma\entails p\of\sigma$ and $p$ 
  $\beta\eta$-reduces  to $p'$.  Then $\Sigma\entails p'\of\sigma$.
\end{lemma}

\begin{lemma} \label{l:shifty'}
  $\Sigma\entails 
  \lam{\vec{x}}p\of (\vec{\sigma})\to\Tallyl{\ell}$
  if and only if 
  $\Sigma,\vec{x}\of\vec{\sigma}\entails 
  p\of\Tallyl{\ell}$.
\end{lemma}

\begin{lemma}[Label Soundness]\label{l:labelsoundness} 
  Suppose $\Sigma\entails p\of\sigma$ has a derivation in
  which the only types assigned by contexts are from
  $\set{\Natl{\varepsilon}}\cup
  \set{(\Natl{\Prg}^k)\to\Natl{\Orl\Prg}\suchthat k>0}$.
  Then $\depth(p)\leq\depth(\sigma)$.
\end{lemma}

\begin{lemma}\label{l:0gamma}
  $\tally0_\gamma$ is the least element of $\semlen{\gamma}$ 
  under the pointwise ordering.
\end{lemma}

\subsection{Semantic troubles}

The na\"{\i}ve (and \emph{false!}) $\atr$-analogue of 
Proposition~\ref{p:bc:size:polybnd} is:  
\begin{center}\em
  For each $\Gamma;\Delta \entails e\of\sigma$, there is a $p_e$ 
  that bounds the size of $e$   with respect to $\Gamma;\Delta$.
\end{center}  
Example~\ref{e:disasters} illustrates the problems with this.
\textbf{N.B.} If the definition of $\bcl$ had allowed 
unrestricted free variables of type-level 1, the problems of 
Example~\ref{e:disasters} would have occurred in that setting too.

\begin{figure}[t]\small
\newcommand{\Com}[1]{\hfill\emph{// \makebox[5cm][l]{#1}}}
\begin{verse}
$e_1\of \Natl{\varepsilon} \to\Natl{\Prg} \;  =\;$  
\Com{Assume $g_1\of\Natl{\Prg}\to\Natl{\varepsilon}$.}
\\
\Quad1   
  $\lam{w}\Ilet$ $h_1\of\Natl{\Prg} \to \Natl{\Prg}
  \to \Natl{\Prg}$ =   $\lam{x,y}
  \Iif\; x\not=\epsilon \; \Ithen\; (dup\; (g_1\; y)\; (g_1\; y))
  \; \Ielse \; w$ \\
\Quad{3.1} $\Iin$ \ $prn\; h_1\; w$  \\[1ex]
  $e_2\of \Natl{\varepsilon} \to\Natl{\Prg} \;=\; $ 
\Com{Assume $g_2\of\Natl{\Prg}\to\Natl{\Prg}$.}
\\
\Quad1  
  $\lam{w}\Ilet$ $h_2\of\Natl{\Prg} \to \Natl{\Prg}
  \to \Natl{\Prg}$ =   $\lam{x,y}
  \Iif\; x\not=\epsilon \; \Ithen\; (g_2\; y)
  \; \Ielse \; w$ \\
  \Quad{3.1}
    $\Iin$ \ $prn\; h_2\; w$  
\end{verse}
\caption{Two problematic programs}
\label{fig:disasters}
\figrule
\end{figure}

\begin{example} \label{e:disasters} 
Let $e_1$ and $e_2$ be as given
in Figure~\ref{fig:disasters}, let \emph{prn} 
be as in Figure~\ref{fig:atr:prn}, 
and let \emph{dup} be an $\atr$-version of the definition 
in Figure~\ref{fig:bcl:ex}.

(a)  Suppose $\Gamma_1= g_1\of \Natl{\Prg}\to\Natl{\varepsilon}$ 
and $\rho_1=\set{g_1\mapsto\lam{z\in\nat}z}$.
Then $|\semval{e_1}\,\rho_1| =
\lam{n\in\omega}n^{2^n}$.
Note $|\rho_1(g_1)| = \lam{n\in\omega}n$ is a polynomial function.
The problem is that  $\rho_1(g_1)  = \lam{x\in\nat}x$ subverts the 
intent of the type-system by allowing an unrestricted flow of 
information about ``safe'' values into ``normal'' 
values.\footnote{By    \label{fn:bang}
  using a similar trick and the full power of $\crec$, one can write
  nonterminating $\atr$ programs.}

(b) Suppose $\Gamma_2= g_2\of \Natl{\Prg} \to \Natl{\Prg}$ and 
$\rho_2=\set{g_2\mapsto \lam{z\in\nat}z\concat z}$.
Then $|\semval{e_2}\,\rho_2| = 
\lam{n\in\omega}n  2^n$.
Note $|\rho_2(g_2)| = \lam{n\in\omega}2n$ is a polynomial function.  
The problem is that $\rho_2(g_2) = \lam{y\in\nat}y\concat y$ 
subverts the fundamental restriction on the sizes of ``safe'' values
in growth-rate bounds as in Proposition~\ref{p:bc:size:polybnd}(b). 
\end{example}

The problem of Example~\ref{e:disasters}(a) is addressed in 
\S\ref{S:impred} by pruning the $\SL$- and $\SV$-semantics  to 
restrict impredicative-type values.  The problem of 
Example~\ref{e:disasters}(b) is addressed in \S\ref{S:flat} by 
further pruning  to restrict flat-type values.

\section{Impredicative types and nearly well-foundedness}
\label{S:impred}

Failing to restrict impredicative-type values leads to problems 
like the one of Example~\ref{e:disasters}(a).  These problems can 
be avoided by requiring that each impredicative-type value have a 
length that is \emph{nearly well-founded}.

\begin{definition} \label{d:wellfnd}  
  A $t\in\semlen{\gamma}$ is \emph{$\gamma$-well-founded} when
  $\gamma=\Tallyl{\ell}$ or else $\gamma =
  (\sigma_1,\dots,\sigma_k)\to\Tallyl{\ell}$ and, for each $i$ 
  with $tail(\sigma_i)\suptyneq \Tallyl{\ell}$, the function $t$ 
  has no dependence on its $i$-th argument.  A $t$ is \emph{nearly
  $\gamma$-well-founded} when there is a $\gamma$-well-founded
  $t'$ such that $t\leq t'$.
\end{definition}

\begin{remark}\label{r:impred}
Why \emph{nearly} well-founded? The natural sources of
$\atr$-terms with impredicative types are the
$\Iif$-$\Ithen$-$\Ielse$ and $\down$ constructs.  Let
$c=\lam{x,y,z}\allowbreak (\Iif\,x\, \Ithen \,y \, \Ielse \,z)$ and
$d= \lam{x,y}(\down \,x\,y)$, where $\entails
c\of(\Natl{\ell},\Natl{\ell'},\Natl{\ell'})\to\Natl{\ell'}$, $\entails d\of
(\Natl{\ell},\Natl{\ell'}) \to\Natl{\ell'}$, and $\ell>\ell'$.  Thus $|c| \in
\semlen{|(\Natl{\ell},\Natl{\ell'},\Natl{\ell'}) \to\Natl{\ell'}|}$ and $|d| \in
\semlen{|(\Natl{\ell},\Natl{\ell'}) \to\Natl{\ell'}|}$.  
Neither $|c|$ nor $|d|$ is well-founded since 
$|c| = \lam{k,m,n}(m,\, \hbox{ if }k=\tally0;\; n,\, 
\hbox{otherwise})$ and 
$|d| = \lam{k,m}\min(k,m)$.  However, both $|c|$ and
$|d|$ are nearly well-founded as
$|c| \leq \lam{k,m,n} (m\bmax n)$ and
$|d| \leq \lam{k,m}m $.
\end{remark}

\begin{lemma} \label{l:tt:wf}
  Suppose $\Sigma\entails p\of\sigma$, \ $\rho\in\semlen{\Sigma}$,
  and $\rho(x)$ is nearly $\Sigma(x)$-well-founded for each
  $x\in\preimg(\Sigma)$.  Then $\semlen{p}\,\rho$ is nearly
  $\sigma$-well-founded.
\end{lemma}

Lemma~\ref{l:tt:wf} follows by a straightforward induction and
indicates that a semantics for the second-order polynomials based 
on nearly well-foundedness will be well defined.  
\emph{Terminology.}  The \emph{restriction} of 
$f\in (X_1,\dots,X_k)\to Y$ to
$(X_1',\dots,X_k')\to Y$ (where $X_1'\subseteq
X_1,\dots,X_k'\subseteq X_k$) is $\lam{x_1\in X_1',\dots,x_k\in
X_k'}f(x_1,\dots,x_k)$.

\begin{definition}[The nearly well-founded semantics]
\label{d:semlennwf}{\ }

  (a) 
  Inductively define $\semlennwf{\gamma}$ by:
  $\semlennwf{\Tallyl{\ell}}=\omega$.  For
  $\gamma=(\sigma_1,\ldots,\allowbreak \sigma_k)\to\Tallyl{\ell}$, \
  $\semlennwf{\gamma}$ is the restriction to $(\semlennwf{\sigma_1},
  \, \dots,\allowbreak\semlennwf{\sigma_k})\to
  \semlennwf{\Tallyl{\ell}}$ of the $\gamma$-nearly well-founded
  elements of $\semlen{\gamma}$.  Define $\semlennwf{\Sigma}$ and
  $\semlennwf{\Sigma\entails p\of\gamma}$ in the standard way.

  (b)  
  Inductively define $\semvalnwf{\gamma}$ by: $\semvalnwf{\Natl{\ell}}
  = \nat$.  For $\gamma=(\sigma_1,\ldots,\allowbreak\sigma_k)\to
  \Natl{\ell}$, \ $\semvalnwf{\gamma}$ is the restriction to
  $(\semvalnwf{\sigma_1}, \dots, \allowbreak
  \semvalnwf{\sigma_k})\to \semvalnwf{\Natl{\ell}}$ of the $f \in
  \semval{\gamma}$ with $|f|\allowbreak \in\semlennwf{|\gamma|}$.
  Define $\semvalnwf{\Gamma;\Delta}$ and
  $\semvalnwf{\Gamma;\Delta\entails E\of\gamma}$ in the standard
  way.

  (c)  
  We write $p\eqnwf p'$ when $\semlennwf{\Sigma \entails
  p\of\gamma}\,|\rho| = \semlennwf{\Sigma \entails
  p'\of\gamma}\,|\rho|$ for all $|\rho|\in\semlennwf{\Sigma}$.  We
  define $\leqnwf$, $\geq_{\rm nwf}$, \dots \ analogously.
\end{definition}

There is still a problem with impredicative-type values. In deriving
closed-form upper bounds on recursions, we often need a well-founded
upper bound on the value of a variable of an impredicative type.
There is no effective way to obtain such bound.  We thus do the next
best thing: give a canonical such upper bound a name and work with
that name.

\begin{definition} \label{d:bp} 
  We add a new combinator, $\bp$, to the second-order polynomials
  such that
  $\semlennwf{\Sigma\entails(\bp\, p)\of\gamma}\,|\rho| = $ the
  least $\gamma$-well-founded upper bound on
  $\semlennwf{\Sigma\entails p\of\gamma}\,|\rho|$.  (See
  Figure~\ref{fig:sharp} for $\bp$'s typing rule.)  For each
  variable $x$, we abbreviate $(\bp\,x)$ by $\bp_x$.
\end{definition}

The choice $\bp$ makes is analogous to choice of $a$ in the
situation where one knows $f\in O(n)$ and picks the least
$a\in\omega$ such that $f(n) \leq a\cdot (n+1)$ for all
$n\in\omega$.  In most uses, $\bp_x$'s are destined to be
substituted for by concrete, well-founded terms.

To help work with terms involving impredicative types we introduce:

\begin{definition}[Shadowing]\label{d:shadowing}
  Suppose $\Sigma\entails p \of\sigma$.
  An occurrence of a subterm $r$ of $p$ is \emph{shadowed} when the
  occurrence properly appears within another shadowed occurrence or
  else the occurrence has an enclosing subexpression $(t\; r)$ where
  the occurrence of $t$ is of an impredicative type $\sigma \to\tau$
  with $\tail(\sigma) \suptyneq \tail(\tau)$.
  A variable $x$ is a \emph{shadowed free variable} for $p$ when all
  of $x$'s free occurrences in $p$ are shadowed; otherwise $x$ is an
  \emph{unshadowed free variable} for $p$.
\end{definition}

\section{Safe upper bounds}\label{S:safe}

The restriction to the $\SV_{\text{nwf}}$-semantics solves the 
problem with impredicative types, but \emph{not} the problem with 
flat types.  To work towards a solution of this later problem, in
this section we  introduce the notion of a \emph{safe}
second-order polynomial (Definition~\ref{d:strict})  and show that 
any expression (in a simplification of $\atr$) that does not 
involve flat-type variables has a safe upper bound.
The next section proposes a solution to the flat-type problem: 
that each flat-type length must have a safe upper bound. 
Theorem~\ref{t:polybnd}, in \S\ref{S:pbnd}, shows that this 
proposed solution does indeed work.
\emph{Convention:} In this section $\bb$, $\gamma$, $\sigma$, 
and $\tau$ range over size types.  In writing 
$p=(x\,p_1\,\dots\,p_k)$, 
we mean $x$ is a variable and, when $k=0$, $p=x$.  

\begin{definition}[Strictness, chariness, and safety]
  \label{d:strict} 
  Suppose $\Sigma\entails p\of \gamma$.

  (a)  
  We say that $p$ is $\bb$-\emph{strict} with respect to 
  $\Sigma$ when $\tail(\gamma)\subty \bb$ and every 
  unshadowed free-variable occurrence in $p$ has a 
  type with tail $\subtyneq\bb$.
  
  (b) 
  We say that $p$ is \emph{$\bb$-chary} with respect to 
  $\Sigma$  when 
  $\gamma=\bb$ and
  either 
  (i) $p = (x\;q_{1}\cdots q_{k})$ 
  with each $q_i$ $\bb$-strict or
  (ii) $p=p_1\bmax\cdots\bmax p_m$, where each $p_{i}$ 
  satisfies (i).
  (Note that $\tally0$ sneaks in as $\bb$-chary; take $m=0$
  in (ii).)
  
  (c) 
  We say that $p$ is $\gamma$-\emph{safe} with respect to $\Sigma$
  if and only if
  
  \makebox[3em][r]{(i)} when
  $\gamma=\Tallyl{\Orl_d}$, then $p \eqnwf q\bmax r$  where $q$ is 
  $\gamma$-strict and $r$ is $\gamma$-chary,
    
  \makebox[3em][r]{(ii)} when
  $\gamma=\Tallyl{\Prg_d}$, then $p \eqnwf q + r$  where
  $q$ is a $\gamma$-strict 
  and $r$ is $\gamma$-chary $r$, and

  \makebox[3em][r]{(iii)} when $\gamma = \sigma\to\tau$, then
  $(p\,x)$ is $\tau$-safe with respect to $\Sigma,x\of\sigma$.
\end{definition}    

With the above notions, we drop the ``with respect to $\Sigma$'' 
when $\Sigma$ is clear from context.

\emph{Examples:} Recall the bound $p+\bigmax_{j=1}^n |y_j|$
of Proposition~\ref{p:bc:size:polybnd}(b).  In terms of the 
size-types, the subterm $p$ is $\Tallyl{\Prg}$-strict, the 
subterm $\bigmax_{j=1}^n |y_j|$ is $\Tallyl{\Prg}$-chary, and 
hence, $p+\bigmax_{j=1}^n |y_j|$ is $\Tallyl{\Prg}$-safe.
Roughly, Proposition~\ref{p:bc:size:polybnd} implies that
each $\bcl$ expression has a safe size-bound. 
Note that if $f\of\Natl{\Prg}\to\Natl{\Orl\Prg}$ and 
$x\of\Natl{\Prg}$, then  $|f|(|x|)$ is 
$\Tallyl{\Orl\Prg}$-chary, but not
$\Tallyl{\Orl\Prg}$-strict.

Strictness and chariness are syntactic notions,
whereas safety is a semantic notion because of
the use of $\eqnwf$ in Definition~\ref{d:strict}(c).  Thus:

\begin{lemma}
  If $\Sigma\entails p\of\bb$ and
  $p$ is $\bb$-strict or $\bb$-chary, then $p$ is also $\bb$-safe.
\end{lemma}

\proof Since $\tally0$ is both $\bb$-strict and $\bb$-chary
and since $p\eqnwf p\bmax\tally0\eqnwf \tally0\bmax p 
\eqnwf p+\tally0\eqnwf \tally0+p$, the lemma follows.\qed

The next lemma notes a key property of safe second-order
polynomials.

\begin{lemma}[Safe substitution] \label{l:safesub}
  Fix $\Sigma$.
  Given a $\gamma$-safe $p_0$, a $\sigma$-safe $p_1$,
  and a variable $x$ with
  $\Sigma(x)=\sigma$, 
  we can effectively find a $\gamma$-safe $p_0'$ such that
  $p_0[x\gets p_1] \leqnwf p_0'$.
\end{lemma}

\proof
  Except for the case when 
  $p_1$ is a $\lambda$-expression, the argument is a 
  straightforward induction.   When $p_1$ is a $\lambda$-expression, 
  the substitution can trigger a cascade of other substitutions 
  to deal with.  However, as we are working with an applied 
  simply-typed $\lambda$-calculus, strong normalization holds 
  \cite{winskel:book}, 
  and hence, these cascades are finite.  Consequently, to deal with 
  this case we simply use a stronger induction than before, say on 
  the syntactic structure of $p_0$ and $p_1$ and on the length of 
  the longest path of $\beta$-reductions to normal form of 
  $p_0[x\gets p_1]$.  
  This is fairly conventional and left to the 
  reader.\footnote{Alternatively, the lemma's proof could be done 
  through a logical relations induction \cite{winskel:book}.}
\qed

Remark~\ref{r:impred} informally argued  that if $e$,  an 
$\atr$ expression,  does not involve im\-pred\-ica\-tive-type variables, then $|e|$ has a well-founded upper bound.   The 
analogous argument here would be that if $e$  does not involve 
flat-type variables, then $|e|$ has a safe upper bound.   
This assertion is true, but not so interesting because most 
natural $\crec$-expressions have their recursor variable of 
flat type.  To get around this
problem we introduce a little formalism, $\gr$ (for \emph{growth 
rate}) which includes a simple iteration construct that does 
not depend so heavily on flat-type variables and which
captures $\atr$'s growth rate properties \emph{including}
$\atr$'s difficulty  with flat-type values.  
We show in Theorem~\ref{t:safe} that $\gr$ expressions
that do not involve flat-type variables have safe upper bounds. 
\begin{figure}[t]\small
\begin{gather*}
    \rulelabel{{\rm$\suc$}-I}  
    \irule{
        \Sigma\entails s\of\Tallyl{\Prg_d}}{
        \Sigma\entails (\suc\;s)\of\Tallyl{\Prg_d}}
    \Quad3    
    \rulelabel{{$\Rep$}-I} 
    \irule{
        \Sigma_0\entails s_0\of\Tallyl{\ell'}\to\Tallyl{\ell'}
	    \Quad{1.5}
		\Sigma_1\entails s_1\of\Tallyl{\ell}
	    \Quad{1.5}
		\Sigma_2\entails s_2\of\Tallyl{\ell'}}{
        \Sigma_0\cup\Sigma_1\cup\Sigma_2
          \entails (\Rep\;s_0\;s_1\;s_2)\of \Tallyl{\ell'}}
        \sidecond{\parbox{1.1cm}{\centering
        $\ell'=$ \\ $\successor(\ell)$}}
\end{gather*}
\caption{The additional typing rules for $\gr$}
\label{fig:sizeexptyping}
\figrule 
\end{figure}

\begin{definition} \label{d:sizepolys}
  $\gr$'s raw terms are given by:
  $S \is\bo^*$ $|$ $(\bmax\;S\;S)$ $|$
  $(\suc\;S)$ $|$ $(\Rep \;S\;S\;S)$ $|$ $X$
  $|$ $(S\; S)$  $|$ $(\lam{X}S )$. The typing rules 
  for $\gr$ consist of 
  \emph{$\to$-I} and 
  \emph{$\to$-E} from Figure~\ref{fig:pcf:typing}; 
  \emph{Zero-I}, \emph{Const-I}, \emph{Subsumption},
  \emph{Shift}, and \emph{$\bmax$-I} from 
  Figure~\ref{fig:sizepolytyping}; and \emph{$\suc$-I}
  and \emph{$\Rep$-I} from  
  Figure~\ref{fig:sizeexptyping}.\footnote{Recall from
  \S \ref{S:atr} that $\successor(\ell) = $ the 
  successor of $\ell$ in the ordering on labels.}  
  The intended interpretations 
  of $\bmax$,  $\suc$, and $\Rep$ are:
  $(\bmax\, \tally{m}\, \tally{n}) = \max(\tally{m},\tally{n})$,
  $(\suc\;\tally{m}) = \tally{m+1}$, and 
  $(\Rep\; f\; \tally{m}\; \tally{n}) = f^{(m)}(\tally{n})$.
\end{definition}

We straightforwardly extend the $\Lennwf$-semantics for 
second-order polynomials to $\gr$.
Note: $\semlennwf{\lam{m,n}(\Rep\; f\; m\; n)}\,\rho$ = 
$\lam{m,n\in\omega}n{2^m}$ when $\rho(f)=\lam{k\in\omega}2k$. So
$\gr$ has familiar problems with 
flat-type values.
We note that the $\gr$ analogues of Lemmas~\ref{l:tt:subred},
\ref{l:shifty'}, and~\ref{l:0gamma} all hold.
\emph{Terminology:} $\Sigma \entails s\of\sigma$ is 
\emph{flat-type-variable free} when no 
variable is explicitly or implicitly assigned a flat type by the 
judgment.

\begin{theorem}\label{t:safe}
  Given a flat-type-variable free $\Sigma\entails s\of\gamma$, 
  we can effectively find a $\gamma$-safe $p_s$ with respect   
  to $\Sigma$ such that $s \leqnwf p_s$.
  Moreover, we can choose $p_s$ so that all free variable
  occurrences are unshadowed.
\end{theorem}

\proof 
Without loss of generality we assume that $s$ is in $\beta$-normal 
form.  The argument is a structural induction on the derivation of 
$\Sigma\entails s\of\gamma$.  We consider the cases of the last 
rule used in the derivation.  Let $d$ range over $\omega$.

\textsc{Case:} \emph{Zero-I}.  
Then $s=\tally0$ and $\gamma=\Tallyl{\varepsilon}$.  So $p_s=
\tally0 $ suffices since $\tally0$ is $\Tallyl{\varepsilon}$-strict.

\textsc{Case:} \emph{Const-I}. 
Then $s=\tally{k}$ and $\gamma=\Tallyl{\Prg}$.  
So $p_s=\tally{k}$ suffices since
$\tally{k}$ is $\Tallyl{\Prg}$-strict.

\textsc{Case:} \emph{Id-I}.  
Then $s=x$, a variable. 
\textsc{Subcase:} $\gamma$ is a base type.
Then $p_s = x$ suffices since $x$ is $\gamma$-chary.
\textsc{Subcase:} $\gamma=(\sigma_0,\dots,\sigma_k)\to \bb$.
(Recall the introduction of $\bp_x$ in Definition~\ref{d:bp}.)
Let $\Sigma'=\Sigma,x_0\of\sigma_0,\dots,x_k\of\sigma_k$ and
$p'= (\bp_x\;p_0\;\dots\;p_k)$ where, for each $i$, \  $p_i=x_i$ 
if $\tail(\sigma_i)\subtyneq \bb$, and $p_i=\tally0_{\sigma_i}$, otherwise.  (Note that since $s$ is 
flat-type-variable free, $\tail(\sigma_i)\not=\bb$ for each $i$.)  
Then $p'$ is $\bb$-chary with respect to $\Sigma'$
and $(x\; x_0\; \dots \; x_k) \leqnwf p'$.  It follows
that $p_s = \lam{x_0,\dots,x_k}p'$ suffices.

\textsc{Case:} \emph{$\to$-I}.  
This case follows by the induction hypothesis and clause (iii) 
in Definition~\ref{d:strict}(c).

\textsc{Case:} \emph{$\to$-E}.  
This case follows by the induction hypothesis and 
Lemma~\ref{l:safesub}.

\textsc{Case:} \emph{Subsumption}. Then by Subsumption we know that 
$\Sigma\entails s\of\gamma'$ where $\gamma'\subty \gamma$.
Without loss of generality, we assume $\gamma'\subtyneq
\gamma$.  By the induction hypothesis  there exists $p$, 
a $\gamma'$-safe size-bound for $s$ with respect to $\Sigma$. 
It follows from Definition~\ref{d:strict} that $p$ is 
$\gamma$-strict with respect to $\Sigma$.  Hence, $p_s=p$
suffices.

\textsc{Case:} \emph{Shift}. Recall that if $(\vec{\sigma})\to\bb
\shiftsto (\vec{\sigma}')\to\bb'$, then, for each $i$, \
$\tail(\sigma_i) \subtyneq \bb$ implies $\tail(\sigma_i')\subtyneq 
\bb'$ and $\tail(\sigma_i) = \bb$ implies $\tail(\sigma_i')= \bb'$.
Thus this case follows from Lemma~\ref{l:shifty'} (in both its
second-order polynomial and $\gr$ versions) and 
Definition~\ref{d:strict}.

\textsc{Case:} \emph{$\suc$-I}.  
Then $s= (\suc\; s_1)$ and $\gamma=\Tallyl{\Prg_d}$.  
So by \emph{$\suc$-I}, we know that $\Sigma\entails s_1\of 
\Tallyl{\Prg_d}$ and by the induction hypothesis we have that 
there is a $\Tallyl{\Prg_d}$-strict $q$ and a 
$\Tallyl{\Prg_d}$-chary $r$ with $s_1 \leqnwf q+r$.  
Thus $p_s = (q+1)+r$ suffices since 
$q+1$ is $\Tallyl{\Prg_d}$-strict.

\textsc{Case:} \emph{$\bmax$-I}.  
Then $s= (\bmax \; s_0\; s_1)$.
\textsc{Subcase:} $\gamma=\Tallyl{\Prg_d}$.
So by \emph{$\bmax$-I} we know that $\Sigma_0\entails s_0\of 
\Tallyl{\Prg_d}$ and $\Sigma_1\entails s_1\of 
\Tallyl{\Prg_d}$, where $\Sigma=\Sigma_0\cup\Sigma_1$.  
By the induction hypothesis, there are
$\Tallyl{\Prg_d}$-strict $q_0$ and $q_1$ and 
$\Tallyl{\Prg_d}$-chary $r_0$ and $r_1$ such that 
$s_0\leqnwf q_0+r_0$ and $s_1\leqnwf q_1+r_1$.
Thus $p_s = (q_0\bmax q_1) + (r_0\bmax r_1)$ suffices
since $s \leqnwf (q_0+ r_0) \bmax (q_1+ r_1)
\leqnwf (q_0\bmax q_1) + (r_0\bmax r_1)$ and since
$(q_0\bmax q_1)$ is $\Tallyl{\Prg_d}$-strict
$(r_0\bmax r_1)$ is $\Tallyl{\Prg_d}$-chary with 
respect to $\Sigma$.
\textsc{Subcase:} $\gamma=\Tallyl{\Orl_d}$.
This follows by an easy modification of the above argument.

\textsc{Case:} \emph{$\Rep$-I}.  
Then $s = (\Rep\; s_0\; s_1\; s_2)$.
\textsc{Subcase:} $\gamma=\Tallyl{\Prg_d}$.
So by \emph{$\Rep$-I}, we have
$\Sigma_0\entails s_0\of\Tallyl{\Prg_d}\to\Tallyl{\Prg_d}$,
$\Sigma_1\entails s_1\of\Tallyl{\Orl_d}$, and
$\Sigma_2\entails s_2\of\Tallyl{\Prg_d}$, where
$\Sigma=\Sigma_0\cup\Sigma_1\cup\Sigma_2$.  
Since $s$ is flat-type-variable free, 
we must have $s_1=\lam{z}s_1'$  where 
$\Sigma_0,z\of\Tallyl{\Prg_d}\entails s_1'\of \Tallyl{\Prg_d}$.
Hence, by the induction hypothesis, there are 
$\Tallyl{\Prg_d}$-strict $q_0$ and $q_2$,
$\Tallyl{\Prg_d}$-chary $r_0$ and $r_2$, and
$\Tallyl{\Orl_d}$-safe $p_1$ such that
$s_1'\leqnwf q_0+r_0$, $s_1\leqnwf p_1$, and $s_2 \leqnwf q_2+r_2$. 
Note that $p_1$ is also $\Tallyl{\Prg_d}$-strict. 
Suppose $z$ has no free occurrences in $q_0+r_0$.  Then it follows
that $s \leqnwf (q_0+r_0)\bmax(q_2+r_2) \leqnwf
(q_0\bmax q_2)+(r_0\bmax r_2)$;  so 
$p_s=(q_0\bmax q_2)+(r_0\bmax r_2)$ suffices.  
Now suppose $z$ does have a free occurrence in $q_0+r_0$. 
Since $q_0$ is $\Tallyl{\Prg_d}$-strict, $z$ cannot occur in $q_0$.
Since $s$ is flat-type-variable free, it follows that 
$r_0 \eqnwf z \bmax r_0'$ where $z$ has no free occurrences in 
$r_0'$ and where $r_0'$ is $\Tallyl{\Prg_d}$-chary.
By the inequality $q+(q'+r')\bmax r \leq (q+q')+ 
r' \bmax r$, it follows that 
$s \leqnwf ( p_1\ast q_0 + q_2) + 
(r_0'\bmax r_2)$.  So,  $p_s = ( p_1\ast q_0 + q_2) + 
(r_0'\bmax r_2)$ suffices.  (Note the parallel to the proof 
of Proposition~\ref{p:bc:size:polybnd}.)
\textsc{Subcase:} $\gamma=\Tallyl{\Orl_d}$.
This follows by an easy modification of the above argument.
\qed

\section{Flat types and well-temperedness}\label{S:flat}

To avoid problems like the one of Example~\ref{e:disasters}(b), 
flat-type values need to be restricted.  The $\gr$ formalism
of the previous section is subject to roughly the same problem 
as that of Example~\ref{e:disasters}(b), but by 
Theorem~\ref{t:safe} flat-type-variable free $\gr$ expressions 
have safe second-order polynomial bounds.  This suggests that a 
solution to the flat-type values problem is to require all 
flat-type values to have safe size-bounds.  We call this property
\emph{well-temperedness}, meaning: all things are in the
right  proportions.

\begin{definition}\label{d:wt}
  A $t\in\semlennwf{\gamma}$ is
  $\gamma$-\emph{well-tempered} when $\gamma$ is strict or
  when $\gamma$ is flat and there is a closed, $\gamma$-safe
  $s$ with $t \leq \semlennwf{s}$.
\end{definition}

\begin{lemma} \label{l:tt:wt}
  Suppose $\Sigma\entails p\of\sigma$, 
  $\rho\in\semlennwf{\Sigma}$, and $\rho(x)$ is 
  $\Sigma(x)$-well-tempered  for each $x\in\preimg(\Sigma)$.
  Then $\semlennwf{p}\,\rho$ is  $\sigma$-well-tempered.
\end{lemma}

Lemma~\ref{l:tt:wt}'s proof is an induction on the
derivation of $\Sigma\entails p\of\sigma$.  Everything is
fairly straightforward except that the
\emph{$\to$-E} case depends critically on Lemma~\ref{l:safesub}.  
Lemma~\ref{l:tt:wt} indicates that a semantics for the second-order 
polynomials based on well-temperedness will be well defined.

\begin{definition}[The well-tempered semantics]
\label{d:semlenwt}{\ }

  (a) 
  Inductively define $\semlenwt{\sigma}$ by:
  $\semlenwt{\Tallyl{\ell}}=\omega$ and,  for 
  $\sigma=(\sigma_1,\ldots,\allowbreak \sigma_k)\to\Tallyl{\ell}$, \
  $\semlenwt{\sigma}$ is the restriction to 
  $(\semlenwt{\sigma_1},\,\dots,\allowbreak 
  \semlenwt{\sigma_k})\to \semlenwt{\Tallyl{\ell}}$ of  
  the  $\sigma$-well-tempered elements of $\semlennwf{\sigma}$.
  $\semlenwt{\Sigma}$ and
  $\semlenwt{\Sigma\entails p\of\sigma}$
  are defined in the standard way.

  (b)  
  Inductively define $\semvalwt{\sigma}$  by:
  $\semvalwt{\Natl{\ell}} = \nat$ and, 
  for $\sigma=(\sigma_1,\ldots,\allowbreak
  \sigma_k)\to \Natl{\ell}$, \
  $\semvalwt{\sigma}$ is 
  the restriction to $(\semvalwt{\sigma_1}, \dots,  \allowbreak 
  \semvalwt{\sigma_k})\to \semvalwt{\Natl{\ell}}$ of 
  the $f \in \semvalnwf{\sigma}$ with
  $|f|\in\semlenwt{|\sigma|}$.  
  $\semvalwt{\Gamma;\Delta}$ and 
  $\semvalwt{\Gamma;\Delta\entails E\of\sigma}$ 
  are defined in the standard way.

  (c)  We write $p\eqwt p'$ when 
  $\semlenwt{\Sigma \entails p\of\sigma}\,|\rho| = 
  \semlenwt{\Sigma \entails p'\of\sigma}\,|\rho|$
  for all $|\rho|\in\semlenwt{\Sigma}$.  We define
  $\leqwt$, $\geq_{\rm wt}$, \dots \ analogously.
\end{definition}

There is still a problem with flat-type values.  To give
closed-form upper bounds on recursions, we sometimes need to
decompose a safe flat-type polynomial into strict and chary
parts.  (Recall that safety is a semantic, not syntactic,
notion.)  For flat-type-variable free safe polynomials this
is easy.  A way of breaking flat-type variables into strict
and chary parts would allow us to extend this decomposition
to all safe polynomials.  We introduce two new combinators
to effect such a decomposition.  Since there is no canonical
way to do this decomposition, we take a different (and
trickier) approach from that of Definition~\ref{d:bp}.
\emph{Terminology:} Let  $(\bb)^\dag = 
(\bb)^\ddag = \bb$,
$((\vec{\sigma})\to\bb)^\dag = 
(\vec{\sigma}')\to\bb$, and 
$((\vec{\sigma})\to\bb)^\ddag = 
(\vec{\sigma}'')\to\bb$, where 
$\vec{\sigma}'=$ the 
subsequence of $\sigma_i$'s in $\vec{\sigma}$ with
$\tail(\sigma_i)\not=\bb$ and 
$\vec{\sigma}''=$ the 
subsequence of $\sigma_i$'s in $\vec{\sigma}$ with
$\sigma_i\not=\bb$.
(Recall: $()\to\bb\;\equiv \; \bb$.)  

\begin{figure}[t]\small
\begin{gather*}
    \rulelabel{$\bp$-I} 
    \irule{
    	    \Sigma\entails p\of\sigma}{
        \Sigma\entails (\bp \,p)\of\sigma}
        \Quad3
    \rulelabel{$\bq$-I} 
    \irule{
        \Sigma\entails p\of\sigma}{
        \Sigma\entails (\bq \, p)\of \sigma^\dag}
        \Quad3
    \rulelabel{$\br$-I} 
    \irule{
        \Sigma\entails p\of\sigma}{
        \Sigma\entails (\br \, p)\of \sigma^\ddag}
\end{gather*}
\caption{Typing rules for the $\bp$, $\bq$, and $\br$ combinators}
\label{fig:sharp}
\figrule 
\end{figure}

\begin{definition}\label{d:qr}  
We add two new combinators, $\bq$ and $\br$, to the
second-order polynomials with typing rules given in 
Figure~\ref{fig:sharp}.   
Suppose $\Sigma=w_1\of\tau_1,\dots,w_n\of\tau_n$, \ 
$\Sigma\entails p\of\gamma$, 
and $\rho\in\semlenwt{\Sigma}$. 
For $\gamma$  strict, define
$\semlenwt{(\bq \, p)}\,\rho  = \tally0_{\gamma^\dag}$
and
$ \semlenwt{(\br \, p)} = \semlenwt{(\bp\, p)}$.
Suppose $\gamma = (\sigma_0,\dots,\sigma_k)\to\bb$ is flat. 
Let
$\zeta = (\vec{\tau},\vec{\sigma})\to\bb$, 
$(\sigma_{i_1}',\dots,\sigma_{i_m}')\to\bb=\gamma^\dag$,   
$(\sigma_{j_1}'',\dots,\sigma_{j_n}'')\to\bb=\gamma^\ddag$,   
$\vec{x} = x_0,\dots,x_k$, 
$\vec{x}'= x_{i_1},\dots,x_{i_m}$,
$\vec{x}''= x_{j_1},\dots,x_{j_n}$,
and 
$\set{z_1,\dots,z_u} = \set{w_i \suchthat \tau_i=\bb}
\cup \set{x_i \suchthat \tau_i=\bb}$ where the $z_i$'s are 
all distinct. 
Define
$\semlenwt{(\bq \, p)} =  \semlenwt{\lam{\vec{x}'}q}$
and
$ \semlenwt{(\br \, p)}  =\semlenwt{\lam{\vec{x}'' } r}$,
where 
(i) 
$q$ is $\bb$-strict with respect to $\Sigma,
\vec{x}\of\vec{\sigma}$, 
(ii)
$r$ is $\bb$-chary with respect to $\Sigma, \vec{x}\of\vec{\sigma}$ 
and $r$ has no occurrence of any $z_i$, and 
(iii) 
\begin{gather*}
 \semlenwt{\entails \lam{\vec{w}}(\bp\,p)\of\zeta}
 \Verb{\leq}
 \begin{cases}
   \semlenwt{\entails \lam{\vec{w},\vec{x}}
   (q+ r\bmax z_1\bmax \dots \bmax z_u)\of
    \zeta}, & \hbox{for computational $\gamma$;}\\[0.5ex]
   \semlenwt{\entails \lam{\vec{w},\vec{x}}
   (q\bmax r\bmax z_1\bmax \dots \bmax z_u)\of
    \zeta}, & \hbox{for oracular $\gamma$.}
  \end{cases}    
\end{gather*} 
For each variable $x$, we abbreviate $(\bq\,x)$ by $\bq_x$ and
$(\br\,x)$ by  $\br_x$.  Also, we take 
$(\bq_x\,\vec{x}')$ as being $\bb$-strict and
$(\br_x\, \vec{x}'')$ as being $\bb$-chary.
\end{definition}

\begin{example}\label{ex:prn:bnd}
By the  definition of \emph{prn} 
given in Figure~\ref{fig:atr:prn} and our proof sketch for 
Proposition~\ref{p:bc:size:polybnd}, it follows that 
$
  |\textit{prn}| 
  \leqwt
  \lam{|e|,|x|}
     \left(\;(|x|+1)\ast \bq_{|e|}(|x|) + \br_{|e|}(|x|)\;\right)$.
\end{example}

By Definitions~\ref{d:wt} and~\ref{d:semlenwt}, $q$ and $r$
as in Definition~\ref{d:qr} must exist.  By the axiom of
choice, there are functions that pick out particular $q$ and
$r$.
\textbf{N.B.}  \emph{The choices of $q$ and $r$ are
\textbf{arbitrary} subject to satisfying conditions (i), (ii), 
and (iii) of Definition~\ref{d:qr}.}  The semantics for the
second-order polynomials is thus parameterized by the functions 
that pick out the required $q$'s and $r$'s.  The choices  $\bq$
and $\br$ make are analogous to the choices of $a$ and
$b\in\omega$ in the situation were one knows that $f\in
O(n)$ and picks some arbitrary $a$ and $b$ such that $f(n)
\leq a\cdot n + b$ for all $n$.  Such $a$ and $b$ can be
used in constructing algebraic upper bounds on expressions
involving $f$.  If later we determine concrete $a_0$ and
$b_0$ such that $f(n)\leq a_0\cdot n + b_0$ for all $n$,
then said algebraic upper bounds are still valid after the
substitution $[a\gets a_0, b\gets b_0]$ since the choices of
$a$ and $b$ were arbitrary.

\begin{definition}\label{d:manisafe}
  Suppose $\Sigma\entails p \of \gamma$, where
  $\set{y_1,\dots,y_k}=\set{y \suchthat 
  \Sigma(y) = \tail(\gamma)}$. 
  We say that $p$ is \emph{manifestly $\gamma$-safe with 
  respect to $\Sigma$} if and only if
  the only applications of the $\bp$, $\bq$, and $\br$ 
  combinators are to variables, and:
  
  (a) when $\gamma=\Tallyl{\Orl_d}$, then $p$ 
  is of one of the forms: $q$, $r\bmax y_{i_1}\bmax \dots \bmax
  y_{i_n}$, and $q\bmax r \bmax y_{i_1}\bmax \dots \bmax
  y_{i_n}$, where $q$ is $\gamma$-strict,
  $r$ is $\gamma$-chary with no occurrences of any of the $y_i$'s, 
  and $\set{y_{i_1},\dots,y_{i_n}}$ is a (possibly empty) 
  subset of $\set{y_1,\dots,y_k}$; 

  (b) when $\gamma=\Tallyl{\Prg_d}$, then $p$ 
  is of one of the forms: $q$, $r\bmax y_{i_1}\bmax \dots \bmax
  y_{i_n}$, and $q+ r \bmax y_{i_1}\bmax \dots \bmax
  y_{i_n}$, where $q$, $r$, and $\set{y_{i_1},\dots,y_{i_n}}$ 
  are as in (a); and
  
  (c) when $\gamma= (\sigma_0,\dots,\sigma_m)\to\bb$, 
  then the $\beta$-normal form of $(p \;\vec{x})$ is
  manifestly $\bb$-safe with respect to $\Sigma,x_0\of\sigma_0,
  \dots,x_m\of\sigma_m$.
\end{definition}

\begin{lemma}[Manifestly safe substitution] \label{l:safesub'}
  Fix $\Sigma$.
  Given a manifestly $\gamma$-safe $p_0$, a 
  manifestly $\sigma$-safe $p_1$,
  and a variable $x$ with
  $\Sigma(x)=\sigma$, 
  we can effectively find a manifestly $\gamma$-safe $p_0'$ 
  such that
  $p_0[x\gets p_1] \leqnwf p_0'$.
\end{lemma}

\proof This is a straightforward adaptation of the 
proof of Lemma~\ref{l:safesub}.
\qed

We now have a reasonable semantics for $\atr$ and the tools to 
work with this semantics to establish (in Theorem~\ref{t:polybnd}) 
a \emph{safe polynomial boundedness} result for $\atr$, where:

\begin{definition} \label{d:safebnd}
  Suppose $\Gamma;\Delta\entails e\of \sigma$.
  We say that $p$ 
  is a \emph{$|\sigma|$-safe polynomial size-bound}
  for $e$ with respect to $\Gamma;\Delta$ when $p$ is a 
  $|\sigma|$-safe second-order polynomial with respect to
  $|\Gamma;\Delta|$ and $|\semvalwt{e}\,\rho| \leq \semlenwt{p}\,|
  \rho|$
  for all $\rho\in \semvalwt{\Gamma;\Delta}$; if in addition $p$ 
  is manifestly $|\sigma|$-safe with respect to $\Gamma;\Delta$,
  we say that $p$ 
  is a \emph{manifestly $|\sigma|$-safe  polynomial 
  size-bound} for $e$ with respect to $\Gamma;\Delta$.
  (The ``with respect to'' clause is dropped when it is clear 
  from context.)
\end{definition}

\section{Polynomial size-boundedness}\label{S:pbnd}

\begin{theorem}[Polynomial Boundedness]\label{t:polybnd}
  Given $\Gamma;\Delta\entails e\of \gamma$,
  we can effectively find $p_{e}$, a manifestly 
  $|\gamma|$-safe polynomial
  size-bound for $e$ with respect to $\Gamma;\Delta$.
\end{theorem}

\proof
The argument is a structural induction on the derivation of 
$\Gamma; \Delta \entails e\of \gamma$.  We consider the cases of 
the last rule used in the derivation.  Excluding the $\crec$
case, everything is fairly straightforward.  Fix 
$\rho\in\semvalwt{\Gamma;\Delta}$.
Note that $|\semvalwt{\,\cdot\,}\,\rho|$  is invariant under 
$\beta$- and $\eta$-equivalence.   So without loss of generality, 
we assume that $e$ is in $\beta$-normal form.

\textsc{Cases:} \textit{Int-Id-I} and \textit{Aff-Id-I}. 
Then $e=x$, a variable.  
\emph{Subcase:} $\gamma$ is strict. Then $p_e= |x|$ clearly 
suffices.  
\emph{Subcase:} $\gamma$ is flat.  
Hence, $\level(\gamma)=1$.  
Let $(\bb_0,\dots,\bb_k)\to\bb=\gamma$.
Then by Definitions~\ref{d:semlenwt} and~\ref{d:qr}, 
\begin{gather*}
    p_e \Verb{=}
      \lam{|x_0|,\dots,|x_k|}
         \left(
             (\bq_{|x|}\; \longvec{|x'|}) 
             \odot 
             (\br_{|x|}\; \longvec{|x'|})
             \bmax |y_1| \bmax \dots \bmax |y_\ell|
         \right)
\end{gather*}
suffices, where $\longvec{|x'|}=$ the subsequence of 
the $|x_i|$'s with $\bb_i\not=\bb$ and
$\set{y_1,\dots,y_\ell} = \set{y \suchthat 
(\Gamma,x_0\of\bb_0,\dots,x_k\of\bb_k;\Delta)(y)=\bb}$,
and where
$\odot=+$, if $\gamma$ is computational, and $\odot=\bmax$, if
$\gamma$ is oracular.

\textsc{Case:}  \textit{Zero-I}.
Then $e=\epsilon$ and $\gamma=\Natl{\varepsilon}$.
Clearly $p_e = \tally0$ suffices.

\textsc{Case:}  \textit{Const-I}.
Then $e=$ some constant $k$ and $\gamma=\Natl{\Prg}$.
Clearly $p_e$ = $|k|$ suffices.

\textsc{Case:}  $\tsta$\introrule.
So $\gamma=\Natl{\Prg_d}$ for some $d$.
Clearly  $p_e = \tally{1}$ suffices.

\textsc{Case:} $\cdr$\textit{-I}.
Then $e=(\cdr\;e')$ for some $e'$ and $\gamma=\Natl{\Prg_d}$ 
for some $d$.  By the induction hypothesis, there is $p_{e'}$, 
a manifestly $\Tallyl{\Prg_d}$-safe polynomial size-bound for $e'$  with respect to 
$\Gamma;\Delta$.   Clearly $p_e = p_{e'}$ suffices.

\textsc{Case:} $\down$\textit{-I}.
Then $e=(\down\;e_{0}\;e_{1})$ with $\Gamma;\Delta\entails 
e_0\of\bb_0$, \ $\Gamma;\Delta\entails e_1\of \bb_1$, and 
$\gamma=\bb_1$.  By the induction hypothesis, there is a 
$p_{e_{1}}$, a manifestly $|\bb_{1}|$-safe polynomial size-bound for $e_{1}$  
with respect to $\Gamma;\Delta$.   Clearly $p_e = p_{e_{1}}$ 
suffices.

\textsc{Case:} $\consa$\textit{-I}.
Then $e=(\consa\;e')$ for some $e'$ and $\gamma=\Natl{\Prg_d}$ 
for some $d$.  By the induction hypothesis, there is $p_{e'}$, 
a manifestly $\Tallyl{\Prg_d}$-safe polynomial size-bound for $e'$ 
with respect to $\Gamma;\Delta$.  Clearly $p_e = \tally1+p_{e'}$ suffices.

\textsc{Cases:} \textit{Subsumption} and \emph{Shift}.
These follow as in the proof of Theorem~\ref{t:safe}.

\emph{Aside:} For the arguments for the  \textit{$\to$-I} and
\textit{$\to$-E} cases below,  recall from 
\S\ref{S:defs:length} that \refeq{e:flen} and \refeq{e:flen2} provide the 
definition of length for elements of $\TC$ of type-level 1 and  
type-level 2, respectively, and that higher-type lengths
are  pointwise monotone nondecreasing.

\textsc{Case:} \textit{$\to$-I}.  
Then $e = \lam{x}e'$ and $\gamma = \sigma\to\tau$.  By our induction hypothesis, there is a $p_{e'}$, a manifestly $|\tau|$-safe polynomial
size bound for  $e'$ with respect to $\Gamma, x\of\sigma; \Delta$. 
Let $p_e = \lam{|x|}p_{e'}$.
By Definition~\ref{d:strict}(c), $p_{e}$ is manifestly $|\gamma|$-safe with 
respect to  $|\Gamma; \Delta|$.  Let  $v$ range over  
$\semvalwt{\sigma}$.
Then, for each $t\in\semlenwt{|\sigma|}$,
we have the chain of bounds of Figure~\ref{fig:toI}.
\begin{figure}[t] \small
\begin{displaymath}\begin{array}{rcll}
\lefteqn{\left|\strut\semvalwt{\lam{x}e'}
  \rho\right|(t)}\\[1ex]
    &=& 
    \max\set{|(\semvalwt{\lam{x}e'}\,\rho)(v)|
       \suchthat |v|\leq t} 
    &  \hbox{(by \refeq{e:flen} and \refeq{e:flen2})}
            \\[1ex] 
    &=& 
    \max\set{|(\semvalwt{e'}\,(\rho\cup\set{x\mapsto v})|
       \suchthat  |v|\leq t} 
            &    \hbox{(by the 
              $\mathcal{V}_{\rm wt}$-in\-ter\-pret\-at\-ion
              of $\lambda$-terms)}
            \\[1ex] 
    &\leq& 
    \max\set{(\semlenwt{p_{e'}}\,
     (|\rho|\cup\set{|x|\mapsto |v|})
       \suchthat |v|\leq t}
        &  \hbox{(by the choice of $p_{e'}$)}
            \\[1ex] 
    &\leq& 
    (\semlenwt{p_{e'}}\,(|\rho|\cup\set{|x|\mapsto t}) 
            &\hbox{(by monotonicity)}
            \\[1ex] 
    &=& 
    (\semlenwt{\lam{|x|}p_{e'}}\,|\rho|)(t) 
            & \hbox{(by the 
              $\mathcal{L}_{\rm wt}$-in\-ter\-pret\-at\-ion
              of $\lambda$-terms)}            
            \\[1ex] 
    &=& 
    (\semlenwt{p_{e}}\,|\rho|)(t) 
            &\hbox{(by the choice of $p_{e}$).}
\end{array}\end{displaymath}
\caption{Bounds for the $\to$-\emph{I} case}\label{fig:toI}
\figrule
\end{figure}
Clearly this $p_{e}$ suffices.

\textsc{Case:} \textit{$\to$-E}. 
Then $e= (e_0 \; e_1)$ and for some $\sigma$ we have that  
$\Gamma;\Delta\entails e_0 \of \sigma\to\gamma$ and 
$\Gamma;\emptycont\entails e_1 \of \sigma$.  
By the induction hypothesis, there are $p_{e_{0}}$ and $p_{e_{1}}$ 
such that $p_{e_{0}}$ is a manifestly $(|\sigma|\to|\tau|)$-safe polynomial 
size bound 
for $e_{0}$ and $p_{e_{1}}$ is a manifestly 
$|\sigma|$-safe polynomial size-bound for 
$e_{1}$.  By Lemma~\ref{l:safesub'} we can 
effectively find  a manifestly $\gamma$-safe $p_e$ such that 
$(p_{e_0}\;p_{e_1}) \lwtleq p_e$.  
Then we have the chain of bounds of Figure~\ref{fig:toE}.
\begin{figure}[t] \small
\begin{displaymath}\begin{array}{rcll}
\lefteqn{\left|\strut\semvalwt{(e_{0}\;e_{1})}\rho\right|}\\[1ex]
    &=& \left|(\semvalwt{e_{0}}\,\rho ) 
        \left(\semvalwt{e_{1}}|\,\rho\strut\right)\right|
        &   \hbox{(by the 
              $\mathcal{V}_{\rm wt}$-interpret\-at\-ion
              of application)}
        \\[1ex]     
    &\leq &(|\semvalwt{e_{0}}\rho| ) 
           \left( \strut |\semvalwt{e_{1}} \rho|\right)
            &   \hbox{(by \refeq{e:flen} and \refeq{e:flen2})}
            \\[1ex]    
    &\leq &(\semlenwt{p_{e_0}}\,|\rho| ) 
           \left( \strut \semlenwt{p_{e_1}}\,|\rho|\right)
            &   \hbox{(by 
                monotonicity and the choices of 
                $p_{e_0}$ and $p_{e_1}$)}
            \\[1ex]    
    &= &\semlenwt{(p_{e_0}\,p_{e_1})}\,|\rho| 
            &   \hbox{(by the 
                  $\Lenwt$-interpretation
              of application)}
            \\[1ex]   
    &= &\semlenwt{p_{e}}\,|\rho| 
            &   \hbox{(by the choice of $p_{e}$).}
\end{array}\end{displaymath}
\caption{Bounds for the $\to$-\emph{E} case}\label{fig:toE}
\figrule
\end{figure}
Clearly this $p_{e}$ suffices.

\textsc{Case:} \textit{If-I}.  
Then $e=(\Iif \;e_0 \; \Ithen \;e_1 \; \Ielse\; e_2)$.  By the
induction hypothesis, there are $p_{e_1}$ and $p_{e_2}$, manifestly
$|\gamma|$-safe polynomial size-bounds for $e_{1}$ and $e_{2}$ respectively.
Clearly $p_e = p_{e_1} \bmax p_{e_2}$ suffices.

\bigskip
We have just one case left, but now the real work 
starts.
\bigskip

\textsc{Case:} $\crec$-\textit{I}.  
Then $\gamma= (\bb_1, \ldots, \bb_{k}) \to \bb_0 \in\SR$, so
$\bb_1=\Natl{\Orl_{d_1}}$ for some $d_1$, and
$e=(\crec \;a\;(\lamr{f} A))$  with $a\in\bo^*$,
$\Gamma;f\of\gamma\entails A\of\gamma$, and $\TailPos(f,A)$.
(Recall: $\TailPos$ is defined in Figure~\ref{fig:ltr:types}.)
For simplicity we assume $\set{\bb_1, \ldots, \bb_{k}} = 
\set{\Natl{\varepsilon}, \ldots, \Natl{\Orl_{d-1}}, 
\Natl{\Orl_{d_1}}} \allowbreak \cup \set{\bb \suchthat 
\Natl{\Orl_{d_1}} \subtyneq \bb \subty \bb_{\text{max}}}$  
for some $\bb_{\text{max}}$.  Without loss of generality we suppose:
\begin{gather} \label{e:AB}
  A \;\;=\;\; \lam{x_1 , \ldots, x_k}B,
\end{gather}
where $\hatGamma;f\of\gamma\entails B\of \bb_{0}$ for 
$\hatGamma = \Gamma,x_1\of\bb_1,\allowbreak \ldots,x_k\of\bb_k$,
\ $B$ is in $\beta$-normal form, and $\TailPos(f,B)$.

\emph{Aside:} 
To find $p_e$ for this case,  we analyze  
$e$'s tail recursion and determine size bounds on 
how large the tail-recursion's arguments can grow.  
In particular, we  show that there is a polynomial 
bound beyond which the first argument \emph{cannot} grow;
hence, by  \refeq{e:lbfix:red}, this polynomial bounds
the depth of $e$'s tail recursion.  From this bound
on recursion depth and from the size bounds on 
the tail-recursion arguments, constructing $p_e$ is straightforward.
To derive these bounds, we proceed a little informally 
and work with unfolded versions of~$e$. 

Consider the occurrences of $f$ in $B$.  Since we have
$\TailPos(f,B)$
and $\hatGamma;f\of\gamma\entails B\of \bb_{0}$, 
these occurrences must have enclosing expressions of the form 
$(f\;e_1 \;\ldots \; e_k)$, where $\hatGamma; \emptycont 
\entails e_{1}\of \bb_{1},\dots \hatGamma; \emptycont \entails 
e_{k}\of \bb_{k}$.  For a given such subexpression of $B$, we know 
by the induction hypothesis that, for each $i=1,\dots,k$, there is 
a  $p_i$,  a manifestly $\bb_{i}$-safe polynomial size-bound for $e_i$ 
with respect to 
$\hatGamma;\emptycont$.  
Since $f$ occurs but finitely many times in $B$, we may choose $p_1, 
\ldots, p_k$ so that they bound the size of the corresponding 
argument expressions for every $f$-application in $B$.  Without 
loss of generality, we assume that if $\bb_{i}=\bb_{j}$, then 
$p_{i}=p_{j}$.

Using the $\crec$ reduction rule \refeq{e:lbfix:red}, we expand out 
one-level of $e$'s $\crec$-recursion and, by using $\beta$- and 
$\eta$-reductions, clean things up to obtain 
\begin{align*}
  e^{(1)}&\;\;=\;\; \lam{\vec{x}} 
    \Iif \; |a|\leq|x_1|
        \;\Ithen \; \widehat{B} \;\;
    \Ielse \;\; 
                \epsilon, \quad \hbox{where}\\
  \widehat{B} &\;\;=\;\; 
    \hbox{the $\beta\eta$-normal form of } 
    B \,[f\gets (\crec \;(\bo\concat a)\; 
    (\lamr{f} A)) ]).
\end{align*}
Clearly, $\semvalwt{e}=\semvalwt{e^{(1)}}$.  
Let $\xi $ denote the substitution $[|x_1|\gets p_1,\dots, 
\allowbreak |x_{k}|\gets p_{k}]$.  From our choices of the 
$p_{i}$'s and $\widehat{B}$ it follows that $(p_1\,\xi ), 
\ldots, \allowbreak (p_k\,\xi )$ bound the size of the 
corresponding argument expressions for every $f$-application 
in $\widehat{B}$.  For each $i$, \ $(p_{i}\,\xi)$  can be 
equivalently expressed in terms of ${p_{i}}$ as follows.
\emph{Terminology:} An $r$ is \emph{strictly $\bb$-chary}
when $r$ is $\bb$-chary and contains no occurrences of 
type-$\bb$ variables.\footnote{Such
  an $r$ \emph{may} contain occurrences of variables 
  of types of the form $(\sigma_0,\dots,\sigma_k)\to\bb$.}
  
(\emph{Note:} In working through the proofs of Lemmas~\ref{l:onestep}
and~\ref{l:iterbnd} 
below, the reader many want to consider the case of: 
$\gamma=(\Natl{\Orl_1},\Natl{\Orl_0},\Natl{\Prg_1})\to\Natl{\Prg_1}$, 
$f$ has but one occurrence in $A$,
$p_1(|x_1|,|x_2|,|x_3|) = |g|(|x_2|) \bmax |x_1|$,
$p_2(|x_1|,|x_2|,|x_3|) = |x_2|$, and
$p_3(|x_1|,|x_2|,|x_3|) = q_3(|x_1|,|x_2|) + |x_3|$,
where $g\of\Natl{\Prg_0}\to\Natl{\Orl_1}$ and where
$q_3$ is an ordinary polynomial.)

\begin{lemma}[The one step lemma] \label{l:onestep}
  Each $p_{i}$ can be taken so that:

  (a)
  If $\bb_i\subty\Natl{\Orl_{d_1}}$, then ${p_{i}\,\xi } \;
  \lwtequiv \;  {p_{i}}$.

  (b)
  If $\Natl{\Orl_{d_1}} \subtyneq \bb_i=\Nat_{\Orl_{d}}$, 
  then there is 
  a $\bb_{i}$-strict $q_{i}$ and
  a strictly $\bb_{i}$-chary $r_{i}$ 
  such that 
  $p_{i}\,\xi \lwtequiv  
  q_{i}\,\xi\bmax r_{i}\,\xi\bmax p_{i}$.

  (c)
  If $\Natl{\Orl_{d_1}} \subtyneq \bb_i=\Nat_{\Prg_{d}}$, 
  then there is 
  a $\bb_{i}$-strict $q_{i}$ and
  a strictly $\bb_{i}$-chary $r_{i}$
  such that 
  ${p_{i}\,\xi }\lwtequiv {q_{i}\,\xi +
  r_{i}\,\xi\bmax p_{i}}$.
\end{lemma}

\proof 
For each $d$, let:
\begin{align*}
  \set{u_0^{d},\ldots,u_{b_{d}}^{d}}
     &\defeq \set{u\suchthat \Gamma(u)= 
        \Natl{\Orl_{d}}}. 
     &\set{w_0^{d},\ldots,w_{b'_{d}}^{d}}  
     &\defeq\set{u\suchthat \Gamma(u)= 
     \Natl{\Prg_{d}}}.
  \\
  \set{\baru_0^{d}, 
       \ldots, \baru_{c_{d}}^{d}}  
     &\defeq \set{x_{i} \suchthat \bb_{i} = 
     \Natl{\Orl_{d}}}.  
     & \set{\barw_0^{d}, 
       \ldots, \barw_{c'_{d}}^{d}}
     &\defeq \set{x_{i} \suchthat \bb_{i} = 
     \Natl{\Prg_{d}}}.
\end{align*}
(The $\baru$'s and $\barw$'s correspond to the  arguments of the
recursion while the $u$'s and $w$'s correspond to the other parameters.)

For part (a), we inductively consider the cases of $\bb_{i} =
\Natl{\varepsilon},\; \Natl{\Orl_1}, \ldots, 
\Natl{\Orl_{d_1}}$ in
turn.  
  
\textsc{Case:} $\bb_{i}=\Natl{\varepsilon}$.  
By the induction hypothesis, we may take 
$p_i$ to be  $q\bmax r \bmax \hatt\bmax\bart$, where 
$q$ is $\Tallyl{\Orl_0}$-strict,
$r$ is strictly $\Tallyl{\Orl_0}$-chary,
$\hatt = \bigmax_{a=0}^{b_0} |u_a^0|$, and
$\bart = \bigmax_{a=0}^{c_{0}} |\baru_{a}^{0}|$.
It follows from the size typing rules that the only
$\Tallyl{\Orl_0}$-\emph{strict} terms are 
$\lwtequiv \tally0$.   So, it suffices 
to take $p_{i}= r \bmax \hatt \bmax \bart$.
Note that $r = r\,\xi$  and 
$\hatt=\hatt\,\xi$ since neither $r$ nor $\hatt$ have any
occurrences of any  $\baru_{a}^{0}$. 
Also recall that we are assuming that if
$\bb_{i}=\bb_{j}$, then $p_{i}=p_{j}$.  
Thus,
for each $a$, \ $|\baru_{a}^{0}|\,\xi =p_{a} = p_{i}$.  So,
$\bart\,\xi \lwtequiv p_{i}=r\bmax\hatt\bmax\bart$.
Consequently,
\begin{gather*}\textstyle
  {p_{i}\,\xi} 
    \;\lwtequiv\; 
  {(r\bmax\hatt\bmax\bart)\,\xi} 
    \;\lwtequiv\;
  r\,\xi \bmax \hatt\,\xi \bmax \bart\,\xi 
    \;\lwtequiv\; 
  r\bmax \hatt \bmax(r\bmax\hatt \bmax \bart) 
    \;\lwtequiv\;
  r\bmax \hatt \bmax \bart 
    \;\lwtequiv\; 
  p_{i}.
\end{gather*}
Hence,  our choice of $p_{i}$  suffices for this case.

\textsc{Case:} $\bb_{i}=\Natl{\Orl_1}$.  
By the induction hypothesis,  we can take 
$p_i$ to be of the form $q\bmax r\bmax \hatt\bmax \bart$, where 
$q$ is $\Tallyl{\Orl_1}$-strict, 
$r$ is strictly $\Tallyl{\Orl_1}$-chary,
$\hatt = \bigmax_{a=0}^{b_1} |u_a^1|$, and 
$\bart = \bigmax_{a=0}^{c_{1}} |\baru_{a}^{1}|$.
We first consider $q$.  Since $\Gamma$ does not assign any of
$x_{1},\dots,x_{k}$ the type $\Natl{\Prg_0}$, the only variables
from $x_1,\ldots,x_k$ whose lengths can occur in $q$ are those assigned type $\Natl{\varepsilon}$.  
Let $\hatq = q\,\xi $, where for each $i'$ with
$\bb_{i'}=\Natl{\varepsilon}$, we take $p_{i'}$ to satisfy part (a).  Hence,
it follows that ${\hatq \, \xi } \lwtequiv {\hatq}$.  Also, by
the monotonicity of everything in sight, we have that $q
\lwtleq \hatq$.  
By the same argument,
for $\hatr=r\,\xi$  we have that ${\hatr \, \xi } \lwtequiv {\hatr}$
and $r \lwtleq \hatr$.
So, it suffices to take  $p_{i}=\hatq\bmax \hatr\bmax \hatt\bmax \bart$.  
Note that $\hatt=\hatt\,\xi $ since
$\hatt$ has no occurrence of any  $\baru_{a}^{d}$. 
Also recall that we are assuming that if $\bb_{i}=\bb_{j}$, then 
$p_{i}=p_{j}$.  Thus, for each $a$, \ 
$|\baru_{a}^{1}|\,\xi =p_{a}= p_{i}$.  So, 
$\bart\,\xi\lwtequiv p_{i}=\hatq\bmax\hatr\bmax\hatt\bmax\bart$.
Consequently,
\begin{gather*}
 {p_{i}\,\xi } 
    \;\;\lwtequiv\;\; 
 \left(\hatq \bmax \hatr \bmax \hatt \bmax \bart  \right) \,\xi 
    \;\lwtequiv\; 
 \hatq\,\xi \bmax  \hatr\,\xi \bmax \hatt\,\xi \bmax \bart\,\xi 
    \;\lwtequiv\; \\
 \hatq \bmax  \hatr \bmax \hatt 
      \bmax (\hatq \bmax  \hatr \bmax\hatt\bmax\bart) 
    \;\lwtequiv\;
 \hatq \bmax  \hatr \bmax\hatt\bmax\bart
    \;\lwtequiv\;
 p_{i}.
\end{gather*}
Hence,  our choice of $p_{i}$  suffices for this case.

\textsc{Cases:} $\bb_{i}=\Natl{\Orl_2}, \dots, \Natl{\Orl_{d_1}}$.  
These cases follow from essentially the same as argument given for 
the $\bb_{i}=\Natl{\Orl_1}$ case.

Therefore, part (a) follows.  

We henceforth assume that  $p_{i}$ satisfies part (a)
for each $i$ with $\bb_{i} \subty \Natl{\Orl_{d_1}}$.

\medskip

For parts (b) and (c), consider the
cases of $\bb_{i}=\Natl{\Prg_{d_1}},\,
\Natl{\Orl_{d_1+1}}, \ldots,
\bb_{\text{max}}$ in turn.

\textsc{Case:} $\bb_{i}=\Natl{\Prg_{d_1}}$.  
By the induction hypothesis, we may take
$p_{i}$ to be of the form 
$q+r\bmax \hatt\bmax \bart$, where 
$q$ is $\Tallyl{\Prg_{d_1}}$-strict, 
$r$ is strictly $\Tallyl{\Prg_{d_1}}$-chary,
$\hatt =\bigmax_{a=0}^{b'_{d_1}} |w_a^{d_1}|$, and
$\bart = \bigmax_{a=0}^{c'_{d_1}}|\barw_{a}^{d_1}|$.  
Note that as in the previous cases, 
$\hatt=\hatt\,\xi $.  Also recall that we 
are assuming that if $\bb_{i}=\bb_{j}$, then $p_{i}=p_{j}$.  
Thus,
for each $a$, \ $|\barw_{a}^{d_1}|\,\xi =p_{a}= p_{i}$.  So,
$\bart\,\xi \lwtequiv p_{i}=q+r\bmax\hatt\bmax\bart$.
Consequently,
\begin{gather*}
 {p_{i}\,\xi } 
 \;\;\lwtequiv\;\; 
 {(q+r\bmax\hatt\bmax \bart ) \,\xi \,} 
 \;\;\lwtequiv\;\; 
 q\,\xi +r\,\xi \bmax \hatt\,\xi \bmax \bart\,\xi
 \;\;\lwtequiv\;\; \\
 q\,\xi +r\,\xi \bmax \hatt \bmax (q+r\bmax \hatt\bmax\bart)
 \;\;\lwtequiv\;\; 
 {q\,\xi  +r\,\xi \bmax(q+r\bmax\hatt\bmax \bart) } 
 \;\;\lwtequiv\;\; 
 {q\,\xi + r\,\xi \bmax p_{i}}.
\end{gather*}
Hence, taking $q_{i}=q$ and  $r_{i}=r$ suffices for this case.

\textsc{Case:} $\bb_{i}=\Natl{\Orl_{d_1+1}}$.  
By the induction hypothesis, we may take
$p_{i}$ to be of the form 
$q\bmax r\bmax \hatt\bmax \bart$, where 
$q$ is $\Tallyl{\Orl_{{d_1}+1}}$-strict, 
$r$ is strictly $\Tallyl{\Orl_{{d_1}+1}}$-chary,
$\hatt =\bigmax_{a=0}^{b'_{d_1+1}} |u_a^{d_1+1}|$, and
$\bart = \bigmax_{a=0}^{c'_{d_1+1}}|\baru_{a}^{d_1+1}|$.    
By an argument similar to
the one for the previous case it follows
that taking $q_{i}=q$ and $r_{i}=r$ suffices for this case too.

\textsc{Cases:}  $\bb_{i}= \Natl{\Prg_{d_1+1}}, \dots, 
\bb_{\text{max}}$.  These cases follow from essentially the same 
as arguments as  given for the previous two cases.
\Qed{Lemma~\ref{l:onestep}}

Henceforth we assume that each $p_{i}$ is as in Lemma~\ref{l:onestep}
and, in the cases where $\Natl{\Orl_{d_1}}\subtyneq \bb_{i}$, $q_{i}$ and 
$r_{i}$ are as in that lemma too.
For each $n\in\omega$,  define 
\begin{gather}\label{e:En}
  e^{(n)} \;\;=\;\;
  \hbox{the $\beta$-normal form of the 
  $n$-level unfolding of $e$'s $\crec$-recursion,}
\end{gather}
where  $\beta$- and $\eta$-reductions are used to neaten up things
as in the definition of $e^{(1)}$.
So, $e^{(0)}=e$ and $e^{(1)}=$ our prior definition of $e^{(1)}$.
Let $\xi^{(0)}=$ the empty substitution
and $\xi ^{(n+1)} = \xi\circ \xi ^{(n)} =$ the $(n+1)$-fold 
composition $\xi $.  It follows that, with respect to 
$\hatGamma;\emptycont$, for each $i$ and $n$, \ $(p_{i}\,\xi^{(n)})$ 
is a size bound for $i$-th argument expression of every 
$f$-application in $e^{(n)}$.  

\begin{lemma}[The $n$ step lemma] \label{l:iterbnd}
For each $i$ and $n$:

(a) 
$p_{i}\,\xi^{(n)}\; \lwtequiv \; p_{i}$ when 
$\bb_{i}\subty\Natl{\Orl_{d_1}}$.

(b) 
$p_{i}\,\xi^{(n)}\; \lwtleq\; (q_{i}\bmax r_{i})\,\xi^{(n)} 
\bmax p_{i}$ when $\Natl{\Orl_{d_1}}\subtyneq\bb_{i} 
=\Natl{\Orl_{d}}$.

(c) 
$p_{i}\,\xi^{(n)} \;\lwtleq\; n\ast (q_{i}\,\xi^{(n)}) +  (r_{i}\, \xi^{(n)})\bmax p_{i}$ when 
$\Natl{\Orl_{d_1}}\subtyneq \bb_{i}=\Nat_{\Prg_{d}}$.
\end{lemma}

\proof
Part (a) follows directly from Lemma~\ref{l:onestep}(a).
For parts (b) and (c) we first note that by  monotonicity we have that, for all $k$ and $i$, \ $(q_{i}\bmax r_{i})\xi^{(k)} \lwtleq 
(q_{i} \bmax r_{i})\,\xi^{(k+1)}$. Now, for part (b), it follows 
immediately from Lemma~\ref{l:onestep}(b) that, for each $n$ and 
$i$, we have $p_{i}\,\xi^{(n)} \lwtequiv  \left(\bigmax_{j=0}^{n} 
(q_{i}\bmax r_{i}) \,\xi^{(j)}\right) \bmax p_{i}$.  Hence by 
the noted monotonicity  of $(q_i\bmax r_i)\xi^{(\cdot)}$, part~(b) follows.  For part (c), 
first fix $i$ such that $\bb_{i} = \Natl{\Prg_{d}}$ with $d\geq d_1$.   
It follows from an easy 
induction that for all $n$, \  $p_i\,\xi^{(n)} 
\lwtleq (\sum_{j=1}^{n}q_{i}\,\xi^{(j)}) + 
(\bigmax_{j=1}^{n}r_{i}\,\xi^{(j)})\bmax p_{i}$;
note the parallel to the argument for the $\prn$-case of
Proposition~\ref{p:bc:size:polybnd}.
Hence by monotonicity of $q_i\xi^{(\cdot)}$
and $r_i\xi^{(\cdot)}$, part (c) follows.
\Qed{Lemma~\ref{l:iterbnd}}

By Lemma~\ref{l:iterbnd}(a) and \refeq{e:lbfix:red} we have
\begin{lemma}[Termination]\label{l:term}
  $\semlenwt{p_{0}} \, |\rho|\;\geq$ the 
  maximum depth of $e$'s $\crec$-recursion.
\end{lemma}     

For each $i$ with $\bb_{i}\subty\Natl{\Orl_{d_1}}$, 
let $p_{i}'=p_{i}$.
For $\sigma=\Natl{\Prg_{d_1}},\ldots,\bb_{\text{max}}$
in turn, we inductively define $\theta_{\sigma}$ to be the 
substitution $[x_{j}\gets p_{j}' \suchthat \bb_{j}\subtyneq \sigma]$ and also define, 
for each $i$ with $\bb_{i}=\sigma$:
\begin{gather*}
  p_{i}' \;\;\defeq\;\;
  \begin{cases}
    (r_{i}\,\theta_{\sigma}) \bmax p_{i},
      & \hbox{if $\sigma$ is oracular;}\\
    p_{0}'\cdot (q_{i}\,\theta_{\sigma}) +
     (r_{i}\,\theta_{\sigma})\bmax p_{i},
      & \hbox{if $\sigma$ is computational.}
  \end{cases}
\end{gather*}   
By Lemma~\ref{l:safesub'}, for each $i$, we can effectively
find a manifestly $\bb_i$-safe $p_i''$ with
$p_i'\leqwt p_i''$.

\begin{lemma}[Final sizes] \label{l:finalb}
  For each $i$,  $p_{i}''$ is a manifestly $\bb_i$-safe 
  polynomial size-bound on the $i$-th argument 
  expression in the final step of the $\crec$-recursion in $e$.
\end{lemma}     

\proof  For each $i$ with $\bb_{i}\subty \Natl{\Orl_{d_1}}$, 
the conclusion follows from Lemma~\ref{l:iterbnd}(a).  
For the $\sigma = \Natl{\Prg_{d_1}}$ case, fix an $i$ with 
$\bb_{i} = \Natl{\Prg_{d_1}}$.  Then the bound for this case 
follows from 
Lemmas~\ref{l:iterbnd}(c) and~\ref{l:term}. The 
$\Natl{\Orl_{d_1+1}}$ through $\bb_{\text{max}}$ 
cases follow similarly.  
\Qed{Lemma~\ref{l:finalb}}

By the induction hypothesis, there exists $p_{B}$, a manifestly
$|\bb_{0}|$-safe polynomial size-bound for $B$ (as in \refeq{e:AB}) 
with respect to $\Gamma;f\of\gamma$.  By Lemma~\ref{l:safesub'},
we can effectively find a manifestly $\bb_0$-safe $\hatp$ such that 
$
  p_{B}\left[\strut
  |f|\gets \tally{0}_{\gamma},\,
  |x_{1}|\gets p_{1}',\dots,|x_{k}|\gets p_{k}'\right]
  \leqwt \hatp$.
The effect of the substitution on $p_{B}$ is to trivialize $|f|$ 
and replace each $|x_{i}|$ with the final size bound from
Lemma~\ref{l:finalb}.  It follows that 
$\hatp$ is a manifestly $\bb_0$-safe 
size bound for the value returned by final step of the 
$\crec$-recursion.  Since $\TailPos(f,A)$, 
$\hatp$ is also a size bound on the value returned by the entire 
(tail) recursion.  Thus, $p_{E}= \lam{|x_1|, \dots |x_{k}|} \hatp$ suffices 
for the $\crec$ case. 
\Qed{Theorem~\ref{t:polybnd}}

\section{An abstract machine}\label{S:cek}

\nocite{FF87}
\nocite{FF03}
\nocite{Plotkin75}
\nocite{EOPL:2}

Our next major goal is to show that every $\atr$ expression is
computable within a second-order polynomial time-bound
(Theorem~\ref{t:ptimebnd}).  Before formalizing time
bounds, we first need to make precise what is being bounded. 
Below we set out the abstract machine that provides 
the operational semantics of  $\PCF$, $\bcl$, and $\atr$ and, based 
on this, \S \ref{S:cek:cost} introduces and justifies our notion
of the time cost of an expression evaluation.

\subsection{The CEK machine} \label{S:cek:mach} 

The operational semantics for $\PCF$, $\bcl$, and $\atr$ are
provided by the abstract machine whose rules are given in
Figure~\ref{fig:cek}.
\begin{figure}[t]\small
\begin{eqnarray*}
&\begin{array}{rcll}
  \state{{(B\;e),\hat{\rho}},\kappa}
     &\to&
     \state{{e,\hat{\rho}},\pair{\bop,B,\kappa}}
     &\foocnt
     \\[0.75ex]   
  \state{{v,\hat{\rho}},\pair{\bop,B,\kappa}}
     &\to&
     \state{{\delta_{1}(B,v),\emptyenv},\kappa}
     &\foocnt
     \\[2ex]   
  \state{{(\down\;e\;e'),\hat{\rho}},\kappa}
     &\to&
     \state{{e,\hat{\rho}},\pair{\dn,e',\hat{\rho},\kappa}}
     &\foocnt
     \\[0.75ex]     
  \state{{v,\hat{\rho}},\pair{\dn,e',\hat{\rho}',\kappa}}
     &\to&
     \state{{e',\hat{\rho}'},\pair{\dn',v,\kappa}}
     &\foocnt
     \\[0.75ex]          
  \state{{v',\hat{\rho}'},\pair{\dn',v,\kappa}}
     &\to&
     \state{{\delta_{2}(v,v'),\emptyenv},\kappa}
     &\foocnt \label{e:delta2app}
     \\[2ex] 
  \state{{x,\hat{\rho}},\kappa}
     &\to& \state{v,\hat{\rho}',\kappa}, \;
     \hbox{where $\pair{v,\hat{\rho}'} = \hat{\rho}(x)$}
     &\foocnt \label{e:envapp}
     \\[2ex]
  \state{{(e\;e'), \hat{\rho}}, \kappa}
     &\to& \state{{e,\hat{\rho}},\pair{\arg,e',\hat{\rho},\kappa}}
     &\foocnt
     \\[0.75ex]
  \state{{v,\hat{\rho}},\pair{\arg,e',\hat{\rho}',\kappa}}
     &\to&
     \state{{e',\hat{\rho}'},\pair{\fun,v,\hat{\rho},\kappa}}
     &\foocnt
     \\[0.75ex]
  \state{{v',\hat{\rho}'},
         \pair{\fun,(\lam{x}e),\hat{\rho},\kappa}}
     &\to& \state{{e,\hat{\rho}[x\mapsto \pair{v',\hat{\rho}']}},\kappa}
     &\foocnt
     \\[0.75ex]
  \state{{v',\hat{\rho}'},
         \pair{\fun,O,\hat{\rho},\kappa}} 
     &\to& \state{{O(v'),\emptyenv},\kappa}  
     &\foocnt \label{e:oracleapp}
     \\[2ex]
  \state{ (\Iif\,e_{?}\,\Ithen\,e_{t}\,\Ielse\,e_{f}) ,\hat{\rho},\kappa} 
     &\to& \state{{e_{?},\hat{\rho}},\pair{\test,e_{t},e_{f},\hat{\rho},\kappa}} 
     &\foocnt
     \\[0.75ex]
  \state{{v_{?}, \hat{\rho}'},  
       \pair{\test,e_{t},e_{f},\hat{\rho},\kappa}}
     &\to& 
     \begin{cases}
       \state{{e_{t},\hat{\rho}},\kappa}, 
            & \hbox{if $v_{?}\not=\epsilon$;}
         \\[0.75ex]
       \state{{e_{f},\hat{\rho}},\kappa}, 
            & \hbox{if $v_{?}=\epsilon$.}
     \end{cases}
     &\foocnt
     \\[5ex]
  \state{(\fix\,(\lam{x}e)),\hat{\rho},\kappa}
     &\to& \state{{e[x\gets (\fix\,(\lam{x}e))],\hat{\rho}},
     \kappa}
     &\foocnt
     \\[2ex]
  \state{(\prn\,e),\hat{\rho},\kappa}
     &\to&        \state{e',\hat{\rho},\kappa}, \quad \hbox{where}
     &\foocnt
     \\
     \lefteqn{\hspace*{-8em}
    e' \;\;=\;\; 
    \lam{y}(\Iif\; y\not=\epsilon \; \Ithen \; 
    (e \; y\; (\prn \; e\; (\cdr\; y)))
     \; \Ielse\; (e\;\epsilon\; \epsilon))}\\[2ex]
   \state{(\crec\;c\;(\lamr{x}e)),\hat{\rho},\kappa} 
     &\to&   \state{{e',\hat{\rho}},\kappa}, \quad \hbox{where}
     &\foocnt     
\end{array}\\
&\begin{array}{rcl}       
 \hbox{ } \Quad3 e' &=&  (\lam{\vec{v}}
       (\Iif\,
         |v_{0}|\leq |c|\,
         \Ithen\,\epsilon\,
         \Ielse\, e'')\quad\hbox{and}\\[0.5ex]
   e'' &=&  (e[x\gets \crec \;\; \conso(c)\;
                (\lamr{x}e) ]\;\vec{v}\,)        
\end{array}
\end{eqnarray*}
\caption{The CEK-rewrite rules} 
\label{fig:cek}
\figrule
\end{figure}
The machine is based on Felleisen and Friedman's CEK-machine
\cite{FF87} as presented by Felleisen and Flatt \cite{FF03}.
\emph{States} in this machine are triples consisting of: (i) an
expression to be reduced or else a value, (ii) an environment, and (iii) 
a continuation.
\emph{CEK-environments}, \emph{closures}, and \emph{values} are
defined recursively by:
\begin{eqnarray*}
  \textrm{CEK-Environments} &=& 
     \textrm{Variables} \stackrel{\text{finite}}{\to} \textrm{Closures}. \\
  \textrm{Closures} &=& 
     (\textrm{Terms}\cup\textrm{Values})\times \textrm{CEK-Environments}.\\
  \textrm{Values} &=& \textrm{Strings} 
         			   \cup \textrm{Oracles}   
                       \cup \textrm{$\lambda$-Terms}.
\end{eqnarray*}
An \emph{oracle} is just an element of $\bigcup_{k>0}
\TC_{(\Nat^{k})\to\Nat}$.  Note that the result of applying an
oracle value $O\in \TC_{(\Nat^{k+1})\to\Nat}$ to a $v\in\Nat$ is the
oracle value $O(v)\in \TC_{(\Nat^{k})\to\Nat}$, where $k>0$.  The
\emph{continuations} should be self-explanatory  from the
rules---and if not, see \cite{EOPL:2}.

The CEK rules use the following variables (plain and decorated) with
indicated ranges.
$B$:Basic-Operations (i.e., $\conso$, $\consl$, $\cdr$, 
$\tsto$, and $\tstl$);
$\kappa$:Contin\-uat\-ions;
$e$:Terms; 
$O$:Oracles;
$\hat{\rho}$:CEK-Environments;
$v$:Values; and
$x$:Variables.  
Also, $\delta_{1}(B,v)$ returns the value of the given
basic-operation on the given value and $\delta_{2}(v,v')$ returns
$\mathit{down}(v,v')$.  
For each expression $e$ and CEK-environment $\hat{\rho}$
with $FV(e) \subseteq \preimg(\hat\rho)$, 
\begin{eqnarray*}
  \cekeval(e,\hat{\rho}) &\defeq& \begin{cases}
     v, & \hbox{if $\state{{e,\hat{\rho}},\halt} \to^{*} 
     \state{{v,\hat{\rho}'},\halt}$};\\
     \hbox{undefined}, & \hbox{if there is no such $v$ and $\hat{\rho}'$.}
     \end{cases}
\end{eqnarray*}
For each ordinary environment $\rho =
\set{x_1\mapsto v_1,\dots, x_k\mapsto v_k}$, let $\rho^*$ be 
the corresponding CEK-environment, i.e, 
$\set{x_1\mapsto (v_1,\emptyenv), \dots, x_k\mapsto
  (v_k,\emptyenv)}$,
and let $\cekeval(e,\rho)$ $\defeq$ $\cekeval(e,\rho^*)$.

\subsection{The CEK cost model} \label{S:cek:cost}
We assume that
the underlying model of computation is along the lines of
Kolmogorov and  Uspenskii's \cite{KolmUspenAlg}
``pointer machines'' or 
Sch\"{o}nhage's \emph{storage modification machines}
\cite{schonhage80}.
A string is represented by a linked list of $\bo$'s and
$\bl$'s.  We take the cost of evaluating an expression $e$ to be the
sum of the cost of the steps involved in evaluating $e$ on the CEK
machine.  We charge unit cost for for CEK-steps that do not involve
operations on strings or else carry out operations that work on just
the fronts of strings (e.g., $\consa$, $\cdr$, and $\tsta$).  For steps 
that involve copying or examining the entirety of arbitrary strings 
(rules \refeq{e:delta2app}, \refeq{e:envapp}, and \refeq{e:oracleapp}), 
our charge involves the sum of the lengths of the
strings involved.  Specifically:

\emph{\refeq{e:oracleapp} Oracle application.}  
Applying this rule has cost $\tally1 \bmax |O(v)|$ when $O(v)$ is of
base type and $\tally1$ otherwise.  (When $O(v)$ is of base type,
an application of the oracle pops into memory a string of length
$|O(v)|$.  We view the action of entering this string in memory,
character-by-character, as observable.)

\emph{\refeq{e:delta2app} $\delta_{2}$ application.}  
Applying this rule has cost $\tally1 + |v| + |v'|$.  ($\down$ looks
at the entirety of its arguments.)

\emph{\refeq{e:envapp} Environment application.} 
Applying this rule has cost $\tally1 \bmax |\hat{\rho}(x)|$ when
$\hat{\rho}(x)$ is of a base type and $\tally1$ otherwise.  (Since
our CEK machine starts with an arbitrary environment, the environment
is essentially another oracle.)

Given this assignments of costs, we introduce:

\begin{definition} \label{d:cekcosts} 
For each expression $e$ and CEK-environment $\hat{\rho}$, 
\begin{gather*}
  \cekcost(e,\hat{\rho}) \;\defeq\;
   \begin{cases}
     s, & \hbox{if }\parbox[t]{8cm}{$\cekeval(e,\hat{\rho})$ 
               is defined,
               where $s$ is the sum of the costs of the steps in this
               CEK-computation;}\\ 
     \hbox{undefined}, & \hbox{otherwise.}
     \end{cases}
\end{gather*}
and for each ordinary environment $\rho$, \ 
$\cekcost(e,\rho)$ $\defeq$ $\cekcost(e,\rho^*)$.
\end{definition}

We note that the standard proof that storage modification machines 
and Turing machines are polynomially-related models of computation 
\cite{schonhage80} straightforwardly extends to show that,
at type-levels 1 and 2,
our CEK model of computation and cost is 
(second-order) polynomially related
to Kapron and Cook's oracle Turing machines under their
answer-length cost model \cite{KapronCook:mach}.
\section{Time bounds}\label{S:time}

As the next step towards showing polynomial time-boundedness for 
$\atr$, the present section sets up a formal framework for working 
with time bounds. We start by noting the obvious:
  \textsl{Run time is \textbf{not} an extensional property of
  programs}.
That is, $\Valwt$-equivalent expressions can have quite distinct 
run time properties.  Because of this we introduce $\ST$, a new
semantics for $\atr$ that provides upper bounds on the time
complexity of expressions.

\subsection*{The setting}

Our framework for time complexities uses the following simple 
setting.

\emph{CEK costs.}  
Time costs are assigned to 
$\atr$-computations via 
the CEK cost model. 

\emph{Worst-case bounds.}
$\semtime{e}$ will provide a worst-case upper bound on the CEK 
cost of evaluating $e$, but not necessarily a tight upper bound.

\emph{No free lunch.} 
All evaluations have positive costs.  This even applies to 
``immediately evaluating'' expressions (e.g., 
$\lambda$-expressions), since checking whether something
``immediate-evaluates'' counts as a computation with costs.

\emph{Inputs as oracles.}  
We treat each type-level~1 input $f$ as an oracle.  In a
time-complexity context this means that $f$ is thought of
answering any query in one time step, or equivalently, any
computation involved in determining the reply to a query happens
unobserved off-stage.  Thus the cost of a query to $f$ involves only
(i) the time to write down a query $v$, and (ii) the time to read
the reply $f(v)$.  The times (i) and (ii) are 
bounded by roughly $|v|$ and
$|f|(|v|)$, respectively.  Thus our time bounds will ultimately be
expressed in terms of the \emph{lengths} of the values of free and 
input variables.

\subsection*{Currying and time complexity}

In common usage, ``the time complexity of $e$'' can mean one of two
things.  When $e$ is of base type, the phrase usually refers
to the time required to compute the value of $e$.  We might think of
this as \emph{time past}---the time it took to arrive at $e$'s
value.  When $e$ is of an arrow type
and thus describes a procedure, the phrase usually
refers to the function that, given the sizes of arguments, returns
the maximum time the procedure will take when run on
arguments of the specified sizes.  We might think of this as
\emph{time in possible futures} in which $e$'s value is applied.  An
expression can have both a past and futures of interest.  Consider
$(e_{0}\,e_{1})$ where $e_{0}$ is of type $\Natl{\varepsilon} \to
\Natl{\varepsilon} \to\Natl{\Prg}$ and $e_{1}$ is of type
$\Natl{\varepsilon}$.  Then $(e_{0}\;e_{1})$ has a time complexity
in the first sense as it took time to evaluate the expression, and,
since $(e_{0}\,e_{1})$ is of type
$\Natl{\varepsilon}\to\Natl{\Prg}$, it also has a time complexity in
the second sense.  Now consider just $e_{0}$ itself.  It too can
have a nontrivial time complexity in the first sense and the
potential/futures part of $e_{0}$'s time complexity must account for
the multiple senses of time complexity just attributed to 
$(e_{0} \, e_{1})$.  Type-level-2 expressions add further twists 
to the story.  Our treatment of time complexity takes into account
these extended senses.

\subsection*{Costs and potentials}

In the following the time complexity of an expression $e$ always has
two components: a \emph{cost} and a \emph{potential}.  A cost is
always a positive (tally) integer and is intended to be an upper
bound on the time it takes to evaluate $e$.  The form of a
potential depends on the type of $e$.  Suppose $e$ is of a base
(i.e., string) type. Then $e$'s potential is intended to be an
upper bound on the length of its value, an element of $\omega$.  The
length of $e$'s value describes the potential of $e$ in the sense
that when $e$'s value is used, its length is the only facet of the
value that plays a role in determining time complexities.  Now
suppose $e$ is of type, say, $\Natl{\varepsilon} \to \Natl{\Prg}$.
Then $e$'s potential will be an $f_{e}\in(\omega \to
\omega\times\omega)$ that maps a $p\in\omega$ (the length/potential
of the value of an argument of $e$) to a $(c_{r},
p_{r})\in\omega\times\omega$ where $c_{r}$ is the cost of applying
the value of $e$ to something of length $p$ and $p_{r}$ is the
length/potential of the result.  Note that $(c_{r}, p_{r})$ is a
time complexity for something of base type.  Generalizing from this,
our motto will be:

\begin{quote} \sl
  The potential of a type-$(\sigma\to\tau)$ thing is a map from
  potentials of type-$\sigma$ things to time complexities of
  type-$\tau$ things.\footnote{In a more general setting
  (e.g., call-by-name), a $(\sigma\to\tau)$ potential is 
  a map from $\sigma$-time-complexities to $\tau$-time-complexities, 
  as an operator may be applied to an unevaluated operand.}
\end{quote}
Our first task in making good on this motto is to situate time
complexities in a suitable semantic model.\footnote{\textbf{N.B.} 
  The time-complexity cost/potential distinction  
  appears in prior work \cite{sands:thesis, Shultis, VanStone}.
  Remark~\ref{rem:costPot} below discusses this prior work and 
  how it relates to ours.}

\subsection*{A model for time complexities}

The \emph{time types} are the result of the following translations
($\tcx{\,\cdot\,}$ and $\pot{\,\cdot\,}$) of $\atr$ types:
\begin{gather*}
  \tcx{\sigma} \;\defeq\; \Tally\times\pot{\sigma}.  
    \qquad
  \pot{\Natl{\ell}} \;\defeq\;  \Tallyl{\ell}. 
    \qquad
  \pot{\sigma\to\tau} \;\defeq\; 
  \pot{\sigma}\to\tcx{\tau}.
\end{gather*}
So,
$
\tcx{\Natl{\ell_1}\to\Natl{\ell_2}\to\Natl{\ell_0}} 
 =
    \Tally \times (\Tallyl{\ell_1} \to\Tally \times (\Tallyl{\ell_2} \to 
       \Tally\times\Tallyl{\ell_0}))$ and
$\Vert(\Natl{\ell_1}\to\Natl{\ell_2})\to\allowbreak \Natl{\ell_0}\Vert 
  = 
    \Tally \times ((\Tallyl{\ell_1} \to\Tally \times \Tallyl{\ell_2}) 
    \to \Tally\times\Tallyl{\ell_0})$. 
The time types are thus a subset of the simple product types over
$\set{\Tally,\, \Tallyl{\epsilon},\, \Tallyl{\Prg},\,
\Tallyl{\Orl\Prg}, \dots}$.  The intent is that $\Tally$ is the
type of costs, the $\Tallyl{\ell}$'s help describe lengths,
$\tcx{\gamma}$ is the type of complexity bounds of type-$\gamma$
objects, and $\pot{\gamma}$ is the type of potentials of
type-$\gamma$ objects.  (Note: $\pot{ \sigma \to \tau}$'s definition
parallels the motto.)

Our proof of polynomial time-boundedness for $\atr$
(Theorem~\ref{t:ptimebnd}) needs to intertwine the size estimates
implicit in potentials and the size bounds of
Theorem~\ref{t:polybnd}. The semantics for the time types thus needs
to be an extension of the $\SLwt$-semantics.  To define this
extension we use a combinator, $\basepot$, defined in
Definition~\ref{d:bigproj} below.  For the moment it is enough to
know that, for each $\atr$-type $\sigma$ and $p \in
\semlenwt{\pot{\sigma}}$, \ $\basepot(p) \in \semlenwt{|\sigma|}$ is
a canonical projection of $p$ to a type-$|\sigma|$ size bound.
Following the definition of $\basepot$, Lemma~\ref{l:loose}
notes that  all of the
notions introduced between here and there mesh properly.  

\begin{definition}[$\SLwt$ extended to the time types]
\label{d:lenwt:time}
  Suppose $\sigma$ and $\tau$ are $\atr$ types.  Then
  $\semlenwt{\tcx{\sigma}} \defeq \omega\times
  \semlenwt{\pot{\sigma}}$ and $\semlenwt{\pot{\sigma}}$ is
  inductively defined by $\semlenwt{\pot{\Natl{\ell}}}\defeq \omega$
  and $\semlenwt{\pot{\sigma\to\tau}}$ $\defeq$ the 
  set of
  all  monotone Kleene-Kreisel functionals  
  $f\of \semlenwt{\pot{\sigma}} \to \semlenwt{\tcx{\tau}}$
  such that:
  (i) $ \basepot(f) \in\semlenwt{|\sigma\to\tau|}$ and 
  (ii) $\basepot(f(p_1))=\basepot(f(p_2))$ whenever 
  $\basepot(p_1)=\basepot(p_2)$. 
\end{definition}

Condition (i) above restricts $\semlenwt{\pot{\sigma\to\tau}}$ so
that the projection $\basepot$ acts as advertised.  Condition (ii)
restricts each $f\in \semlenwt{\pot{\sigma\to\tau}}$ so that the
size information in $f(p)$ depends only on the size information in
$p$.

We can now define the $\ST$ (time-complexity) and $\SP$ (potential) 
interpretations of the $\atr$ types. (The $\SP$-interpretation 
is a notational convenience.)

\begin{definition}\label{d:tsigma}
  Suppose $\sigma$ is an $\atr$-type.  Then $\semtime{\sigma}$
  $\defeq$ $\semlenwt{\tcx{\sigma}}$ and $\sempot{\sigma}$ $\defeq$
  $\semlenwt{\pot{\sigma}}$.
\end{definition}

\subsection*{The $\Time$-interpretation of constants and oracles}

The following two definitions introduce a translation from the
$\Valwt$ model into the $\Time$ model.  We use this translation to
assign time complexities to program inputs: 
string constants and oracles.

\begin{definition}\label{d:tcxa}
  Let $\tcx{a}\defeq(\tally1\bmax|a|,\pot{a})$ and
  $\pot{a}\defeq|a|$ for each $a\in\semvalwt{\Natl{\ell}}$.
\end{definition}

By Lemma~\ref{l:loose}(a) below, $\tcx{a}\in\semtime{\Natl{\ell}}$.  We
view $\tcx{a}$ as the time complexity of the string/integer constant
$a$.  The interpretation of the cost component of $\tcx{a}$ is that
the cost of evaluating the constant $a$ is the cost of writing down 
$a$ character by character.  (When $a=\epsilon$, we still charge
$\tally1$.)

\begin{definition}\label{d:tcxf}
 Let
 $\tcx{f}\defeq (\tally1,\pot{f})$ and  
  $\pot{f}\defeq\allowbreak
        \lam{p\in\sempot{\sigma}}
        \max\Set{ \tcx{(f\, v)} \suchthat \pot{v}\leq p}$ for
        each $f\in\semvalwt{\sigma\to\tau}$.
\end{definition}

By Lemma~\ref{l:loose}(a) below, $\tcx{f} \in \semtime{\sigma \to
\tau}$.  We view $\tcx{f}$ as the time complexity of $f$ as an
oracle: the only time costs associated with applying $f$ are those
involved in setting up applications of $f$ and reading off the
results. Recall that under call-by-value, a $\lambda$-expression
immediately evaluates to itself.  The function-symbol $f$ will be 
treated
analogously to a $\lambda$-term. Hence, the cost component of
$\tcx{f}$ is $\tally1$.  The definition of $\pot{f}$ parallels both
our informal discussion of the notion of the potential of a
type-level~1 function and the definition of the length of functions
of type levels~1 and~2 in \S \ref{S:defs:length}.  One can show that when
$f$ is a type-level~2, $\pot{f}$ is total.  (The argument is similar
to the proof of the totality of the type-level 2 notion of length
defined by \refeq{e:flen2} in \S \ref{S:defs:length}.)

Definition~\ref{d:tcxa} and the type-level~1 part of
Definition~\ref{d:tcxf} describe the time complexities of possible
$\atr$ inputs. The following lemma unpacks the definition of
$\pot{f}$ for $f$ of type-level~1.  The proof is a straightforward
induction and hence omitted.

\begin{lemma}\label{l:||f||}
  For  $f\in\semvalwt{(\Natl{\ell_1},\dots,\Natl{\ell_k}) \to\Natl{\ell_0}}$, \ 
  $\pot{f}= q_1$ where  
  $q_i = \lam{p_i\in\omega}\left(\tally1,q_{i+1}\right)$
  (for $1\leq i < k$) and
  $q_{k} = \lam{p_{k}\in\omega}\allowbreak
    \bigl(\tally1 \bmax \allowbreak |f|(p_1,\dots,p_k), \allowbreak
       |f|(p_1,\dots,p_k)\bigr)$.
\end{lemma}

\subsection*{$\Time$-Applications}

\begin{definition}{\ }\label{d:star}

(a) Suppose 
$t_{0}\in\semtime{\sigma\to\tau}$ and 
$t_{1}\in\semtime{\sigma}$, where 
$t_{0}=(c_{0},p_{0})$, 
$t_{1}=(c_{1},p_{1})$, and 
$(c_{r},p_{r}) = p_{0}(p_{1})$.
Then
$t_{0} \aptime t_{1} \defeq (c_{0}+c_{1}+c_{r}+\tally3,\,p_{r})$. 

(b) Suppose 
$t_{0}\in\semtime{(\sigma_{1},\ldots,\sigma_{k})\to\tau}$,
$t_{1}\in\semtime{\sigma_{1}},\ldots,
t_{k}\in\semtime{\sigma_{k}}$. 
Then 
$t_{0}\aptime\vec{t} \;\defeq t_{0}\aptime t_{1}\aptime \dots \aptime t_{k}$.
(The $\aptime$ operation left associates.)
\end{definition}

By Lemma~\ref{l:loose}(b) below, $t_0\aptime t_1 \in\semtime{\tau}$
when $t_0\in\semtime{\sigma\to\tau}$ and $t_1\in\semtime{\sigma}$.
Suppose that $t_{0}$ (respectively, $t_{1}$) is the time complexity
of a type-$(\sigma\to\tau)$ expression $e_0$
(respectively, type-$\sigma$ expression $e_{1}$).
Then $t_{0}\aptime t_{1}$ is intended to be the
time complexity of  $(e_{0}\; e_{1})$.  The cost
component of $t_{0}\aptime t_{1}$ is:
(the cost of evaluating $e_{0}$) +
(the cost of evaluating $e_{1}$) + 
(the cost of applying $e_0$'s value to $e_1$'s value) + 
$\tally3$, where the $\tally3$ is the
CEK-overhead of an application.  The potential component is simply
the potential of the result of the application.  
The next lemma works out of the effect of the $\aptime$ 
operation for type-level 1 oracles.

\begin{lemma}\label{l:time:oracle}
  Suppose 
  $f\in\semvalwt{(\Natl{\ell_1},\dots,\Natl{\ell_k}) \to\Natl{\ell_0}}$,
  $v_{1}\in\semvalwt{\Natl{\ell_1}},\dots$,
  $v_{k}\in\semvalwt{\Natl{\ell_k}}$.  Then 
  \begin{gather}     \label{e:time:oracle}
    \tcx{f}\aptime \longvec{\tcx{v}}
    \Verb{=}
      \left(\!({\textstyle \sum_{i=1}^{k}(\tally1\bmax|v_{i}|})) + 
      \tally1\bmax|f|(\longvec{|v|}) +
      \tally{5k-1},   \; |f|(\longvec{|v|})\right)\!,
  \end{gather}
  where $\longvec{\tcx{v}}$ abbreviates 
  $\tcx{v_{1}},\dots,\tcx{v_{k}}$ and
  $\longvec{|v|}$ abbreviates $|v_{1}|,\dots,|v_{k}|$.
\end{lemma}

The proof is a straightforward calculation.
Equation~\refeq{e:time:oracle} can be interpreted as giving an upper
bound on the time complexity of applying an oracle $f$ to arguments
$v_{1}, \dots, v_{k}$.  Let us consider the cost component of the 
$k=1$ and $k=2$ cases of \refeq{e:time:oracle} in more detail. For
$k=1$, the right-hand side of \refeq{e:time:oracle} simplifies to:
$((\tally1\bmax|v_{1}|) + \tally1\bmax|f|(|v_{1}|) + \tally{4}, \,
|f|(|v_{1}|))$.  Its cost component is broken down in
Figure~\ref{fig:onestar}.
\begin{figure}[t]\small
\begin{center}
\begin{tabular}{lcl}
   \tally1 &  = & the cost of evaluating $f$\\
   $\tally1 \bmax |v_{1}|$ 
           & =& the cost of evaluating $v_{1}$, i.e., the 
           cost of writing down the value $v_{1}$ \\
   $\tally1\bmax|f|(|v_{1}|)$ 
           & =& the cost of applying 
              $f$ to  $v_{1}$, i.e., 
              the cost of writing down
             $f(v_{1})$'s value \\
   $\tally3$ & =& the overhead of the application
\end{tabular}
\end{center}
\caption{Break down of the cost component of $\tcx{f}\aptime\tcx{v_{1}}$} 
\label{fig:onestar}
\figrule
\end{figure}
For $k=2$, the right-hand side of \refeq{e:time:oracle} simplifies to:
$    \left((\tally1\bmax|v_{1}|) + (\tally1\bmax|v_{2}|) + \Strut{2ex}
      \tally1\bmax|f|(|v_{1}|,|v_{2}|) +
      \tally{9}, \;  |f|(|v_{1}|,|v_{2}|)\right)$.
We leave it to the reader to break down its cost component.

\subsection*{$\Time$-Environments}

As a companion to $\Time$-application we shall define an analogue of
currying in $\Time$.  First, we introduce
$\Time$-environments.  Recall that in a call-by-value language,
variables name \emph{values} \cite{Plotkin75}, i.e., the end result
of a (terminating) evaluation. Thus, a value does
not need to be evaluated again, at least no more than an input value
does.  Hence, if a $\Time$-environment maps a variable to a
type-$\gamma$ time complexity $(c,p)$, then $c$ should be: $\tally1
\bmax p$, when $\gamma$ is a base type, and $\tally1$, when $\gamma$
is an arrow type.

\begin{definition}\label{d:env}
Suppose $\sigma$ and $\tau$ vary over $\atr$ types and
$\Gamma;\Delta$ is an $\atr$ is type context.

(a)
$\tcx{\Gamma;\Delta} \, \defeq\,
\Set{ x \mapsto \tcx{\sigma} \suchthat (\Gamma;\Delta)(x)=\sigma}$.

(b)
For $p\in\sempot{\Natl{\ell}}$, \ $\val(p)\, \defeq\, (\tally1\bmax p, p)$.

(c)
For $p\in\sempot{\sigma\to\tau}$, \ $\val(p)\, \defeq \,
(\tally1,p)$.

(d) 
$\semtimeval{\sigma}\, \defeq\, \set{\val(p)\suchthat
p\in\sempot{\sigma}}$. 

(e) 
$\semtime{\Gamma;\Delta}$ is the set of all finite maps of the form
$\set{x_{1}\mapsto t_{1},\dots,x_{k}\mapsto t_{k}}$, where
$\set{x_{1},\dots,x_{k}} = \textrm{preimage}(
\Gamma;\Delta)$, and, for $i=1,\dots, k$, \
$t_{i}\in\semtimeval{(\Gamma;\Delta)(x_{i}))}$.

(f)
For each $\rho \in \semvalwt{\Gamma;\Delta}$, define $\tcx{\rho}
\in\semtime{{\Gamma;\Delta}}$ by $\tcx{\rho}(x) = \tcx{\rho(x)}$.
Such as $\tcx{\rho}$ is called an \emph{oracle environment}.
\end{definition}

\emph{Convention:} 
We use $\varrho$ as a variable over $\Time$-en\-vi\-ron\-ments.
\textbf{N.B.} Not every $\varrho$ of interest is 
an oracle environment.

\subsection*{$\Time$-currying}

Here then is our time-complexity analogue to currying.  Recall that
$\semtime{\Gamma;\Delta\entails e\of\tau}$ will be (when we get
around to defining it) a function from $\semtime{\Gamma;\Delta}$ to
$\semtime{\tau}$.

\begin{definition}\label{d:currystar}
Suppose (i) $\Gamma;\Delta$ is a $\atr$ type context with
$(\Gamma;\Delta)(x_{i})=\sigma_{i}$, for $i=1,\dots,k$; (ii)
$\Gamma';\Delta'$ is the result of removing $x_{1}\of\sigma_{1}$
from $\Gamma;\Delta$; and (iii) $X$ is a function from
$\semtime{\Gamma; \Delta}$ to $\semtime{\tau}$.  Then
$\currytime(x_{1},X)$ is the function from $\semtime{\Gamma'; 
\Delta'}$ to $\semtime{\sigma_{1}\to\tau}$ given by:
\begin{gather}\label{e:Lam}
  \currytime(x_{1},X)\,\varrho' 
    \Verb{\defeq}
    \left(\;\tally1,\,\strut
       \lam{p
       \in\sempot{\sigma_{1}}}
       (X \,(\varrho'\cup
       \set{x_{1}\mapsto \val(p)}))\;\right),
\end{gather}       
where $\varrho'\in\semtime{\Gamma';\Delta'}$.  Also, 
$
  \currytime(x_{1},x_{2},\dots,x_{k},X)
    \defeq
    \currytime(x_{1},\currytime(x_{2},\dots,x_{k},X))
$
when $k>1$.
\end{definition}

Note the complementary roles of $\currytime$ and $\aptime$:
$\currytime$ shifts the past (the cost) into the future (the
potential) and $\aptime$ shifts part of the future (the potential)
into the past (the cost).  This being complexity theory, there are
carrying charges on all this shifting.  This is illustrated in the
next lemma that shows how $\currytime$ and $\aptime$ interact.  First,
we introduce:

\begin{definition}\label{d:dally}
  $\dally(d,(c,p)) \,\defeq\, (c+d,p)$ 
  for $d\in\omega$ and $(c,p)$, a time complexity.  
\end{definition}

\begin{lemma}[Almost the $\eta$-law] \label{l:curry}
  Suppose $\Gamma$, $\Delta$, $X$, $\vec{x}$, $\vec{\sigma}$, and
  $\tau$ are as in Definition~\ref{d:currystar}.  Let
  $\Gamma';\Delta'$ be the result of removing $x_{1}\of\sigma_{1},
  \dots,x_{k}\of\sigma_{k}$ from $\Gamma;\Delta$.  Let
  $\varrho\in\semtime{\Gamma;\Delta}$ and let $\varrho'$ be the
  restriction of $\varrho$ to $\preimg(\Gamma';\Delta')$.  Then
  \begin{gather}\label{e:curry:time}
  \left(\currytime(x_{1},\dots,x_{k},X)\,\varrho'\right) 
  \aptime  \varrho(x_{1}) \aptime \dots \aptime \varrho(x_{k}) 
      \Verb{=} \dally(\tally{5\cdot k+4}
      +{\textstyle\sum_{i=1}^k}c_i,\;X\,\varrho),
  \end{gather}
  where
  $(c_{1},p_{1}) = \varrho(x_{1}),\dots,
  (c_{k},p_{k}) = \varrho(x_{k})$.
\end{lemma}

The lemma's proof is another straightforward calculation.

\subsection*{Projections}

The next definition introduces a way of recovering more conventional 
bounds from time complexities.  Note, by Definitions~\ref{d:tcxa}
and~\ref{d:tcxf}, and Lemmas~\ref{l:||f||} and~\ref{l:time:oracle}, 
when $v$ is a string constant or a type-1 oracle, the value of 
$\tcx{v}$ is a function of the value of $|v|$.  So, by an abuse of 
notation,  we treat $\tcx{v}$ as a function of $|v|$ for such $v$.

\begin{definition}\label{d:bigproj}
Suppose $\sigma$ and $(\sigma_1,\dots,\sigma_k)\to\Natl{\ell}$ are
$\atr$ types.

(a) 
For each $t\in\semtime{\sigma}$, let $\Cost(t) \defeq \pi_1(t)$ and
$\Pot(t)\defeq\pi_{2}(t)$.  \emph{(So, $t = (\Cost(t),\Pot(t))$.)}

(b)
For each $t\in\semtime{\Natl{\ell}}$, let $\basecost(t)=\Cost(t)$ and
$\basepot(t)=\Pot(t)$ and, for each
$t\in\semtime{(\sigma_1,\dots,\sigma_k)\to\Natl{\ell}}$, let:
\begin{gather*}
  \basecost(t)\Verb{\defeq}
     \lam{\longvec{|v|}}\Cost (t\aptime\longvec{\tcx{v}}).
     \Quad2
  \basepot(t)\Verb{\defeq}
     \lam{\longvec{|v|}}\Pot (t\aptime\longvec{\tcx{v}}).
\end{gather*}
where $\longvec{|v|}$ abbreviates
$|v_{1}|\in\semlenwt{\sigma_{1}},\dots, \allowbreak |v_{k}|
\in\semlenwt{\sigma_{k}}$ and $\longvec{\tcx{v}}$ abbreviates
$\tcx{v_{1}},\dots,\tcx{v_{k}}$.  \emph{(So,
$t\aptime\longvec{\tcx{v}} =
(\basecost(t)(\longvec{|v|}),\basepot(t)(\longvec{|v|}))$.)}
We call $\basecost(t)$ and $\basepot(t)$, respectively, the \emph{base cost} 
and \emph{base potential} of $t$.

(c)
For each $p\in\sempot{\sigma}$, let $\basepot(p) \defeq 
\basepot(\,(\tally1,p)\,)$.
\end{definition}

Suppose $t$ is the time complexity of  $e$ of type
$(\vec{\sigma})\to\Natl{\ell}$.  Then both $\basecost(t)$ and
$\basepot(t)$ are functions of the sizes of possible arguments of
$e$.  The intent is that $\basecost(t)(\longvec{|v|})$ is an upper
bound on the time cost of first evaluating $e$ and then applying its
value to arguments of the specified sizes and that $\basepot(t)$ is
an upper bound on the length of $e$'s value.

With $\basepot$'s definition in hand, we make good on the promise to
check that the notions defined between
Definitions~\ref{d:lenwt:time} and~\ref{d:bigproj} make sense.

\begin{lemma}\label{l:loose} 
Suppose $\sigma$ and $\sigma\to\tau$ are $\atr$ types.

(a) 
For each $v\in\semvalwt{\sigma}$, \ $\tcx{v}\in\semtime{\sigma}$ and
$\basepot(v)=|v|$.

(b) 
For each $t_0\in\semtime{\sigma\to\tau}$ and
$t_1\in\semtime{\sigma}$, \ $t_0\aptime t_1 \in\semtime{\tau}$.

(c) 
$\currytime$ is well-defined in the sense that the left-hand side of
\refeq{e:Lam} is in $\semtime{\sigma_1\to\tau}$ as asserted in
Definition~\ref{l:loose}.
\end{lemma}

All three parts follow straightforwardly from the definitions.

\subsection*{Time-complexity polynomials}

To complete the basic time-complexity framework, we define an
extension of the second-order polynomials for the simple product
types over $\Tally$, $\Tallyl{\varepsilon}$, $\Tallyl{\Prg},
\dots\;$ under the $\SL$-semantics.  The restriction of these to the
time types under the $\SLwt$-semantics are the
\emph{time-complexity polynomials}.  First we extend the grammar for
raw expressions to include: $
  P \;\;\is \;\; (P,P) \synsep \pi_{1}(P) \synsep \pi_{2}(P). 
$
Then we add the following new typing rules for second-order 
polynomials:
\begin{gather*}
    \irule{
                \Sigma\entails p\of\sigma_{1}\times\sigma_{2}}{
                \Sigma\entails \pi_{i}(p)\of\sigma_{i} }
    \Quad{1.8}
    \irule{\Sigma_{1}\entails p_{1}\of\sigma_{1} \Quad{1.25} 
           \Sigma_{2}\entails p_{2}\of\sigma_{2}
    }{\Sigma_{1}\cup\Sigma_{2}\entails 
        (p_{1},p_{2})\of\sigma_{1}\times\sigma_{2}}  
        \Quad{1.8}      
    \irule{\Sigma_{1}\entails p_{1}\of\sigma \Quad{1.25} 
           \Sigma_{2}\entails p_{2}\of\sigma
    }{\Sigma_{1}\cup\Sigma_{2}\entails 
        p_{1}\odot p_{2}\of\sigma}      
\end{gather*}
where $\sigma$, $\sigma_1$, and $\sigma_2$ simple product types over
$\Tally$, $\Tallyl{\varepsilon}$, $\Tallyl{\Prg}, \dots\;$ and
$\odot$ stands for any of $\ast$, $+$, or $\bmax$.  Next we extend
the arithmetic operations to all types by recursively defining, for
each $\gamma$ and each $u,v\in\semlen{\gamma}$:
\begin{gather}\label{e:arth:tc}
    u\odot v \Verb{\defeq}
    \begin{cases}
      \hbox{the standard thing}, 
         & \hbox{if }\gamma=\Tally;\\
      (\pi_{1}(u)\odot \pi_{1}(v),\pi_{2}(u)\odot \pi_{2}(v)),
         & \hbox{if }\gamma=\sigma\times\tau;\\
     \lam{z\in\semlen{\sigma}}(u(z)\odot v(z)),
         & \hbox{if } \gamma=\sigma\to\tau.
     \Quad2
\end{cases}\end{gather}  
Finally, the $\SL$-interpretation of the polynomials is just the
standard definition.

\begin{remark}\label{rem:tcpolys}
  Note that $q_1$ of Lemma~\ref{l:||f||} and the right-hand sides of 
  \refeq{e:time:oracle} and \refeq{e:curry:time} are well-typed, 
  time-complexity polynomials.  Also note that by 
  Definition~\ref{d:star}(a), if $q_1$ and $q_2$ are 
  time-complexity
  polynomials with $\tcx{\Gamma;\Delta}\entails q_1\of
  \tcx{\sigma\to\tau}$ and $\tcx{\Gamma;\Delta}\entails q_2\of
  \tcx{\sigma}$, then $q_1 \aptime q_2$ is a time-complexity polynomial
  with $\tcx{\Gamma;\Delta}\entails q_1\aptime q_2\of \tcx{\tau}$.
\end{remark}

\section{The time-complexity interpretation of $\atr^{-}$} 
\label{S:tcminus}

Here we establish a polynomial time-boundedness result for
$\atr^{-}$, the subsystem of $\atr$ obtained by dropping the $\crec$
construct.  Definition~\ref{d:semtime:1} introduces the
$\mathcal{T}$-interpretation of $\atr^{-}$ and the proof of
Theorem~\ref{t:ptimebnd-} shows that $\atr^{-}$-expressions have
time complexities that are polynomial bounded and well-behaved in
other ways.  All of this turns out to be pleasantly straightforward.
The hard work comes in the following two sections: \S \ref{S:decomp}
establishes a key time-complexity decomposition property concerning
the affine types and \S \ref{S:tcplus} uses this decomposition to
define the $\Time$-interpretation of $\crec$ expressions and to
prove a polynomial boundedness theorem for $\atr$ time complexities.

\emph{Convention:} Through out this section suppose that $\gamma$,
$\sigma$, and $\tau$ are $\atr$ types and $\Gamma;\Delta$ is an
$\atr$ type context.

\begin{figure}[t]\small
\begin{align*}
  \semtime{k} \,\varrho 
   &\Verb{\defeq}  
     \tcx{k}   
  &
  \semtime{(\consa\;e_{0})}\,\varrho 
     &\Verb{\defeq} 
     (c_{0}+\tally2,\, p_{0}+\tally1).
  \\
  \semtime{(\tsta\;e_{0})}\,\varrho 
     &\Verb{\defeq}
     (c_{0}+\tally2,\, \tally1).  
  &
  \semtime{(\cdr\;e_{0})}\,\varrho 
     &\Verb{\defeq} 
     (c_{0}+\tally2,\, (p_{0}-1)\bmax\tally0). 
  \\[1ex]
  \semtime{v} \,\varrho 
      &\Verb{\defeq} 
      \varrho(v). 
  &
  \semtime{(\down\;e_0\;e_1)}\,\varrho  
      &\Verb{\defeq}   
      (c_{0}+c_{1}+p_{0}+p_{1}+\tally3,\, \min(p_{0},p_{1})).
  \\[1ex]
  \semtime{ (\lam{x} e_{0})} \,\varrho
      &\Verb{\defeq}  
      \currytime(x,\semtime{e_{0}})\,\varrho.
  &
  \semtime{ (e_{0}\;e_{1})} \,\varrho 
      &\Verb{\defeq}
      (\semtime{e_{0}} \,\varrho)
      \aptime (\semtime{e_{1}}\,\varrho).
\end{align*}
\begin{gather*}
  \semtime{(\Iif \;e_0 \; \Ithen \;e_1 \; \Ielse\; e_2)}\,\varrho
      \Verb{\defeq} 
      (c_{0} +\tally2,\tally0) + (c_{1},p_{1})\bmax(c_{2},p_{2}).
      \\[1ex]
  \hbox{Above: $k$ is a string constant, 
          $\varrho\in\semtime{\Gamma;\Delta}$, and
          $(c_{i},p_{i})=
          \semtime{\Gamma;\Delta\entails e_{i}\of\gamma_i}\,\varrho$ 
          for $i=0,1,2$.}    
\end{gather*}
\caption{The $\mathcal{T}$-interpretation of  $\atr^{-}$.} 
\label{fig:ltr:time}
\figrule
\end{figure}
\begin{definition}\label{d:semtime:1}
  Figure~\ref{fig:ltr:time} provides the
  $\mathcal{T}$-interpretation for each $\atr^{-}$
  construct.
\end{definition}

We note that our $\ST$-interpretation of $\atr^-$ is well-defined
in the sense that $\semtime{\Gamma;\Delta\entails e\of\gamma}\,
\varrho\in \semtime{\gamma}$ for each $\atr^-$ judgment $\Gamma;
\Delta \entails e\of\gamma$ and $\varrho\in\semtime{\Gamma;\Delta}$. 
(This follows from Lemma~\ref{l:loose} and some straightforward calculations.)
Here is a simple application of Definition~\ref{d:semtime:1}.
Let $\mathsf{g} = (\lam{y}(\conso\;(\conso\;y)))
\of\allowbreak \Natl{\varepsilon}\to\Natl{\Prg}$ and
$\mathsf{A} = (\lam{f} \allowbreak 
   (\lam{x}(f\;x))\,)\of(\Natl{\varepsilon}\to\Natl{\Prg})
   \to\Natl{\varepsilon} \to\Natl{\Prg}$.
We write $\semtime{e}$ for $\semtime{e}\emptyenv$ to cut 
some clutter.  The reader may check that:
\begin{align*} 
  \semtime{\mathsf{g}} 
  &\Verb{=} 
  (\tally1,\, \lam{p_{y}\in
  \sempot{\Natl{\varepsilon}}}  
    (\tally1\bmax p_y+\tally4, p_{y}+\tally2)). \\
  \semtime{\mathsf{A}} 
  &\Verb{=}
    \left(\tally1,\, 
    \lam{p_{f}\in\sempot{ \Natl{\varepsilon} 
    \to 
    \Natl{\Prg}}}
    \big(\tally1,\lam{p_{x}\in\sempot{\Natl{\varepsilon}}}
    \val(p_f) \aptime \val(p_{x})\big) \right).\\
  \semtime{(\mathsf{A}\;\mathsf{g})} 
  &\Verb{=}
    (\semtime{\mathsf{A}} ) \aptime (\semtime{\mathsf{g}}) 
      \;\;=\;\; \dally({7},\semtime{\mathsf{g}}).
\end{align*}

There are three key things to establish about the time 
complexities assigned by $\Time$, that they are:
not too big,
not too small, and
well-behaved.
``Not too big'' 
means that the time complexities are polynomially-bounded
in the  sense of Definition~\ref{d:polytimebnd} below.  
``Not too small'' 
means that $\cekcost(e,\rho)\leq \Cost(\semtime{e}\tcx{\rho})$ 
and $|\semvalwt{e}\,\rho| \leq \basepot(\semtime{e} \tcx{\rho})$.  
This ``not too small'' property (\emph{soundness}) is introduced in 
Definition~\ref{d:cek-sound}.  Finally,  ``well-behaved'' means that 
the $\Time$-assigned time complexities are \emph{monotone} 
(Definition~\ref{d:tmono}) which requires  that $\semtime{e}\,\varrho \leq 
\semtime{e}\,\varrho'$ when $\varrho \leq \varrho'$ (see 
Definition~\ref{d:tmono}(a)) and that when $\semtime{e}\,\varrho$ is 
a function, it is pointwise, monotone nondecreasing. 
Monotonicity  plays an important role in dealing with
$\crec$.  Theorem~\ref{t:ptimebnd-} establishes that the
$\Time$-interpretation of $\atr^{-}$ satisfies each of these 
properties.
Let $\SF$ range over programming 
formalisms (e.g.,   $\atr^{-}$ or $\atr$) in the following. 

\begin{definition}[Polynomial time-boundedness] 
\label{d:polytimebnd}
  A $\ST$-interpretation of  $\mathcal{F}$ is 
  \emph{polynomial time-bounded} 
  when, given  $\Gamma;\Delta\entails_\SF e\of \gamma$,
  we can effectively find a time-complexity polynomial $p_e$ with
  $|\Gamma;\Delta|\entails p_e\of\tcx{\gamma}$ 
  such that
  $\semtime{e}\tcx{\rho} \,\leq\, \semlenwt{p_e}\,|\rho|$
  for each $\rho\in\semvalwt{\Gamma;\Delta}$.
\end{definition}

\begin{definition}[Soundness] \label{d:cek-sound}
  A $\ST$-interpretation of $\mathcal{F}$
  is \emph{sound}  
  when, for each 
  $\Gamma;\Delta\entails_\SF e\of\gamma$ and each
  $\rho\in\semvalwt{\Gamma;\Delta}$,  we have 
  $\cekcost(e,\rho)$ 
         $\leq$ $\Cost(\semtime{e}\tcx{\rho})$ and
  $\left| \strut \semvalwt{e}\,\rho\right| 
        \leq
        \basepot(\semtime{e}\tcx{\rho})
        $. 
\end{definition}

\begin{definition}[Monotonicity]\label{d:tmono}{\ }

  (a) 
  For $\varrho,\,\varrho'\in\semtime{\Gamma;\Delta}$,
  we write $\varrho \leq \varrho'$ when
  $\varrho(x)\leq   \varrho'(x)$ for each 
  $x\in\preimg(\Gamma;\Delta)$.
  
  (b)
  We say that a $\ST$-interpretation of $\SF$ is 
  \emph{monotone} when, for each
  $\Gamma;\Delta\entails_\SF e\of\gamma$:
  (i)   $\semtime{e}$
  is a pointwise, monotone nondecreasing function from
  $\semtime{\Gamma;\Delta}$ to $\semtime{\gamma}$, and 
  (ii) if $\gamma=(\sigma_{0},\dots,\sigma_{k})\to\bb$, 
  then the function from
  $\semtime{\Gamma;\Delta}\times\semtime{\sigma_{0}}
  \times \dots\times \semtime{\sigma_{k}}$ to $\semtime{\bb}$ 
  given by  $(\varrho,v_{0},\dots,v_{k}) \mapsto 
  ((\semtime{e}\,\varrho)\, v_{0}\, \dots \,v_{k})$ is 
  pointwise, monotone nondecreasing.
\end{definition}

\begin{theorem}\label{t:ptimebnd-}
  The $\Time$-interpretation of $\atr^-$ is
  (a) poly\-nomial time-bounded,
  (b) monotone, and
  (c) sound.
\end{theorem}

The proofs of parts (a) and (b) are straightforward standard 
structural inductions, but the argument for (c) is
a logical-relations arguments \cite{winskel:book}.
Before proving the above we first introduce a few useful 
time-complexity polynomials.

\begin{definition} \label{d:basepolys}
  \textbf{N.B.} The following definitions are purely syntactic.
  Suppose $v\in\nat$. Let $\tcx{v}\defeq
  (\tally1\bmax|v|,|v|)$.
  For each  $\bb$, let
  $\tcx{x}_{\bb}\defeq (\tally1\bmax|x|,|x|)$.
  For each $\gamma=(\bb_{1},\ldots,\bb_{k})\to\bb_{0}$, let
  $\tcx{x}_{\gamma}\defeq (\tally1,q_{1})$ where
  $
  q_{i}=\lam{p_{i}}
    \left(\tally1,q_{i+1})\right)$, 
  for each $i$ with $1\leq i< k$, and
  $q_{k}= \lam{p_{k}}
    \left(\tally1 \bmax \allowbreak |x|(p_{1},\dots,p_{k}), \strut
       |x|(p_{1},\ldots,p_{k})\right)$.
   (Recall Lemma~\ref{l:||f||}.)    
\end{definition}

Note that if  $\Gamma;\Delta\entails x\of\gamma$ where $x$ 
is a variable, then $|\Gamma;\Delta|\entails \tcx{x}\of\tcx{\gamma}$.

\ProoF{of Theorem~\ref{t:ptimebnd-}(a): Polynomial time-boundedness}
Fix an $\atr^-$-judgment $\Gamma;\Delta\entails e\of\gamma$.  Let
$\rho$ range over $\semvalwt{\Gamma;\Delta}$.  
We have to effectively construct a t.c.~polynomial $q_e$ as required 
by Definition~\ref{d:polytimebnd}.
The argument is yet another a structural
induction on the derivation of $\Gamma;\Delta\entails e\of \gamma$.
We consider the cases of the last rule used in the derivation.

\textsc{Cases:} \textit{Zero-I} and \textit{Const-I}.  Then
$e=v\in\nat$ and $\gamma$ is a base type.  Let $q_e = \tcx{v}$.  By
Definition~\ref{d:semtime:1}, $\semtime{v}\tcx{\rho}= (\tally1 \bmax
|v|,|v|) = \semlenwt{q_{e}}\,|\rho|$ and thus $q_e$ suffices.

\textsc{Cases:} \textit{Int-Id-I} and \textit{Aff-Id-I}.  Then
$e=x$, a variable.  Then by Definition~\ref{d:semtime:1},
$\semtime{x}\tcx{\rho}= \tcx{\rho}(x) = \tcx{\rho(x)}$.  Let $q_{e}=
\tcx{x}_{\gamma}$.
\emph{Subcase:} $\gamma$ is a base type.  By
Definition~\ref{d:basepolys}(b), $q_{e} = (\tally1 \bmax |x|,|x|)$.
So by Definition~\ref{d:tcxa}, $\tcx{\rho(x)} = \semlenwt{q_{e}} \,
|\rho|$ and thus $q_e$ suffices.
\emph{Subcase:} $\gamma = (\bb_{1},\dots,\bb_{k})\to\bb_{0}$.  By
Definition~\ref{d:basepolys}(c), $q_{e} = (\tally1,q_{1})$, where
$q_{1},\ldots,q_{k}$ are as in that definition.  By
Lemma~\ref{l:||f||}, $\tcx{\rho(x)} = \semlenwt{q_{e}}|\rho|$ 
and thus $q_{e}$ suffices.

\textsc{Case:} $\consa$\textit{-I}, where $\ba\in\set{\bo,\bl}$.
Then $e=(\consa\;e_{0})$ for some $e_{0}$ and $\gamma=\Natl{\Prg_d}$
for some $d$.  Let $(c_{0},
p_{0}) = \semtime{e_{0}}\tcx{\rho}$. By
Definition~\ref{d:semtime:1}, $\semtime{e}\tcx{\rho} = (c_{0} +
\tally2,p_{0}+\tally1)$.  By the induction hypothesis, we can construct $q_{e_{0}}$ with
$|\Gamma;\Delta|\entails q_{e_0}\of \Natl{\Prg_d}$ such that
$\semtime{e_{0}}\tcx{\rho} \leq \semlenwt{q_{e_{0}}}\,|\rho|$.
Thus, $q_{e}=q_{e_{0}}+(\tally2,\,\tally1)$ suffices.

\textsc{Cases:} $\tsto$\introrule, $\tstl$\introrule,
$\down$\introrule, $\cdr$\textit{-I}, \textit{$\to$-e}, and
\textit{If-I}.  These follow by arguments analogous to the proof for
the $\consa$\textit{-I} case.

\textsc{Cases:} \textit{Subsumption} and \textit{Shifting}.  
There is nothing to prove here.

\textsc{Case:} \textit{$\to$-E}.  Then
$e= (e_{0}\; e_{1})$ for some $e_{0}$ and $e_1$ 
with $\Gamma;\Delta\entails e_{0}\of\tau\to\gamma$ and
$\Gamma;\emptycont\entails e_{1}\of\tau$.
By the induction hypothesis, we can construct 
$q_0$ and $q_1$, bounding time-complexity polynomials for 
$\semtime{e_0}$ and $\semtime{e_1}$, respectively. 
Let $q_e = q_0 \aptime q_1$.  
By Remark~\ref{rem:tcpolys}, $q_e$ is a time-complexity polynomial
and it follows from the monotonicity of $\aptime$ that 
$\semtime{(e_{0}\; e_{1})}\,\tcx{\rho} = 
(\semtime{e_{0}}\,\tcx{\rho}) \aptime (\semtime{e_{1}}\,\tcx{\rho}) 
\leq (q_0\,\tcx{\rho}) \aptime (q_1\,\tcx{\rho}) = 
q_e \,\tcx{\rho}$.  Thus, $q_e$ suffices.

\textsc{Case:} \textit{$\to$-I}.  Then $\gamma=\sigma\to\tau$ and
$e= (\lam{x} e_{0})$ for some $e_{0}$ with $\Gamma, x\of\sigma;
\Delta\entails e_{0}\of\tau$.  By Definitions~\ref{d:currystar}
and~\ref{d:semtime:1} we thus have
$\semtime{e}$ = $\currytime(x,\semtime{e_{0}})$.
By the induction hypothesis, we can construct $q_{e_0}$ with
$|\Gamma,x\of\sigma;\Delta|\entails q_{e_0} \of \tcx{\tau}$ with
$\semtime{e_0}\,\tcx{\rho'} \leq \semlenwt{q_{e_0}}\,|\rho'|$ for
each $\rho'\in \semvalwt{\Gamma,x\of\sigma;\Delta}$.
\emph{Subcase:} $\sigma$ is a base type.  So, $\pot{\sigma} =
\Tally\times \sigma$.  Let $q_{e}=(\tally1,\lam{|x|} q_{e_{0}})$.  A
straightforward argument shows that $q_{e}$ suffices for the
polynomial bound.
\emph{Subcase:} $\sigma=(\sigma_1,\dots,\sigma_k)\to\bb$.  Let $p'$
be the expression $\lam{\longvec{|y|}} \pi_2\big(\, (\tally1, p)
\aptime \longvec{\tcx{y}}\,\big)$, where $\longvec{|y|}
=|y_1|,\dots,|y_k|$ and $\longvec{\tcx{y}} =
(\tally1\bmax|y_1|,|y_1|),\dots,(\tally1\bmax|y_k|,|y_k|)$.  (See
Definition~\ref{d:bigproj}(b).)  Let $p''$ be the expansion of $p'$
in which $p$ is treated as being of type $\pot{\sigma}$ and the
$\Time$-applications are expanded out per Definition~\ref{d:star}.
It follows that $p''$ is a time complexity polynomial with
$|\Gamma|,p\of\pot{\sigma};|\Delta|\entails p'' \of |\sigma|$.  Let
$q_e = (\tally1,\, \lam{p} q_{e_0}[|x|\gets p''])$.  Again, a
straightforward argument shows that $q_{e}$ suffices for the
polynomial bound. \Qed{Theorem~\ref{t:ptimebnd-}(a)}

\ProoF{of Theorem~\ref{t:ptimebnd-}(b): Monotonicity}
This argument follows along the lines of the proof of part (a)
and is left to the reader. 
\Qed{Theorem~\ref{t:ptimebnd-}(b)}

For the proof of soundness, we shall first define a logical relation 
$\tcapprox{\gamma}$
between CEK-closures and time-complexities.  Roughly, 
$e\hat\rho \tcapprox{\gamma} (c,p)$ says that the time complexity
$(c,p)$ bounds the cost of evaluating the closure $e\hat\rho$.
\emph{Conventions on CEK-closures:} 
CEK-closures are written $e\hat\rho$.
(We always assume $FV(e)\subseteq \preimg(\hat\rho)$.)
A CEK-closure $e\hat{\rho}$ is called a \emph{value} when
$e$ is a CEK-value. $e\hat\rho \evalsto v\hat\rho'$ means that
starting from $(e,\hat\rho,\halt)$, the CEK-machine 
eventually ends up with $(v,\hat\rho',\halt)$, where
$v\hat\rho'$ is a value.
Below, $v$ ranges over CEK-values and 
$p$ and $q$ range over potentials.

\begin{definition}\label{d:lr}    \ 

(a) For each $\atr$-type $\gamma$ we 
define a relation $\tcapprox{\gamma}$
between type-$\gamma$ CEK-closures and time complexities 
and a second relation $\potapprox{\gamma}$ between type-$\gamma$ 
CEK-closures and potentials as follows.
\begin{itemize}
  \item
    $e\hat{\rho} \tcapprox{\gamma} (c,p)$ 
    $\equiv_{\text{def}}$
    $\cekcost(e,\hat{\rho}) \leq c$ $\&$     
    $v\hat{\rho}' \potapprox{\gamma} p$, 
    where $e\hat{\rho} \evalsto v\hat{\rho}'$.
  \item
    $v\hat{\rho} \potapprox{\bb} p $ $\equiv_{\text{def}}$
    $|v\hat{\rho}| \leq p$. 
  \item 
    $(\lam{x}e)\hat{\rho} \potapprox{\sigma\to\tau} p $ 
    $\equiv_{\text{def}}$
    for all $v\hat{\rho}'$ and
    all $q$ with $v\hat{\rho}' \potapprox{\sigma} q$, \
    $e (\hat{\rho}[x\mapsto v\hat{\rho}'] )
    \tcapprox{\tau} p(q)$.       
  \item 
    $O\hat{\rho} \potapprox{\gamma\to\tau} p$ 
    $\equiv_{\text{def}}$
    for all  $v\hat{\rho}'$ and
    all  $q$ with $v\hat{\rho}' \potapprox{\gamma} q$, \
    $O (v\hat{\rho}') \emptyenv \tcapprox{\tau} p(q)$.       
\end{itemize}

	(b) Suppose $\varrho\in\semtime{\Gamma;\Delta}$.
	We write $\hat{\rho} \sqsubseteq \varrho$ when, 
	for each  $x\in\preimg(\hat{\rho})$, \ 
	$x\hat{\rho} \tcapprox{\gamma}\varrho(x)$, where 
	$\gamma = (\Gamma;\Delta)(x)$.

	(c) Suppose $\Gamma;\Delta \entails e\of\gamma$ and
	$X\of\semtime{\Gamma;\Delta}\to\semtime{\gamma}$.  We write
	$e\tcapprox{\gamma} X$ when, for all CEK-environments 
	$\hat\rho$ and 
	$\varrho\in\semtime{\Gamma;\Delta}$ with 
	$\hat\rho\sqsubseteq \varrho$, 
	 \ $e\hat\rho\tcapprox{\gamma} X\varrho$.
\end{definition}

\begin{lemma} \label{l:soundness:atr-} \

(a) Suppose $x$ is a variable and $v\hat\rho \potapprox{\gamma} q$. 
Then
$x(\hat\rho' \cup \set{x\mapsto v\hat\rho})
\tcapprox{\gamma} \val(q)$.

(b)  Suppose $\Gamma;\Delta \entails e\of\gamma$. Then
$e \sqsubseteq^{\text{tc}}_{\gamma}
\semtime{e}$.

(c) Suppose $e\hat\rho$ is a type-$\gamma$ CEK-closure
  and $t$ and $t'$ are type-$\gamma$ time complexities
  with $e\hat\rho \tcapprox{\gamma} t$ and $t\leq t'$.
  Then $e\hat\rho \tcapprox{\gamma} t'$.
\end{lemma}

\proof \emph{Part (a).}  Since 
$v\hat\rho \potapprox{\gamma} q = \potprj(\val(q))$, we just need to
show that $\cekcost(x(\hat\rho' \cup \set{x\mapsto v\hat\rho})) \leq
\costprj(\val(q))$.  If $\gamma$ is a base type, then
$|v| \leq q$, hence
$\cekcost(x(\hat\rho' \cup \set{x\mapsto v\hat\rho})) 
= \tally1 \bmax |v| \leq \tally1 \bmax q = 
\costprj(\val(q))$.

\emph{Part (b).} The argument is a structural
induction on the derivation of $\Gamma;\Delta\entails e\of \gamma$.
We consider the cases of the last rule used in the derivation.
Fix a CEK-environment $\hat\rho$ and a 
$\varrho\in\semtime{\Gamma;\Delta}$ with
$\hat{\rho}\sqsubseteq \varrho$.

\textsc{Case:} \textit{Zero-I} and \textit{Const-I}.
Then $e=v$, a string constant.  So, 
$\semtime{e}\,\varrho=(\tally1\bmax |v|,|v|)$,
$\cekcost(e,\hat{\rho})=\tally1 \leq \tally1\bmax |v|$, and 
$|e\,\varrho|=|v|$.
Hence, $e$ is as required.

\textsc{Case:} \textit{Int-Id-I} and \textit{Aff-Id-I}.
Then $e=x$, a variable.  Since 
$\hat{\rho}\sqsubseteq 
\varrho$, we have $x\hat{\rho} \sqsubseteq^{\text{tc}}_{\gamma}
\varrho(x) = \semtime{x} \, \varrho$.
Hence, $e$ is  as required.

\textsc{Case:} $\consa$\textit{-I}, where $\ba\in\set{\bo,\bl}$.
Then $e=(\consa\;e_{0})$ where $\Gamma;\Delta\entails e_0\of\gamma$
and $\gamma$ is a base type.
Let  $(c_{0},p_{0}) = \semtime{e_{0}}\,\varrho$ and 
suppose $e_0\hat{\rho}\evalsto v\hat{\rho}'$.
By the induction hypothesis applied to $e_0$, we know
$\cekcost(e_0,\hat{\rho}) \leq c_0$ and 
$|v\hat{\rho}'| \leq p_0$. By inspection of the CEK machine 
and the
definition of $\cekcost$, 
$\cekcost(\consa\;e_0,\hat{\rho}) = 
\cekcost(e_0,\hat{\rho}) +\tally2 \leq c_0+\tally2$.
It also follows that $(\consa\;e_{0})\hat{\rho} \evalsto
(\ba\oplus v)\hat{\rho}'$ and $|(\ba\oplus v)\hat{\rho}'|
= | v\hat{\rho}'|+\tally1 \leq p_0+\tally1$.  By
Definition~\ref{d:semtime:1}, $\semtime{e}\tcx{\rho} = (c_{0} +
\tally2,p_{0}+\tally1)$.  Hence, $e$ is as required.

\textsc{Cases:} $\tsto$\introrule, $\tstl$\introrule,
$\down$\introrule, $\cdr$\textit{-I}, \textit{$\to$-E}, and
\textit{If-I}.  These follow by arguments analogous to the proof for
the $\consa$\textit{-I} case.

\textsc{Cases:} \textit{Subsumption} and \textit{Shifting}.  
There is nothing to prove here.

\textsc{Case:} $\to$-\emph{I}.  Then $\gamma=\sigma\to\tau$ and
$e= (\lam{x} e_{0})$ for some $e_{0}$ with $\Gamma, x\of\sigma;
\Delta\entails e_{0}\of\tau$.   
So, by Definition~\ref{d:semtime:1}, 
$\costprj(\semtime{\lam{x} e_{0}}\,\varrho) = \tally1$
and $\lam{x} e_{0}\,\varrho$ is itself a value.        
Since  $\cekcost(\lam{x} e_{0},\hat{\rho})=\tally1$, all that
is left to show is that 
$(\lam{x} e_{0})\hat\rho\potapprox{\sigma\to\tau} 
\potprj(\semtime{\lam{x} e_{0}}\,\varrho)$.  
Let $p=\potprj(\semtime{\lam{x} e_{0}}\,\varrho) =
\potprj(\currytime(x,\semtime{e_{0}})\,\varrho) =
\lam{p' \in\sempot{\sigma}} (\semtime{e_{0}} \,(\varrho'\cup
\set{x\mapsto \val(p')}))$, let $v\hat{\rho}$ be an
arbitrary type-$\sigma$ 
value  and let $q$ be an arbitrary potential with 
$v\hat{\rho} \potapprox{\sigma} q$.   
Then establishing
$(\lam{x} e_{0})\hat\rho\potapprox{\sigma\to\tau} 
\semtime{\lam{x} e_{0}}\,\varrho$ is equivalent to showing
$e_0 (\hat{\rho}[x\mapsto v\hat{\rho}'] )
       \tcapprox{\tau}   p(q)$
By part (a), $x(\hat\rho\cup\set{x\mapsto v\rho} 
\tcapprox{\sigma} \val(q)$.  Hence, 
$\hat\rho\cup\set{x\mapsto v\rho} 
\sqsubseteq \varrho\cup\set{x\mapsto\val(q)}$.
Thus, by the induction hypothesis on $e_0$, \
$e_0 (\hat{\rho}[x\mapsto v\hat{\rho}'] )
       \tcapprox{\tau} \semtime{e_{0}}(\varrho'\cup
       \set{x\mapsto \val(q)}) = p(q)$.  
Hence, $e$ is as required.

\textsc{Case:} \textit{$\to$-E}.  Then
$e= (e_{0}\; e_{1})$ for some $e_{0}$ and $e_1$ 
with $\Gamma;\Delta\entails e_{0}\of\sigma\to\gamma$ and
$\Gamma;\emptycont\entails e_{1}\of\sigma$.
Suppose 
$e_{0}\hat\rho \evalsto v_0\hat\rho_0$, \
$e_{1}\hat\rho \evalsto v_1\hat\rho_1$, \
$(e_{0}\; e_{1})\hat\rho \evalsto v_r\hat\rho_r$,\ 
$(c_0,p_0)=\semtime{e_0}\varrho$, \
$(c_1,p_1)=\semtime{e_1}\varrho$, and
$(c_r,p_r) = p_0(p_1)$. 
By the induction hypothesis on $e_0$ and $e_1$:
\begin{align} \label{e:sound0}
& \text{(a) } \cekcost(e_0,\hat\rho) 
   \leq c_0. & 
& \text{(b) }  v_0\hat\rho_0 
   \potapprox{\sigma} p_0. \\
   \label{e:sound1}
& \text{(a) } \cekcost(e_1,\hat\rho) 
   \leq c_1. & 
& \text{(b) }  v_1\hat\rho_1 
   \potapprox{\sigma} p_1. 
\end{align}
There are two subcases to consider based on the form of $v_0$.
\emph{Subcase:} $v_0=\lam{x}e_0'$ for some $\Gamma,x\of\sigma;\Delta
\entails e_0'\of \gamma$.  Then (\ref{e:sound0}b) means that,
for all type-$\tau$ values $v\hat\rho''$ and all  $q$
with $v\hat\rho'' \potapprox{\sigma} q$, we have
$e_0'\,(\hat\rho'\cup\set{x\mapsto v\hat\rho''})
\tcapprox{\gamma} p_0(q)$.
So by (\ref{e:sound1}b), 
$e_0'\hat\rho'_0 \tcapprox{\gamma}  (c_r,p_r)$, where $\hat\rho'_0 =  
\hat\rho'\cup\set{x\mapsto v_1\hat\rho_1}$. Now
\begin{align*}
{\cekcost((e_0\;e_1),\hat\rho) }
  & \; = \;
    \lefteqn{\cekcost(e_0,\hat\rho) + \cekcost(e_1,\hat\rho) 
       + \cekcost(e_0',\hat\rho'_0 ) + \tally3}\\
  &&& \hbox{(by Figure~\ref{fig:cek}
  \& Definition~\ref{d:cekcosts})}
  \\
  &\; \leq \; 
    c_0 + c_1 + c_r  + \tally3
  && \hbox{(by (\ref{e:sound0}a), (\ref{e:sound1}a), 
  \& $e_0'\hat\rho'_0 \tcapprox{\gamma}  (c_r,p_r)$)}
  \\
  &\; = \;
    \costprj(\semtime{(e_0\,e_1)}\, \varrho)
  && \hbox{(by Definition~\ref{d:semtime:1})}.  
\end{align*}
Note that $e_0'\hat\rho'_0 \evalsto v_r\hat\rho_r$.
So by $e_0'\hat\rho'_0 \tcapprox{\gamma}  (c_r,p_r)$,
$v_r\hat\rho_r \potapprox{\gamma} p_r = 
\potprj(\semtime{(e_0\,e_1)}\,\varrho)$.  Hence, 
in this subcase $e$ is as required.
\emph{Subcase:} $v_0$ is an oracle. 
The argument here is a repeat, \emph{mutatis mutandis}, of the proof 
of previous subcase.

\emph{Part (c).}   The argument follows along the 
lines of the proof of (b).
\Qed{Lemma~\ref{l:soundness:atr-}}

\ProoF{of Theorem~\ref{t:ptimebnd-}(c): Soundness}
This follows straightforwardly from Lemma~\ref{l:soundness:atr-}(b) 
and Definition~\ref{d:bigproj}.
\qed

\begin{scholium}\label{sch:why:T}
  The $\Time$-interpretation of $\atr^{-}$ (and later, $\atr$)
  sits in-between the actual costs of evaluating expressions on 
  our CEK machine and the sought-after polynomial time-bounds 
  on these costs.  \emph{Why is 
  working with $\Time$-interpretations preferable to
  working directly with executions of CEK machines and their costs?}
  Part of the reason is that $\Time$-interpretations have
  built-in to them the cost-potential aspects expressions.
  One would somehow have to replicate these in working directly 
  with CEK-computations.  Another part of the reason is that 
  $\Time$-interpretations collapse the many possible paths of 
  a CEK-computation into a single time-complexity.  
  The $\Time$-interpretation of $\Iif$-$\Ithen$-$\Ielse$ is 
  chiefly responsible for these collapses. 
  Scholium~\ref{sch:whyCEKcost} notes that these collapses are a source 
  of some trouble in dealing with $\crec$-expressions.  
\end{scholium}

\section{An affine decomposition of time complexities}
\label{S:decomp}

When analyzing the time complexity of a program, one often
needs to decompose its time complexity into pieces that may
have little to do with the program's apparent syntactic structure.
Theorem~\ref{t:lin:time} below is a general time-complexity
decomposition result for  $\atr$ expressions.  The $\atr$ 
typing rules for affinely restricted variables are critical in 
ensuring this time-complexity decomposition.  The decomposition
is used in the next section to obtain the recurrences for
the analysis of the time complexity of $\crec$ expressions.  Note
that the theorem presupposes that that $\semtime{\,\cdot\,}$
is defined on $\crec$ expressions.  However, since no
affinely restricted variable can occur free in a well-typed
$\crec$ expression and since the application of the theorem 
will be within a structural induction, this presupposition does 
not add any difficulties.   

\begin{remark}\label{rem:handwave}
  In fact, the time-complexity of a 
  $\crec$ expression $e$ will be defined in terms of 
  time-complexities of expressions built up from subexpressions 
  of $e$ using term constructors other than $\crec$.  Thus a 
  completely standard structural induction for establishing 
  soundness does not quite work.  A fully formal proof would have 
  first established results such as  
  ``if $e_0 \tcapprox{\sigma\to\tau} X_0$ and
    $e_1 \tcapprox{\sigma} X_1$, then
	$(e_0\,e_1) \tcapprox{\tau} X_0\star X_1$'' 
  where the $X_i$'s are general mappings from $\Time$-environments
  to time complexities.  These lemmas would then be used to 
  carry out the induction steps of a structural induction which, in 
  all but the $\crec$ case, would just quote the relevant lemma.  
  Rather than impose this additional level of detail on the reader,  
  we have opted for a less formal approach here and will assume 
  that if we inductively have soundness for a subterm $e$, then 
  we also have it for terms built up from $e$ without $\crec$.
\end{remark}

To help in the statement and proof of the Affine Decomposition Theorem, we introduce the following definitions and conventions.

\begin{definition} \label{d:uplus} {\ }

(a) 
$(c_{1},p_{1})\uplus (c_{2},p_{2}) \defeq
(c_{1}+c_{2},p_{1}\vee p_{2})$, where
$(c_{1},p_{1})$, $(c_{2},p_{2}) \in \semtime{\gamma}$.
\emph{(Clearly, $(c_{1},p_{1})\uplus (c_{2},p_{h2}) 
\in\semtime{\gamma}$.)}

(b)   
For each $\atr$-type $\gamma$, define
$\epsilon_\gamma$ inductively by:
$\epsilon_{\Natl{\ell}} = \epsilon$ and 
$\epsilon_{\sigma\to\tau} = \lam{x} \epsilon_{\tau}$.  
\emph{(Clearly, $\entails \epsilon_{\gamma}\of\gamma$ and 
$|\semvalwt{\epsilon_\gamma}\, \emptyenv|=\tally0_{|\gamma|}$.)}

(c) Given $f \of 
(\sigma_{1},\dots,\sigma_{k})\to\Natl{\ell}$, an expression of the  form $(f\;e_{1}\;\dots\; e_{k})$  is called a
\emph{full application} of $f$.
\end{definition}

\emph{Conventions on factoring out environments:}
Suppose $\odot$ is a binary operation on time complexities.
We often write $\semtime{e_{0}} \,{\odot}\,
\semtime{e_{1}}$ for $\varrho \mapsto
(\semtime{e_{0}}\,\varrho) \,\odot\,
(\semtime{e_{1}})\,\varrho)$.  For example:
$
 (\semtime{e_{0}} \uplus\semtime{e_{1}}) \,\varrho
   = (\semtime{e_{0}}\,\varrho) \uplus 
       (\semtime{e_{1}}) \,\varrho)$ and
$ (\semtime{e_{0}} \aptime \semtime{e_{1}}) \,\varrho 
   = (\semtime{e_{0}}\,\varrho) \aptime (\semtime{e_{1}}) \,\varrho)$.  
We extend this convention to $n$-ary operations.
For example: $ \val(\semtime{e}) \,\varrho = \val(\semtime{e} \,\varrho)$ and
$
 (\semtime{e_{0}} \aptime \dots \aptime 
    \semtime{e_{k}}) \,\varrho
    =
    (\semtime{e_{0}} \,\varrho) \aptime  \dots \aptime 
    (\semtime{e_{k}}) \,\varrho)$. 
We also generalize this last equality as follows.  Suppose
$X$ is a map from $\semtime{\Gamma;\Delta}$ to
$\semtime{(\sigma_{1},\dots,\sigma_{k})\to\Natl{\ell}}$ and, for
$i=1,\dots,k$, \ $Y_{i}$ is a map from
$\semtime{\Gamma;\Delta}$ to $\semtime{\sigma_{i}}$.  Then
$X\aptime \vec{Y}$ denotes the map $\semtime{\Gamma;\Delta}$
to $\semtime{\Natl{\ell}}$ given by: $(X\aptime\vec{Y})\varrho = (X
\varrho) \aptime (Y_{1}\varrho) \aptime
\dots\aptime (Y_{k}\varrho)$.

\begin{theorem}[Affine decomposition]\label{t:lin:time}
  Suppose $\Gamma;f\of\gamma\entails e\of\Natl{\ell_0}$, 
  where   
  $\gamma=(\Natl{\ell_{1}},\dots,\Natl{\ell_k})\allowbreak
  \to\Natl{\ell_0}\in\mathcal{R}$
  and $\TailPos(f,e)$.
  Let $\zeta$ denote the substitution $[f\gets \epsilon_{\gamma}]$.
  Then
  \begin{gather} \label{e:linbnd}
    \semtime{e}
     \Verb{\leq}
       \strut \semtime{ e\,\zeta} 
                \;\uplus\;   
       (\semtime{f}
       \aptime \vec{t}\;) ,
  \end{gather} 
  where
  $(f\;e_{1}^{1}\;\dots\;e_{k}^{1}),\ldots, 
  (f\;e_{1}^{m}\;\dots\;\allowbreak e_{k}^{m})$ 
  are the full applications  of $f$ occurring in 
  $e$ and
  $t_{j} = \bigmax_{i=1}^{m}\val(\semtime{e_{j}^{i}})$
  for $j=1,\ldots,k$.
\end{theorem}

By Lemma~\ref{l:1use} we know that there is at most one use of an 
affinely restricted variable in an expression.  In terms of costs, 
one can thus interpret \refeq{e:linbnd} as saying that the cost 
of evaluating $e$ can be bounded by the sum of: (i) the cost of 
evaluating $e\,\zeta$, which includes the all of the costs of $e$ 
except for the possible application of the value of $f$ to the 
values of its arguments, and (ii) $\Cost( (\semtime{f} \aptime 
\vec{t}\;)\,\varrho)$, which clearly bounds the cost of any such 
$f$ application. In terms of potentials, one can interpret 
\refeq{e:linbnd} as saying that the size of the value of $e$ is 
bounded by the maximum of (i) the size of the value of $e\,\zeta$, 
which covers all the cases where $f$ is not applied, and (ii) 
$\Pot((\semtime{f} \aptime \vec{t}\;)\,\varrho)$, which covers all 
the cases where $f$ is applied.

If \refeq{e:linbnd} solely concerned CEK costs, the
above remarks would almost constitute a proof.  However,
\refeq{e:linbnd} is about $\Time$-interpretations of
expressions and $\semtime{e}$ is an approximation to the
true time complexities involved in evaluating $e$.  The
theorem asserts that our $\Time$-interpretation of
$\atr$ is verisimilar enough to capture this property of time
complexities. This later requires a little work.

\ProoF{of Theorem~\ref{t:lin:time}}
Fix $\varrho\in \semtime{\Gamma;f\of\gamma}$.  Without loss
of generality, we assume there are no bound occurrences of
$f$ in $e$.  
We argue by structural induction that for each $A$, a
subterm of $e$ with $\Gamma;f\of\gamma\entails A\of\Natl{\ell_0}$,
we have
\begin{gather}\label{e:lin:goal}
  \semtime{A}\,\varrho 
  \Verb{\leq}
    \semtime{ A\,\zeta} \,\varrho \;\uplus\; \left(\semtime{f}
   \aptime \vec{t}\;  \right) \varrho, 
\end{gather}   
where the $\vec{t}\,$'s are as in the lemma's statement.
It follows from  $\TailPos(f,e)$ that the following three cases
are the only ones to consider.

\emph{Case 1:} $f$ fails to occur  in $A$. 
Then \refeq{e:lin:goal} follows immediately.   

\emph{Case 2:} $A=(f\;e_1 \;\ldots \; e_k)$, where $\Gamma; 
\emptycont \entails e_{1}\of \Natl{\ell_1},\ldots, \Gamma; 
\emptycont \entails e_{k}\of \Natl{\ell_k}$.  
By the monotonicity of $\semtime{f}$ and
the $\Time$-interpretation of application from
Figure~\ref{fig:ltr:time}, it follows that
\refeq{e:lin:goal} holds for $A$.

\emph{Case 3:} $A=(\Iif \;A_0 \; \Ithen \;A_1 \; \Ielse\; A_2)$
where $f$ occurs in $A_{1}$ or $A_{2}$ or both.  
By Definitions~\ref{d:semtime:1} and~\ref{t:lin:time}(c), 
$\semtime{A}\,\varrho$  =  
$(\Cost(\semtime{A_{0}}\,\varrho)+\tally2,\,\tally0)
        \;\uplus\; 
        {\textstyle \bigmax_{i=1}^{2}}\semtime{A_{i}}\,\varrho.
$
\begin{figure}[t] \small
\begin{align*}
\semtime{A}\,\varrho 
   &\Verb{=}
   (\Cost(\semtime{A_{0}}\,\varrho)+\tally2,\,\tally0)
        \;\uplus\; 
        {\textstyle \bigmax_{i=1}^{2}}\semtime{A_{i}}\,\varrho
        \\
   &\Verb{\leq}
   (\Cost(\semtime{A_{0}}\,\varrho)+\tally2,\,\tally0)
       \;\uplus\;
       {\textstyle 
       \bigmax_{i=1}^{2}}\left(\semtime{A_{i}\,\zeta}\,\varrho  \Strut{2.5ex}
       \;\uplus\;  
       (\semtime{f} \aptime \vec{t}\, )\,\varrho \right)
       \\
   &\Verb{\leq}
   (\Cost(\semtime{A_{0}}\,\varrho)+\tally2,\,\tally0)
       \;\uplus\;
       \left({\textstyle \bigmax_{i=1}^{2}}\semtime{A_{i}\,\zeta}\,\varrho
       \right) 
       \;\uplus\; 
       (\semtime{f} \aptime \vec{t}\, )\,\varrho
       \\[-0.25ex]
   &\Verb{\leq}
   \left((\Cost(\semtime{A_{0}\,\zeta}\,\varrho)+\tally2,\,\tally0)
       \;\uplus\;
       {\textstyle \bigmax_{i=1}^{2}}\semtime{A_{i}\,\zeta}\,\varrho
       \right) 
       \;\uplus\; 
       (\semtime{f} \aptime \vec{t}\, )\,\varrho
       \\[-0.25ex]
   &\Verb{=} \semtime{A\,\zeta}\,\varrho 
       \;\uplus\;  (\semtime{f} \aptime \vec{t}\,  )\,\varrho .
\end{align*}
\caption{The decomposition for $\Iif$-$\Ithen$-$\Ielse$ expressions}
\label{fig:iftc}
\figrule
\end{figure}
Note: $A_{0}=A_{0}\,\zeta$ since $f$ cannot appear in
$A_{0}$.  By the induction hypothesis applied to $A_{1}$ and
$A_{2}$, $\semtime{A_{i}}\,\varrho \leq \semtime{
A_{i}\,\zeta} \,\varrho \;\uplus\; (\semtime{f}
\aptime\vec{t}\; )\,\varrho$ for $i=1,2$.  Thus we have the
chain of bounds of Figure~\ref{fig:iftc}.
\qed

\begin{scholium}\label{sch:decomp}
  As demonstrated in \cite{Danner:Royer:twoalg}, 
  handling forms of recursion beyond tail recursion
  requires  notions of decomposition more sophisticated than
  \refeq{e:linbnd}.  Moreover, if explicit $\lollipop$-types 
  were added to  $\atr$, then the decomposition also 
  becomes more involved than \refeq{e:linbnd}. 
\end{scholium}

For the analysis of $\crec$ expressions we need the following 
corollary to Theorem~\ref{t:lin:time}.  We leave its proof
to the reader who should be mindful of Remark~\ref{rem:handwave}
above.

\begin{corollary}\label{c:patch}
  Suppose $\Gamma;f\of\gamma\entails A\of\gamma$, 
  where 
  $\gamma=(\Natl{\ell_{1}},\dots,\Natl{\ell_k})\allowbreak
  \to\Natl{\ell_0}\in\mathcal{R}$,
  $A=\lam{u_1,\dots,u_k}B$, \  
  $\TailPos(f,A)$, 
  $\Gamma(x_1)=\Natl{\ell_1},\dots,\Gamma(x_k)=\Natl{\ell_k}$, and  
  $\zeta$ is as before. 
  Then
  $\semtime{(A\;\vec{x})}$
  $\leq$
  $\semtime{ (A\;\vec{x})\,\zeta}$ $\uplus$
  $(\semtime{f} \aptime \vec{t}\;)$,
  where
  $(f\;e_{1}^{1}\;\dots\;e_{k}^{1}),\ldots, 
  (f\;e_{1}^{m}\;\dots\;\allowbreak e_{k}^{m})$ 
  are the full applications  of $f$ occurring in 
  $B$ and
  $t_{j} = (\bigmax_{i=1}^{m}\val(\semtime{e_{j}^{i}}))[\vec{u}\gets\vec{x}]$
  for $j=1,\ldots,k$.
\end{corollary}

\section{The time-complexity interpretation of $\atr$}
\label{S:tcplus}

We are now in a position to consider the time complexity properties
of $\crec$ expressions. Remark~\ref{r:lbrec} below motivates the
$\Time$-interpretation of $\crec$ expressions given in
Definition~\ref{d:lbrec:time}. The remark's analysis will be reused
in establishing soundness and polynomial time-boundedness for $\atr$.

\begin{remark}
\label{r:lbrec}
Suppose $\Gamma;f\of\gamma\entails A\of\gamma$, where $\gamma =
(\vec{\bb})\to\bb_{0}\in\mathcal{R}$ and $\TailPos(f,A)$.  
For each $a\in\nat$, let $e_{a} = (\crec \;a\; (\lamr{f}A))$.
Thus, $\Gamma;\emptycont\entails e_{a} \of 
\gamma$.  Our goal is to express $\semtime{e_{a}}$ in terms of 
$\semtime{e_{\bo\concat a}}$ so as to later
extract recurrences, the solutions of which
will provide a closed form polynomial time-bound for $e_{a}$.  So
suppose in the following that $\semtime{e_{\bo\concat a}}$ has a
settled value and that $\Time$-soundness holds for all proper
subterms of $e_{a}$ and their expansions below.  
In a CEK evaluation
of $e_{a}$, in one step $e_a$ is rewritten to
$\lam{\vec{x}}B_a$, where
\begin{gather*}
  B_{a} \Verb{=} 
    (\Iif\,|a|\leq |x_{1}|\,\Ithen\,C_a\,
     \Ielse\, \epsilon) \Verb{ \hbox{ and }}
  C_a \Verb{=} (A\;\vec{x}) [f\gets e_{\bo\concat a}].   
\end{gather*}      
Let $\barGamma=\Gamma,\vec{x}\of\vec{\bb}$.  So,
$\barGamma;f\of\gamma\entails B_{a} \of\bb_{0}$ and
$\barGamma;f\of\gamma\entails C_a\of\bb_{0}$.  
Fix a CEK-environment $\hat\rho$ and a 
$\varrho\in\semtime{\barGamma;f\of\gamma}$ with $\hat\rho
\sqsubseteq \varrho$.  From Figure~\ref{fig:cek} and 
Definition~\ref{d:cekcosts} it follows that
\begin{align}\label{e:ba:cost1}
\cekcost(B_{a}, \hat\rho) &\Verb{\leq}
  \tally2\cdot|\hat\rho(x_{1})| +\tally2\cdot |a| + \tally5 +\;
     \begin{cases}
      \cekcost(C_a,\hat\rho),
         & \hbox{if $|a| \leq |\hat\rho(x_{1})|$;}\\[1ex]
      \tally1, 
         & \hbox{otherwise.}
     \end{cases}  
\end{align}
By our $\Time$-soundness assumptions, 
\begin{gather}\label{e:ca:snd1}
  C_a \Verb{\tcapprox{\bb_0}} \semtime{C_a}.
\end{gather}
Let $\zeta$ be the substitution $[f\gets \epsilon_{\gamma}]$.  By
Corollary~\ref{c:patch} applied to $(A\;\vec{x})$: $
  \semtime{(A\;\vec{x})}
  \leq
  \semtime{(A\;\vec{x})\,\zeta} 
       \;\uplus\;  
       (\semtime{f}\aptime \vec{t}\, )$,
where $t_1,\dots,t_k$ are as in Theorem~\ref{t:lin:time}. 
Let $\xi$ be the substitution $[f\gets e_{\bo\concat a}]$.
Since  $f$ has no occurrence in 
$\vec{t}$, we have that 
$
  \semtime{(A\;\vec{x})\,\xi}\allowbreak
  \leq
  \semtime{(A\;\vec{x})\,\zeta\,\xi} 
       \;\uplus\;  
       (\semtime{f\,\xi}\aptime \vec{t}\, )
$
which can be restated as:
\begin{gather}
  \semtime{C_a} \label{e:catime:bnd}
  \Verb{\leq}
  \semtime{(A\;\vec{x})\,\zeta}
       \;\uplus\;  
       (\semtime{e_{\bo\concat a}} \aptime \vec{t}\, ). 
\end{gather}
Since $\hat\rho\sqsubseteq \varrho$, \ 
$|\hat\rho(x_{1})| \leq
\potprj(\semtime{x_{1}}\varrho)$.  So, 
by Lemma~\ref{l:soundness:atr-}(c), \refeq{e:ba:cost1},
\refeq{e:ca:snd1},  and \refeq{e:catime:bnd}, 
$B_a\hat\rho\tcapprox{\bb_0} X_a\varrho$, 
where $X_{a}\of \semtime{\barGamma;f\of\gamma}\to
\semtime{\bb_{0}}$ is given by
\begin{align*}
  X_a \,\varrho'
    &\Verb{=} \
     \begin{cases}
       \dally(\tally{c},\semtime{(A\;\vec{x})\,\zeta}\,\varrho')
          \;\uplus\;    
       (\semtime{e_{\bo\concat a}}
       \aptime\vec{t}\, )\,\varrho',
           & \hbox{if $|a|\leq p_{1}$;}\\[1ex]
       (\tally{c+1},\,\tally0), 
         & \hbox{otherwise};\Quad2
     \end{cases}\\[0.5ex]
     & \notag
     \Quad2
     \hbox{where 
     $p_{1} = \Pot(\semtime{x_{1}}\,\varrho')$,
     $c = 2\cdot p_{1} + 2\cdot|a|+5$, and 
     $\vec{t}$ is as before.}
\end{align*}
By the analysis for the $\to$-\emph{I} case in
Theorem~\ref{t:ptimebnd-}'s proof,  
$(\lam{\vec{x}}B_a)\tcapprox{\gamma} 
\currytime(\vec{x},X_{a}) $.
As 
$\cekcost(\lam{\vec{x}}B_a,\underline{\ })=1$, 
we have that $e_a\tcapprox{\gamma} Y_a$, where 
$Y_{a} \;\defeq\;
\dally(1,\currytime(\vec{x},X_{a}))$.
\end{remark}

\begin{definition}[The $\Time$-interpretation of $\atr$] 
\label{d:lbrec:time} 
$\semtime{ \Gamma;\emptycont \entails (\crec\;a\;
   (\lamr{f}A))\of\gamma} \allowbreak \defeq  Y_{a}$, 
where $Y_{a}$ is as above.  Figure~\ref{fig:ltr:time} provides the
the $\Time$-interpretations for the other $\atr$ constructs.
\end{definition}

The well-definedness of 
 $\semtime{ \Gamma;\emptycont \entails (\crec\;a\;   (\lamr{f}A))\of\gamma}$ 
is part of:

\begin{theorem} \label{t:ptimebnd}
  The $\Time$-interpretation of $\atr$ is
  (a) poly\-nomial time-bounded,
  (b) monotone, and
  (c) sound, as well as 
  (d) well-defined.
\end{theorem}

\ProoF{sketch}
All the parts are shown simultaneously by a structural
induction on the derivation of $\Gamma;\Delta\entails e\of \gamma$.
Along with parts (a)--(d)  we also show:

\emph{Claim:} For all $\varrho \in\semtime{\Gamma;\Delta}$, \
$\basepot(\semtime{e}\,\varrho) \leq
\semtime{\widetilde{p_e}}\,\varrho$, where $p_e$ is the 
polynomial size-bound for $e$ from Theorem~\ref{t:polybnd}
and $\widetilde{p_e}$ is the result of replacing each 
\emph{occurrence} of each variable $|x|$ in $p_e$ with $\basepot(x)$.  
(E.g., if $p=\lam{|z|}(\tally2*|g|(|z|)+\tally1)$, then 
$\widetilde{p_e}= \lam{|z|}(\tally2*\basepot(g)(\basepot(|z|))+\tally1)$.)  

Intuitively, $\widetilde{p_e}$ is the version of $p_e$ that is
over base potentials (Definition~\ref{d:bigproj}(b))
instead of lengths and the Claim says that the upper bound on
size that is implicit in our $\Time$-interpretation, is at least
as good as the size bounds of Theorem~\ref{t:polybnd}.
Through the Claim  we are able to make use, 
in a time-complexity context, of the 
polynomial bound on the depth of $\crec$-recursions from 
the proof of Theorem~\ref{t:polybnd}. 

Here, then, is the induction.

For each case, except the $\crec$ one,  parts 
(a), (b), and (c) are as in the proof of Theorem~\ref{t:ptimebnd-};
part (d) is evident, and the Claim  follows from an inspection
of the bounds assigned in the proof of Theorem~\ref{t:polybnd}
and Definition~\ref{d:semtime:1}.
We thus consider the case of $e=(\crec\;a\, (\lamr{f} A))$ where
$\Gamma;f\of\gamma\entails A\of\gamma$,\ $\gamma = (\sigma_{1},
\dots, \sigma_{k}) \to \bb_{0} \in \mathcal{R}$, and
$\TailPos(f,A)$.  Without loss of generality, we assume $a$ is a
tally string $\tally{n}$.  So, $\bo\concat a= \tally{n+1}$.

We first import the notation from Remark~\ref{r:lbrec}.  So,
$e=e_{\tally{n}}$, where $e_{\tally{n}}$ is as in 
Remark~\ref{r:lbrec} with $a=\tally{n}$. 
Also let $\longvec{\semtime{x}}$ denote
$\semtime{x_1},\ldots,\semtime{x_k}$, $\varrho
\in\semtime{\barGamma; \emptycont}$, and let $m$ range over
$\set{n,n+1,\dots}$. 
Then, by Remark~\ref{r:lbrec}, Definition~\ref{d:lbrec:time},
and   Lemma~\ref{l:curry} we have:
$( \semtime{e_{\tally{m}}}\aptime \longvec{\semtime{x}})\varrho$
=
$(Y_{\tally{m}}\, \varrho ) \aptime \longvec{\varrho(x)}$
=
$(\dally(1,\currytime(\vec{x},X_{\tally{m}}) )\varrho)
    \aptime \longvec{\varrho(x)}$
=
$\semtime{r_{0}}\,\varrho \;\uplus \; X_{\tally{m}}\varrho$,
where $r_0 = (\tally{5k+4} + \costprj(\varrho(x_1))+\dots+
\costprj(\varrho(x_k)),\tally0)$.
Let 
$r_{1,m}=r_{0}\uplus(\tally2\cdot 
\potprj(\varrho(x_{1})) + \tally2 \cdot \tally{m} +
\tally6,\tally0)$ and 
$r_{2,m}=r_{0}\uplus(\tally2\cdot \potprj(\varrho(x_{1})) + \tally2 \cdot \tally{m} + 
\tally5,\tally0)$.
Then, by the definition of $X_{\tally{n}}$ in Remark~\ref{r:lbrec},
\begin{gather} \label{e:recur}
( \semtime{e_{\tally{m}}}\aptime \longvec{\semtime{x}})\varrho
\Verb{=} 
    \begin{cases}
      \semtime{r_{1,m}}\,\varrho, 
       \Quad{9}
        &
        \hbox{if 
           $\potprj(\varrho(x_1))\leq \tally{n}$;}
           \\[1ex]
      \lefteqn{\semtime{r_{2,m}}\,\varrho
        \;\uplus\;
        \semtime{(A\,\vec{x})\zeta}\varrho
        \;\uplus\;  (\semtime{e_{\tally{m+1}}} \aptime \vec{t}\,) 
            \varrho,} \\
        & 
        \hbox{otherwise.}
      \end{cases}  
\end{gather}

Now let us import some notation from the proof of
Theorem~\ref{t:polybnd}: Let $p_{1},\dots, p_{k}$ be the 
manifestly safe polynomials
that bound the sizes of the arguments of $f$ in $A$ and let
$p_{1}',\ldots,p_{k}'$ be the polynomials that bound the
\emph{final} sizes of said arguments.

\emph{Part (d): Well-definedness.}  
Let $\varrho_n=\varrho$. 
Combine the $m=n$ and $m=n+1$ versions of \refeq{e:recur} to 
express $( \semtime{e_{\tally{n}\,}}\aptime \longvec{\semtime{x}})\varrho_n$
in terms of $\semtime{e_{\tally{n+2}}}$ and
$\varrho_{n+1}$ = the
update to $\varrho_n$ produced by
the application $(\semtime{e_{\tally{n+1}}} \aptime \vec{t}\,) 
\varrho_n$.
It follows from the Claim that
$\potprj(\varrho_{n+1}(x_1)) \leq \semtime{\widetilde{p_1'}}\varrho$.
We can keep repeating this process, for $m=n+2,n+3,\dots$, 
to 
express $( \semtime{e_{\tally{n}\,}}\aptime \longvec{\semtime{x}})\varrho_n$
in terms of of $\semtime{e_{\tally{m+1}}}$ and
$\varrho_{m}$ = the update to $\varrho_{m-1}$ produced by
the application $(\semtime{e_{\tally{m}\,}} \aptime \vec{t}\,) 
\varrho_{m-1}$.  The Claim still tells us that 
$\potprj(\varrho_{n+1}(x_1)) \leq \semtime{\widetilde{p_1'}}\varrho$.
Hence, the otherwise clause of \refeq{e:recur} can
hold only finitely many $m$.
Thus, it follows that, 
$\semtime{e_{\tally{n}\,}}$ is defined and total.
\smallskip

\emph{Part (b): Monotonicity.} Note that the terms 
$\semlenwt{r_{1,m}}$ and $\semlenwt{r_{2,m}}$ clearly satisfy 
monotonicity.  It follows from the induction hypothesis
that  the terms 
$\semtime{(A\vec{x})\zeta}$ and $t_{1},\dots,t_{k}$
also satisfy monotonicity.  It follows from \refeq{e:recur} that if,
for a particular $m$, the $\semtime{e_{\tally{m+1}}}$ term satisfies
monotonicity, then so does $\semtime{e_{\tally{m}\,}}$.  
Hence, by the finiteness of the expansion it
follows that $\semtime{e_{\tally{n}\,}}$
satisfies monotonicity.

\smallskip 

\emph{Part (c) and the Claim.}  By arguments along the 
lines of the one just given for monotonicity, one can establish 
soundness and the Claim for $e_{\tally{n}}$.

\smallskip
\emph{Part (a): Polynomial time-boundedness.}  Recall from
Definition~\ref{d:polytimebnd}, the definition of 
polynomial time-boundedness, the key inequality to be shown is
$\semtime{e}\tcx{\rho} \,\leq\, \semlenwt{p_e}\,|\rho|$
for each $\rho\in\semvalwt{\Gamma;\Delta}$.
So if $\rho$ is an 
$\atr$-en\-viron\-ment and 
$x$ is a variable
with a string or oracle value, then $\semtime{\basepot(x)}\tcx{\rho}
= \semlenwt{|x|}|\rho|$.  Thus, for each $e'$, \ 
$\semtime{\widetilde{p_{e'}}}
\tcx{\rho} = \semlenwt{p_{e'}} |\rho|$. 

Now, it follows from the induction 
hypothesis that there
is an $\barr$, a polynomial time-bound for $(A\,\vec{x})\zeta$
relative to $\Gamma;\emptycont$.  Let $r_1'=r_{1,p_1'}$ and 
$r_2'=r_{2,p_1'}$.  
Let $\xi$ and $\xi'$ respectively
denote the substitutions $[|x_{1}|\gets p_{1},\dots,|x_{k}| \gets
p_{k}]$ and $[|x_{1}|\gets p_{1}',\dots,|x_{k}|\gets p_{k}']$, where
$p_{1},\dots,p_{k}, p_{1}',\dots,p_{k}'$ are the polynomials from
Theorem~\ref{t:polybnd} introduced before.  Note that $\tcx{x_i} \xi
= (\tally1\bmax p_i,p_i)$.  By the Claim, for each $j =
1,\dots,k$, $ \semtime{t_{j}} \tcx{\barrho} \leq \semlenwt{(\tally1
\bmax p_{j},p_{j})}\, |\barrho| = \semlenwt{(\tcx{x_{j}}
\xi}\,|\barrho|$.  Hence, assuming 
$\potprj(\semlenwt{|x_{1}|}\,|\barrho|) >
\tally{n}$,
\begin{align*}
   (\semtime{e_{\tally{n+1}}} \aptime \vec{t}) \tcx{\barrho} 
   &\Verb{\leq}
     (\semtime{e_{\tally{n+1}}} \aptime 
     (\longvec{\tcx{x}}\xi) ) \tcx{\barrho} 
     && \hbox{(by monotonicity)}\\
   &\Verb{\leq}
     \semlenwt{(r_{2}'\uplus\barr)\xi}\,|\barrho| \uplus
     (\semlen{e_{\tally{n+2}}}\aptime (\vec{t} \,\xi) )\tcx{\barrho} 
     && \hbox{(by \refeq{e:recur})} \\
   &\Verb{\leq}
     \semlenwt{(r_{2,m}\uplus\barr)\xi}\,|\barrho| \uplus
     (\semlen{e_{\tally{n+2}}}\aptime (\vec{t} \,\xi) )\tcx{\barrho} 
     && \hbox{(by monotonicity)}.
\end{align*}
Clearly, we can repeat the above expansion $(p_1 - n)$-many times
(i.e., until termination), collect terms, and produce the desired
polynomial bound.  Here is the algebra.  Let
   $s = r_{1}'\xi' \,\uplus \allowbreak \;
    {\textstyle  \biguplus_{m=0}^{\,p_{1}'-n}
    (r_{2}'\,\uplus\, \barr)\xi^{(m)}}$,
   $s_{1} = \Cost(r_{1}'\xi') + 
          p_{1}'\cdot \left(
             \Cost(r_{2}'\xi') + 
             \Cost(\barr\xi')
          \right)$, and 
   $s_{2} = \Pot(\, (r_1'\bmax r_2'\bmax \barr)\xi'\,)$.
Then
$
( \semtime{e_{\tally{n}}} \aptime \longvec{\tcx{x}})\,
     \tcx{\barrho} 
     \leq  \semlenwt{s}\,|\barrho| 
     \leq  \semlenwt{(s_{1},s_{2})}\,|\barrho|$
by a straightforward argument.
Thus, $\lam{\longvec{|x|}}(s_{1},s_{2})$ suffices as the polynomial
time bound for $e_{\tally{n}}$.
\qed

\begin{scholium} \label{sch:whyCEKcost}
  Note that we resorted to reasoning directly about CEK-costs
  to obtain \refeq{e:ba:cost1}.  This is because if we had used 
  Definition~\ref{d:semtime:1}'s
  $\Time$-interpretation of $\Iif$-$\Ithen$-$\Ielse$, 
  then we would have be left 
  without a base case in our recursive unfoldings
  of $\crec$-expressions.
\end{scholium}

We note that as a consequence of parts (a) and (c) of 
Theorem~\ref{t:ptimebnd} we have:

\begin{corollary}\label{c:ptbnd}
  For each $\Gamma;\Delta\entails e\of\gamma$, 
  there is a second-order polynomial $q_e$ with
  $|\Gamma;\Delta|\entails q_e\of|\gamma|$ such that
  $\cekcost(e,\rho) \leq \semlenwt{q_e}|\rho|$
  for each $\rho\in\semvalwt{\Gamma;\Delta}$.  
\end{corollary}

\begin{remark}[Related work]\label{rem:costPot}
The time-complexity cost/potential distinction  
appears in prior work.  A version of this distinction 
can be found in Sands' Ph.D.~thesis \cite{sands:thesis}. Shultis 
\cite{Shultis} sketched how to use the distinction in order to 
give time-complexity semantics for reasoning about the run-time 
programs that involve higher types.  Van~Stone \cite{VanStone} 
gives a much more detailed and sophisticated semantics 
for a variant of $\PCF$ using the cost/potential distinction.  
Very roughly, Shultis and Van Stone were focused on giving static 
analyses to extract time-bounds for functional programs that 
compute first-order functions. 
The time-complexity semantics of this paper was developed 
independently of Shultis' and Van~Stone's work.
\nocite{Gurr:th:90}
We also note that Benzinger's work \cite{benzinger:01,benzinger04}
on automatically inferring the complexity of Nuprl programs 
made extensive use of higher-type recurrence equations.
\end{remark}

\section{Complexity-theoretic completeness}\label{S:comp}

Our final result on $\atr$ is that each type-1 and type-2 BFF is
$\atr$ computable.
\emph{Conventions:} In this section, let $\sigma =
(\sigma_1,\dots,\sigma_k)\to\Nat$ range over simple types over
$\Nat$ of levels 1 or 2, and let $\gamma$, $\gamma_0$,
$\gamma_1,\dots$ range over $\atr$ types.
Recall from \S\ref{S:defs:bff} that $f\in\semval{\sigma}$ is
\emph{basic feasible} when there is a closed type-$\sigma$,
$\PCF$-expression $e_f$ and a second-order polynomial function $q_f$ such
that $\semval{e_f}=f$ and, for all $v_i \in \semval{\sigma_1},
\dots, v_k \in \semval{\sigma_k}$, \ $\cektime(e_f,v_1,\allowbreak
\dots,\allowbreak v_k) \;\leq\; q_f(|v_1|,\dots,|v_k|)$.  Let
$\BFF_\sigma$ = the class of all type-$\sigma$ BFFs.

\begin{definition} 
  We say that each base type is \emph{unhindered} and that
  $(\gamma_1,\dots,\gamma_k)\to\Natl{\ell}$ is \emph{unhindered} when
  $(\gamma_1,\dots,\gamma_k)\to\Natl{\ell}$ is strict, predicative and
  each $\gamma_i$ unhindered.
\end{definition}

Note that $\semvalwt{\gamma} = \semval{\shape(\gamma)}$ if and only
if $\gamma$ is unhindered.

\begin{theorem}\label{t:bff}
  $\BFF_{\sigma} = \{\,\semvalwt{\entails e\of\gamma}\suchthat
  \sigma=\shape(\gamma)$ $\&$ $\gamma \hbox{ is unhindered}\,\}$ for
  each $\sigma$.
\end{theorem}

\begin{proof}
Fix $\sigma$ and let $\SU_\sigma = \{\,\semvalwt{\entails
e\of\gamma}\suchthat \sigma=\shape(\gamma)$ $\&$ $\gamma \hbox{
is unhindered}\,\}$.

\emph{Claim 1:} $\SU_\sigma\subseteq \BFF_\sigma$.  
\emph{Proof:}
It is straightforward to express a $\crec$-recursion with $\PCF$'s
$\fix$-construct with only polynomially-much over head on the cost
of the simulation.  Hence, the claim follows from
Theorem~\ref{t:ptimebnd}.

\emph{Claim 2:} $\BFF_\sigma\subseteq \SU_\sigma$.
\emph{Proof:}
Kapron and Cook \cite{KapronCook:mach} showed that the type-2 basic
feasible functionals are characterized by the functions computable
in second-order polynomial time-bounded oracle Turing machines
(OTMs).  Proposition 18 from \cite{IKR:I} shows how to simulate any
second-order polynomial time-bounded oracle Turing machine using
that paper's $\textsf{ITLP}_2$ programming formalism.  That
simulation is easily adapted to $\atr$.  Hence, the claim follows.
\end{proof}

\emph{Note:} The proof's two claims are constructive in that:
(i)
    given a closed $\atr$-expression $e$ of unhindered type, one 
    can construct an equivalent $\PCF$ expression $e'$
    and a second-order
    polynomial $p_e$ that bounds the run time of $e'$, and 
(ii)
    given an OTM $\BM$ and a second-order polynomial $p$ that
    bounds the run time of $\BM$, one can construct
    an $\atr$-expression that computes the same function 
    as $\BM$. 
    
Claim 2 can be extended beyond unhindered types as follows. 
For each $\atr$ arrow-type $\gamma= (\gamma_1,\dots,\gamma_k)
\to\Natl{\ell}$, and each type-$\shape(\gamma)$ OTM $\BM$,
we say that $\BM$ computes a $\BFF_\gamma$-function
when there is a type-$|\gamma|$ polynomial $p$
such that the run time of $\BM$ on $(\vec{v})$ is
bounded by $p(|v_1|,\dots,|v_k|)$.  The proof of Claim 2 
lifts to show: \emph{for all $\atr$ arrow-types $\gamma$,
each $\BFF_\gamma$-function is $\atr$ computable.}

\section{Conclusions}\label{S:finis}

$\atr$ is a small functional language, based on $\PCF$, which has
the property that each $\atr$ program has a second-order
polynomial time-bound.  The $\atr$-computable functions include the
basic feasible functionals at type-levels~1 and~2.  However, the
$\atr$-computable functions contain other functions, such as 
$\mathit{prn}$,
that are \emph{not} basic feasible in the original sense of Cook and
Urquhart \cite{CookUrquhart:feasConstrArith}.  $\atr$ is able to
express such functions thanks to its type system and supporting
semantics that work together to control growth rates and time
complexities.  Without some such controls feasible recursion schemes,
such as $\mathit{prn}$, cannot be first-class objects of a 
programming language.  

The $\atr$ type-system and semantics were crafted so that 
$\atr$'s complexity properties could be established through
adaptations of standard tools
for the analysis of conventional programming languages (e.g.,
intuitionistic and affine types, denotational semantics for $\atr$
and its time complexity, and an abstract machine that provides both
an operational semantics for $\atr$ and a basis for the
time-complexity semantics).  As $\atr$ is based on $\PCF$ (a
theoretical first-cousin of both ML and Haskell), our results
suggest that one might be able to craft ``feasible sublanguages'' of
ML and Haskell that are both theoretically well-supported and
tolerable for programmers.

\medskip

$\atr$ and its semantic and analytic frameworks are certainly not
the final word on any issue.  Here we discuss several possible
extensions of our work.

\topic{More general recursions} 
In \cite{Danner:Royer:twoalg} we consider an expansion of $\atr$ 
that allows a fairly wide range of affine (one-use) recursions. 
In particular, the expanded $\atr$ can fairly naturally express
the classic insertion- and selection-sort algorithms.
Handling this larger set of recursions requires some  
nontrivial extensions of our framework for analyzing time-complexities. 

Dealing with nonlinear recursions (e.g., the standard quicksort
algorithm) is trickier to handle because there must be independent
clocks on each branch of the recursion that together guarantee
certain global upper bounds.  

\topic{Recursions with type-level 1 parameters} 
Another possible extension of $\atr$ would be
to allow type-level 1 parameters in
$\crec$-recursions so that, for example, one could give a
continuation-passing-style definition of \emph{prn}. Because type-1
parameters in recursions act to recursively define functions, these
parameters must be affinely restricted just like principle recursor
variables of $\crec$-expressions. Consequently, such an extension
must also include explicit $\lollipop$-types to restrict these
parameters.  However, along with the $\lollipop$-types come
(explicitly or implicitly) tensor-products and these cause problems
in analyzing $\crec$-recursions (e.g., one is forced account for all
the possible interactions of the affine parameters in the course of
a recursion and so the na\"{\i}ve ``polynomial'' time-bounds are
exponential in size).

\topic{Lazy evaluation} For a lazy (e.g., call-by-need) version
of $\atr$, one would need to: (i) construct an abstract machine for
this lazy-$\atr$, (ii) modify the $\ST$-semantics a bit to
accommodate the lazy constructs; and (iii) rework the
$\ST$-interpretation of $\atr$ which would then have to be shown
monotone, sound, and constructively polynomial time-bounded.
(Since the well-tempered semantics is extensional, it requires
very few changes for a lazy-$\atr$.)  If our lazy-$\atr$ allowed
infinite strings, then the $\Valwt$-semantics would also have to
be modified.  Note that Sands \cite{sands:thesis} and Van Stone
\cite{VanStone} both consider lazy evaluation in their work.

\topic{Lists and streams} 
There are multiple senses of the ``size'' of a list.  For example,
the run-time of \emph{reverse} should depend on just a list's
length, whereas the run-time of a search depends on both the list's
length and the sizes of the list's elements.  Any useful extension
of $\atr$ that includes lists needs to account for these multiple
senses of size in the type system and the well-tempered and
time-complexity semantics.  If lists are combined with laziness,
then we also have the problem of handling infinite lists.  However,
$\atr$ and its semantics already handle one flavor of infinite
object, i.e., type-level 1 inputs, so handling a second flavor of
infinite object many not be too hard.

\topic{Type checking, type inference, time-bound inference} 
We have not studied the problem of $\atr$ type checking.  But since
$\atr$ is just an applied simply typed lambda calculus with
subtyping, standard type-checking tools should suffice.  Type
inference is a much more interesting problem.  We suspect that a
useful type inference algorithm could be based on Frederiksen and
Jones' \cite{frederiksen-jones} work on applying size-change
analysis to detect whether programs run in polynomial time.  Another
interesting problem would be to start with a well-typed $\atr$
program and then extract reasonably tight size and time bounds
(as opposed to the not-so-tight bounds given by
Theorem~\ref{t:ptimebnd}).

\topic{Beyond type-level 2} 
There are semantic and comp\-lex\-ity-theoretic issues to be
resolved in order to extend the semantics of $\atr$ to type-levels 3
and above.  The key problem is that our definition of the length of
a type-2 function \refeq{e:flen2} does not generalize to type-level
3.  This is because for  
$\Psi\in\MC_{((\Nat\to\Nat)\to\Nat)\to\Nat}$ and 
$G\in \MC_{(\Nat\to\Nat)\to\Nat}$, 
we can have $\sup \set{ |\Psi(F)| \suchthat |F|\leq |G|} = \infty$,
even when $G$ is 0--1 valued.  To fix this problem one can introduce a
different notion of length that incorporates information about a
function's modulus of continuity.  It appears that $\atr$ and the
$\Valwt$- and $\ST$-semantics extend to this new setting.  However,
it also appears that this new notion of length gives us a new notion
of higher-type feasibility that goes beyond the BFFs.  Sorting out
what is going on here should be the source of other adventures.

{\small \bibliography{ops} }

\providecommand{\bysame}{\leavevmode\hbox to3em{\hrulefill}\thinspace}
\providecommand{\MR}{\relax\ifhmode\unskip\space\fi MR }
\providecommand{\MRhref}[2]{%
  \href{http://www.ams.org/mathscinet-getitem?mr=#1}{#2}
}
\providecommand{\href}[2]{#2}
\begin{thebibliography}{O'{H}03}

\bibitem[Bar96]{barber:dill:96}
A.~Barber, \emph{Dual intuitionistic linear logic}, Tech. report, LFCS, Univ of
  Edinburgh, 1996.

\bibitem[BC92]{BellantoniCook}
S.~Bellantoni and S.~Cook, \emph{A new recursion-theoretic characterization of
  the polytime functions}, Computational {C}omplexity \textbf{2} (1992),
  97--110.

\bibitem[Ben01]{benzinger:01}
R.~Benzinger, \emph{Automated complexity analysis of {N}uprl extracted
  programs}, Journal of Functional Programming \textbf{11} (2001), 3--31.

\bibitem[Ben04]{benzinger04}
\bysame, \emph{Automated higher-order complexity analysis}, Theoretical
  {C}omputer {S}cience \textbf{318} (2004), 79--103.

\bibitem[BNS00]{BNS}
S.~Bellantoni, K.-H. Niggl, and H.~Schwichtenberg, \emph{Characterising
  polytime through higher type recursion}, Annals of {P}ure and {A}pplied
  {L}ogic \textbf{104} (2000), 17--30.

\bibitem[BP97]{bp:dill:97}
A.~Barber and G.~Plotkin, \emph{Dual intuitionistic linear logic}, Tech.
  report, LFCS, Univ of Edinburgh, 1997.

\bibitem[CK90]{CookKapron:FM}
S.~Cook and B.~Kapron, \emph{Characterizations of the basic feasible functions
  of finite type}, Feasible {M}athematics: {A} {M}athematical {S}ciences
  {I}nstitute {W}orkshop (S.~Buss and P.~Scott, eds.), Birkh\"auser, 1990,
  pp.~71--95.

\bibitem[Cob65]{Cobham65}
A.~Cobham, \emph{The intrinsic computational difficulty of functions},
  Proceedings of the {I}nternational {C}onference on {L}ogic, {M}ethodology and
  {P}hilosophy (Y.~{Bar Hillel}, ed.), North-Holland, 1965, pp.~24--30.

\bibitem[CU93]{CookUrquhart:feasConstrArith}
S.~Cook and A.~Urquhart, \emph{Functional interpretations of feasibly
  constructive arithmetic}, Annals of {P}ure and {A}pplied {L}ogic \textbf{63}
  (1993), 103--200.

\bibitem[DR06]{DR:ATS:Popl}
N.~Danner and J.~Royer, \emph{Adventures in time and space}, 33th ACM Symposium
  on Principles of Programming Languages (S.~Peyton~Jones, ed.), {ACM} Press,
  2006, pp.~168--179.

\bibitem[DR07]{Danner:Royer:twoalg}
\bysame, \emph{Time-complexity semantics for feasible affine recursions},
  Computation and Logic in the Real World: Third Conference of Computability in
  Europe, CiE 2007 (S.B. Cooper, B.~L\"{o}we, and A.~Sorbi, eds.), Lecture
  Notes in Computer Science, vol. 4497, Springer-Verlag, 2007, to appear.

\bibitem[FF87]{FF87}
M.~Felleisen and D.~Friedman, \emph{Control operators, the {SECD}-machine, and
  the lambda calculus}, Formal Descriptions of Programming Concepts {III},
  1987, pp.~193--217.

\bibitem[FF06]{FF03}
M.~Felleisen and M.~Flatt, \emph{Programming languages and lambda calculi},
  unpublished manuscript, 2006.

\bibitem[FJ04]{frederiksen-jones}
C.~Frederiksen and N.~Jones, \emph{Recognition of polynomial-time programs},
  Tech. Report TOPPS/D-501, DIKU, University of Copenhagen, 2004.

\bibitem[FWH01]{EOPL:2}
D.~Friedman, M.~Wand, and C.~Haynes, \emph{Essentials of programming
  langauges}, second ed., MIT Press, 2001.

\bibitem[Gol01]{Goldreich:I}
O.~Goldreich, \emph{Foundations of cryptography, {V}ol.~{I}: {B}asic tools},
  Cambridge University Press, 2001.

\bibitem[Gur90]{Gurr:th:90}
D.~J. Gurr, \emph{Semantic frameworks for complexity}, Ph.D. thesis, University
  of Edinburgh, 1990.

\bibitem[Hof00]{Hofmann:survey}
M.~Hofmann, \emph{Programming languages capturing complexity classes}, SIG\-ACT
  News \textbf{31} (2000), 31--42.

\bibitem[Hof02]{Hofmann:strength}
\bysame, \emph{The strength of non-size increasing computation}, 29th {ACM}
  {S}ymposium on {P}rinciples of {P}rogramming {L}anguages (J.~Michell, ed.),
  {ACM} Press, 2002, pp.~260--269.

\bibitem[Hof03]{Hofmann03}
\bysame, \emph{Linear types and non-size increasing polynomial time
  computation}, Information and {C}omputation \textbf{183} (2003), 57--85.

\bibitem[IKR01]{IKR:I}
R.~Irwin, B.~Kapron, and J.~Royer, \emph{On characterizations of the basic
  feasible functionals, {P}art {I}}, Journal of Functional Programming
  \textbf{11} (2001), 117--153.

\bibitem[IKR02]{IKR:II}
\bysame, \emph{On characterizations of the basic feasible functionals, {P}art
  {II}}, unpublished manuscript, 2002.

\bibitem[Kap91]{Kapron:thesis}
B.~Kapron, \emph{Feasible computation in higher types}, Ph.D. thesis,
  Department of Computer Science, University of Toronto, 1991.

\bibitem[KC96]{KapronCook:mach}
B.~Kapron and S.~Cook, \emph{A new characterization of type 2 feasibility},
  {SIAM} {J}ournal on {C}omputing \textbf{25} (1996), 117--132.

\bibitem[KU58]{KolmUspenAlg}
A.N. Kolmogorov and V.A. Uspenskii, \emph{On the definition of an algorithm},
  Uspekhi Mat. Nauk \textbf{13} (1958), 2--28.

\bibitem[Lei94]{Leivant:poly}
D.~Leivant, \emph{A foundational delineation of poly-time}, Information and
  {C}omputation \textbf{110} (1994), 391--420.

\bibitem[Lei95]{Leivant:FM2}
\bysame, \emph{Ramified recurrence and computational complexity {I}: {W}ord
  recurrence and poly-time}, Feasible {M}athematics {II} (P.~Clote and
  J.~Remmel, eds.), Birkh\"{a}user, 1995, pp.~320--343.

\bibitem[Lei03]{Leivant03}
\bysame, \emph{Feasible functionals and intersection of ramified types},
  Proceedings of the Second Workshop on Intersection Types and Related Systems,
  Electronic Notes in Theoretical Computer Science, vol.~70, Elsevier Science
  Publishers, 2003, pp.~1--14.

\bibitem[LM93]{LeivantMarion93}
D.~Leivant and J.-Y. Marion, \emph{Lambda calculus characterizations of
  polytime}, Fundament{\ae} Informatic{\ae} \textbf{19} (1993), 167--184.

\bibitem[Lon04]{longley:ubiq}
J.~Longley, \emph{On the ubiquity of certain total type structures ({E}xtended
  abstract)}, Proceedings of the Workshop on Domains {VI} (M.~Escard\'{o} and
  A.~Jung, eds.), Electronic Notes in Theoretical Computer Science, vol.~73,
  Elsevier Science Publishers, 2004, pp.~87--109.

\bibitem[Lon05]{Longley:notions:1}
\bysame, \emph{Notions of computability at higher types {I}}, Logic Colloquium
  2000 (R.~Cori, A.~Razborov, S.~Torcevic, and C.~Wood, eds.), Lecture Notes in
  Logic, vol.~19, A. K. Peters, 2005.

\bibitem[Mar72]{March72}
S.~Marchenkov, \emph{The computable enumerations of families of general
  recursive functions}, Algebra and Logic \textbf{11} (1972), 326--336.

\bibitem[Meh74]{Mehlhorn74}
K.~Mehlhorn, \emph{Polynomial and abstract subrecursive classes}, Proceedings
  of the {S}ixth {A}nnual {ACM} {S}ymposuium on the {T}heory of {C}omputing,
  1974, pp.~96--109.

\bibitem[Meh76]{Mehlhorn76}
\bysame, \emph{Polynomial and abstract subrecursive classes}, Journal of
  {C}omputer and {S}ystem {S}cience \textbf{12} (1976), 147--178.

\bibitem[Nor99]{Normann99}
D.~Normann, \emph{The continuous functionals}, Handbook of Computability Theory
  (E.~R. Griffor, ed.), North-Holland, 1999, pp.~251--275.

\bibitem[O'{H}03]{ohearn:bunched:03}
P.~O'{H}earn, \emph{On bunched typing}, Journal of Functional Programming
  \textbf{13} (2003), 747--796.

\bibitem[Pie02]{Pierce:types}
B.~Pierce, \emph{Types and programming languages}, MIT Press, 2002.

\bibitem[Plo75]{Plotkin75}
G.~Plotkin, \emph{Call-by-name, call-by-value and the $\lambda$-calculus},
  Theoretical {C}omputer {S}cience \textbf{1} (1975), 125--159.

\bibitem[Plo77]{Plotkin:PCF}
\bysame, \emph{{LCF} considered as a programming language}, Theoretical
  {C}omputer {S}cience \textbf{5} (1977), 223--255.

\bibitem[RC94]{RC94}
J.~Royer and J.~Case, \emph{Subrecursive programming systems: {C}omplexity
  {\it\&} succinctness}, Birkh\"{a}user, 1994.

\bibitem[Rey72]{Reynolds72}
J.~Reynolds, \emph{Definitional interpreters for higher-order programming
  languages}, Proceedings of the ACM National Conference, 1972, pp.~717--740.

\bibitem[Rey93]{Reynolds:cont:hist}
J.~Reynolds, \emph{The discoveries of continuations}, Lisp and Symbolic
  Computation \textbf{6} (1993), 233--247.

\bibitem[Rey98]{Reynolds:def2:98}
J.~Reynolds, \emph{Definitional interpreters for higher-order programming
  languages}, Higher-Order and Symbolic Computation \textbf{11} (1998),
  363--397, reprint of \cite{Reynolds72}.

\bibitem[Roy87]{Royer87a}
J.~Royer, \emph{A connotational theory of program structure}, Lecture Notes in
  Computer Science, vol. 273, Springer-Verlag, 1987.

\bibitem[San90]{sands:thesis}
D.~Sands, \emph{Calculi for time analysis of functional programs}, Ph.D.
  thesis, University of London, 1990.

\bibitem[Sch80]{schonhage80}
A.~Sch\"{o}nhage, \emph{Storage modification machines}, {SIAM} {J}ournal on
  {C}omputing \textbf{8} (1980), 490--508.

\bibitem[Sch96]{Schwichtenberg:cont}
H.~Schwichtenberg, \emph{Density and choice for total continuous functionals},
  Kreiseliana (P.~Odifreddi, ed.), A.K. Peters, 1996, pp.~335--362.

\bibitem[Shu85]{Shultis}
J.~Shultis, \emph{On the complexity of higher-order programs}, Tech. Report
  CU-CS-288-85, University of Colorado, Boulder, 1985.

\bibitem[VS03]{VanStone}
K.~Van~Stone, \emph{A denotational approach to measuring complexity in
  functional programs}, Ph.D. thesis, School of Computer Science, Carnegie
  Mellon University, 2003.

\bibitem[Win93]{winskel:book}
G.~Winskel, \emph{Formal semantics}, {MIT} {P}ress, 1993.

\end{thebibliography}

\vskip-55 pt

\end{document}